\documentclass[a4paper,12pt,final]{report}

\usepackage{amsmath,amssymb,graphicx,setspace,fancyhdr}
\usepackage[breaklinks=true]{hyperref}
\usepackage[hyphenbreaks]{breakurl}
\usepackage{soul}
\usepackage{comment}
\usepackage[all]{hypcap}

\newcommand{\be}{\begin{eqnarray}}
\newcommand{\en}{\end{eqnarray}}
\newcommand{\nn}{\nonumber\\}
\newcommand{\ctk}{\chi(t,\mathbf{k})}
\newcommand{\msq}{\alpha^2 v^2}
\newcommand{\noi}{\noindent}

\newcommand{\chir}{\chi_{\rm{R}}}
\newcommand{\phic}{\phi_{\rm{crit}}}
\newcommand{\sigmac}{\sigma_{\rm{crit}}}
\newcommand{\fnl}{f_{\rm{NL}}}
\newcommand{\mpl}{m_{\rm{pl}}}
\newcommand{\ns}{n_{\rm{s}}}

\begin{document}
\thispagestyle{plain}
\begin{titlepage}
\thispagestyle{plain}
\setcounter{page}{1}
\begin{center}
\textsc{\LARGE Cosmological Perturbations From Hybrid Potentials}\\[1cm]
by \\[1cm]
{\large Stefano Orani}\\[3cm]
Submitted in partial fulfillment of the
requirements for the degree of\\
[1cm]
Doctor of Philosophy of Imperial College London\\
[1cm]
and the\\ 
[1cm]Diploma of Imperial College London\vfill

March 2013\\[1cm]

\end{center}

\end{titlepage}

\section*{Declaration}

I, Stefano Orani, hereby declare that the work contained within this thesis
represents entirely my own work. All instances where the work of others has been
consulted or contributions have been used for the formation of this thesis that are
not my own work have been properly attributed.
\vfill
\section*{Acknowledgements}
\setcounter{page}{2}

The research carried during my PhD would not have been possible without the support of many people.\\
\noi In particular, I would like to express my gratitude to my supervisor, Dr.~Arttu Rajantie, who offered invaluable assistance and guidance.\\
\noi Dr.~David Mulryne, for helping me through my first research project and for enriching discussions.\\
\noi My graduate colleagues and friends Dionigi Benincasa, Michel Buck and Andre Coimbra, for creating a stimulating and pleasant working environment.\\
\noi Last but not least, my deepest love and gratitude goes to my family and dear friends.

\newpage

\phantomsection
\addcontentsline{toc}{chapter}{Abstract}
\begin{abstract}
\setcounter{page}{3}
\thispagestyle{plain}

\noi Cosmological inflation is the dominating paradigm to account for observations of the Cosmic Microwave Background (CMB).\\
\noi In this thesis, we study the phenomenology of a class of particularly well motivated models of inflation, known under the generic name of hybrid models. They are characterised by a transition from a valley to a hilltop shaped potential. In particular, we study three limiting regimes of the simplest realisation, hybrid inflation, constraining its parameter space using observational bounds on the spectral index and the non-gaussianity of the primordial perturbations.\\
\noi We find that the model is highly constrained by observations, with large part of the parameter space either ruled out by a blue spectral index ($\ns>1$) or by a large non-gaussianity parameter $\fnl$, two quantities measured with precision by PLANCK. However, there exists regions in parameter space leading to interesting phenomenology compatibly with observational bounds.\\
\noi Also, a version of hybrid inflation with a third light scalar field at horizon crossing is derived from the supersymmetry framework. We find that the model can generate observables within observational bounds.

\end{abstract}

\newpage

\section*{Publications}
\setcounter{page}{4}
\thispagestyle{plain}

\noi The work presented in this thesis contains work published in the following articles \cite{mpHP,mpHPS}.
\begin{itemize}
\item   D.~Mulryne, S.~Orani and A.~Rajantie, '\emph{Non-Gaussianity from the hybrid potential}',
  Phys.Rev. {\bf D84} (2011) 123527,
  [hep-th/11074739].
\item S.~Orani, A.~Rajantie, '\emph{Supersymmetric hybrid inflation with a light scalar}',  
	Phys.Rev. {\bf D88} (2013) 043508,
	[astro-ph.CO/13048041]. 
  \end{itemize}
\setcounter{page}{5}
\begin{spacing}{0.9}
\begin{small}
\cleardoublepage\phantomsection\pdfbookmark{Contents}{contents}
\tableofcontents
\cleardoublepage\phantomsection\pdfbookmark{List of Figures}{figures}
\listoffigures
\end{small}
\end{spacing}
\newpage

\pagestyle{fancy}
\fancyhf{}
\rhead{\textit{\small{\thesection\ \rightmark}}}
\cfoot{\thepage}

\renewcommand{\chaptermark}[1]{\markboth{#1}{}}
\renewcommand{\sectionmark}[1]{\markright{#1}{}}

\chapter{Introduction} 
\label{chap:Intro}

\noi The Cosmic Microwave Background (CMB) is made of radiation traveling through the universe since recombination time, when the decreasing density of the universe first allowed photons to travel freely over cosmological distances. Recent experiments measured precisely its properties \cite{planck, wmap,observations}, revealing that today this radiation is feeble and has an average thermal black body spectrum of temperature $2.725K$. It is highly homogeneous but not exactly so, with an amplitude of fluctuations of order $\frac{\delta T}{T}\approx 10^{-5}$. Current observations describe them as nearly scale-invariant and either Gaussian or nearly so. \\

\noi  Since its introduction in the 1980s, inflation \cite{inflation} became the most promising framework for explaining these features of the observable universe. It is a period of accelerated expansion of space in the primordial universe, driven by negative pressure matter. It explains the homogeneity of the CMB, the observed flatness of space on observable scales and the absence of relics from primordial physics in the sky. Furthermore, the accelerated expansion magnifies quantum fluctuations to large scales, providing a framework for understanding the origin of large scale structures. The simplest realisation, canonical single field inflation, successfully explains the scale invariance and almost gaussianity of the fluctuations \cite{Maldacena:2002vr}. It consists of a scalar field slowly rolling down a potential energy slope. When the field rolls slower than the expansion rate of space, inflation occurs. Eventually, when the potential gets steeper and the the field rolls faster, inflation ends. At this point, the inflaton decays into radiation, creating a thermal bath that reheats the universe. From then on, the hot Big Bang model describes the evolution.\\

\noi Single field inflation is compatible with current observations, in agreement with PLANCK bounds at $95\%$ confidence level on primordial observables \cite{planck}. Nevertheless, it is important to understand the signatures of more complex inflationary models. Those include models with non-canonical kinetic terms and multi-field models. The higher number of degrees of freedom implies a richer phenomenology, allowing for a wider range of predictions for the observables.  For this reason, it is important to have good motivations before studying multi-field models. A promising approach is to look for models embedded in high energy particle physics theories, such as supersymmetric (SUSY) extensions of the Standard Model of particle physics and supergravity (SUGRA). Among those, of particular interest are hybrid models \cite{Linde:1993cn}. First introduced in the early nineties, they are the result of the association of the idea of inflation with that of spontaneous symmetry breaking. The simplest version consists of a two real scalar fields potential containing the canonical single field inflation term together with a self-interacting hilltop shaped term for the symmetry breaking field and an interaction term between the two scalars. Depending on the relative magnitude of the three terms, inflation can be realised in very different ways, as we will see in Chapter~\ref{ch:hi}.\\
\noi The link with particle physics is the Higgs mechanism. Although the properties of the Standard Model Higgs boson make it an unlikely candidate for having a role in inflation dynamics \cite{higgsINF}, other Higgs type scalar fields with suitable properties might exist, perhaps in the form of a supersymmetric partner of a Standard Model particle. In this case, the inflationary potential can be derived from a superpotential, the supersymmetric version of the potential. An example of this approach to inflationary model building is given in Chapter~\ref{ch:susyhi}.\\

\noi The thesis is structured as follows. In Chapter~\ref{ch:iandqf}, the Hot Big Bang model and inflation are introduced. The quantum properties of the inflationary universe and perturbations are discussed in Chapter~\ref{sec:qfandp}. In Chapter~\ref{ch:sepuni} we derive convenient expressions for suitable observables to test models of inflation and discuss the predictions of the simplest models. In Chapter~\ref{ch:hi} we study the simplest realisation of hybrid inflation, with emphasis put on three limiting regimes of its rich phenomenology. The first regime is discussed in Section~\ref{sec:vdhchi} and is the original scenario in which hybrid inflation was considered. It consists of an inflationary phase along the canonical single field direction, with energy density dominated by the hilltop, ending promptly when the symmetry breaking phase transition starts. The other regimes discussed are characterized by a slow phase transition. In Section~\ref{sec:vdlchi}, the energy density is again dominated by the hilltop, with a first phase of inflation analogous to that of the original scenario. However, here the inflationary trajectory rolls very slowly and a considerable amount of accelerated expansion happens during the phase transition. Finally, in Section~\ref{sec:philchi}, the inflationary phase is dominated by the canonical single field's mass term. As in the previous scenario, the accelerated expansion lasts for a considerable time after the phase transition. In this scenario, most of the inflationary phase looks like the simplest single field inflation models, with the difference that the symmetry breaking field typically kicks in at the end of inflation, generating interesting signals.

\noi In Chapter~\ref{ch:susyhi}, we discuss how hybrid inflation models are derived from supersymmetry. In Section~\ref{sec:SUSYhi} we introduce the simplest version of supersymmetric hybrid inflation. In Section~\ref{sec:3HI}, we modify the simplest scenario and study a minimal extension characterised by the presence of a third dynamically relevant scalar field, whose lightness is protected by supersymmetry.

\subsubsection{A note on conventions}

\noi Throughout this thesis we will use the natural units $\mpl=c=k_{\rm B}=1$, where $\mpl=(8\pi G)^{-1/2}=4.342\times10^{-6}g=2.436\times10^{18}GeV$ is the Planck mass, $c= 2.99 792 458\times10^8 ms^{-1}$ is the speed of light and $k_{\rm B}=1.3806488 \times 10^{-23} m^2kg s^{-2}K^{-1}$ is the Boltzmann constant. In figures the Planck mass $\mpl$ appears explicitly although we set it equal to $1$.\\
\noi We use the metric signature (-,+,+,+).

\chapter{Cosmological Inflation}
\label{ch:iandqf}

\section{The hot Big Bang model and its issues}
\label{sec:SMc}

\noi Modern cosmology was born as a corollary of General Relativity (GR) in the $1920$'s (see \cite{Wald} for a complete introduction to the theory of GR). Before, spacetime was understood as a fixed background on which radiation and matter evolved. With the advent of GR, however, our understanding of spacetime drastically changed, becoming itself a dynamical entity, the evolution of which is ruled by the Einstein equations. They are given by
	
\be
G_{\mu\nu} \equiv R_{\mu\nu} - \frac{1}{2}\delta_{\mu\nu}R = T_{\mu\nu}\,,
\label{eq:einstein}
\en
where $G_{\mu\nu}$ is the Einstein tensor and depends solely on the metric $g_{\mu\nu}$, $R_{\mu\nu}$ is the Riemann tensor and $R$ is the Ricci scalar. $T_{\mu\nu}$ is the energy-momentum tensor and contains information about the matter content of the universe. \\

\noi The cornerstone of modern cosmology is the assumption that, on scales larger than the observable universe, the universe is homogeneous and isotropic. A variety of observations confirm this belief, the most striking of which is the high homogeneity of the CMB \cite{wmap,observations}. The mathematical solution of GR equations corresponding to an homogeneous and isotropic spacetime is \cite{FRLW}

\be 
ds^2 = -dt^2 + a(t)^2d\Sigma^2,
\label{FRW}
\en

\noi where $a(t)$ is the scale factor and $d\Sigma^2$ is the spatial surface element. The corresponding Einstein equations give:

\be
H^2&\equiv&\left(\frac{\dot{a}}{a}\right)^2=\frac{\rho}{3}-\frac{k}{a^2}\,,\label{FRW1}\\
p&=&2\frac{\ddot{a}}{a}+\left(\frac{\dot{a}}{a}\right)^2+\frac{k}{a^2}\,,\label{FRW2}\\
\en

\noi Here and henceforth, the overdot indicates differentiation with respect to time.\\
\noi The first Friedmann equation, Eq.~(\ref{FRW1}), relates the energy density $\rho$ to the rate of expansion $H$ and the curvature parameter $k$. Positive, zero or negative $k$ correspond respectively to a closed, flat or open universe.\\
\noi Eq.~(\ref{FRW2}), the second Friedmann equation, relates the pressure $p$ to the expansion rate and acceleration.\\
\noi An equation for the evolution of $\rho$ can be found by combining the first and the second Friedmann equations. It is called the continuity equation and is given by 

\be
\dot{\rho} &=& -3H\left(\rho + 3p\right)\,.\label{conteq}
\en

\noi On smaller scales, however, we observe a very inhomogeneous universe, with clusters of galaxies, galaxies, stars, planets and life. These structures are believed to have originated from a much more homogeneous distribution, as can be seen in the CMB, with irregularities growing over time mainly through gravitational attraction.\\ 

\noi When inflation \cite{inflation} was introduced, in the early $1980$'s, cosmological observations were at a very primitive stage. The dominating paradigm of cosmology was the Hot Big Bang model, according to which spacetime originated in a hot and dense state, known as the Big Bang. In the aftermath of the Big Bang, the universe gradually cooled down via the expansion of space. As the temperature dropped, the structures that we observe today emerged. The greatest successes of the hot Big Bang model are the prediction of the existence of the CMB \cite{Alpher:1948ve} and the theory of nucleosynthesis \cite{nucleosyn}, which explains how the lightest elements were produced and correctly accounts for their primordial abundance.\\

\noi Despite its great success in describing the evolution of the universe, the Hot Big Bang model has some major shortcomings. They are related to the initial conditions required for the universe to evolve into what we see today. 

\subsection{Flatness problem}
\label{ssec:fp}

\noi From Eq.~(\ref{FRW1}), we can determine the energy density of a flat universe by taking $k=0$. It is called the critical energy density, $\rho_{\rm c}=3H^2$, and the ratio of the universe's actual energy density $\rho$ to $\rho_{\rm c}$, $\Omega\equiv\frac{\rho}{\rho_{\rm c}}$ is a measure of the curvature of spacetime: $\Omega=1$ corresponds to a flat universe, whereas $\Omega > 1$ and $\Omega < 1$ correspond respectively to a closed and open universe.\\
\noi Eq.~(\ref{FRW1}) can be rearranged as follows:
\be 
\left(\Omega^{-1} - 1 \right) = -\frac{3k}{\rho a^2}\label{flatness}.
\en

\noi As the universe expands, the scale factor $a$ increases, whereas the density $\rho$ decreases as the energy gets diluted. According to the Hot Big Bang model , the universe contains mainly matter and radiation throughout its history. For radiation $\rho\propto a^{-4}$ and for matter $\rho\propto a^{-3}$, meaning that the right hand side term of Eq.~(\ref{flatness}) increases in magnitude, the universe evolving away from flatness. Indeed, since the Big Bang, it has increased by a factor of approximately $e^{60}$, as did $(\Omega^{-1} -1)$ to balance the equation.\\

\noi Present measurements indicate that $|\Omega^{-1} -1| < 0.01$, implying that it must have been extremely close to zero in the aftermath of the Big Bang. Without a mechanism for explaining such feature of the early universe, it seems unlikely and fine-tuned to an extremely high degree.  
    
\subsection{Horizon problem}
\label{ssec:hp}

\noi The high homogeneity of the CMB is a striking feature. It suggests that a thermalisation process occurred throughout the observable universe at some stage during its evolution.\\
\noi In the Hot Big Bang model, there is no mechanism allowing for this thermalisation to happen, meaning the homogeneity has to be introduced as an initial condition.\\

\noi To highlight the problem, consider the comoving particle horizon. It is defined as the distance light rays have traveled since the Big Bang until time $t$ and it is given by
\be
d_p(t) = \int_0^t\frac{dt'}{a(t')}.
\label{ph}
\en
\noi Information from events at a distance greater than $d_p(t_0)$, where $t_0$ indicates the present, is not yet available to us.\\
\noi The CMB is made of radiation traveling since recombination time. The ratio of the particle horizon at recombination and the particle horizon today is given by

\be 
\frac{d_p(t_{\rm rec})}{d_p(t_0)} \approx \left(\frac{t_{\rm rec}}{t_0}\right) \approx (10^{-5})^n.\label{ratio}
\en

\noi If matter dominates it is of order $10^{-2}$ and if radiation dominates it is of order $10^{-3}$. Crucially, according to the Hot Big Bang model, this quantity is always smaller than $1$. It means that there exists regions inside our particle horizon today that have never been in causal contact. Therefore, thermalisation could not have happened and the homogeneity of the CMB has to be understood as a coincidence.

\section{Inflation}
\label{sec:i}

\noi The problems outlined in the previous section can be solved by a phase of inflation after the Big Bang and before recombination.\\
\noi Inflation (see \cite{reviews} for complete reviews) is defined as an era of accelerated expansion in a homogeneous and isotropic universe, described by the metric in Eq.~(\ref{FRW}). This implies

\be 
\ddot{a} > 0,
\label{infdef}
\en

\noi or, equivalently

\be 
\frac{d}{dt}\left(aH\right)^{-1} < 0.
\label{infdef2}
\en

\noi A phase of inflation in the early universe allows to solve both the flatness and horizon problem in one go.

\subsubsection{Solution to flatness problem}
\label{ssec:sfp}

\noi In contrast to matter and radiation domination, during inflation the right hand side of Eq.~(\ref{flatness}) decreases in magnitude. Indeed, from Eq.~(\ref{FRW1}) and Eq.~(\ref{infdef2}), it follows that

\be 
\frac{d}{dt}(\rho a^2)^{-1} = \frac{d}{dt} (3a^2H^2+3k)^{-1} &=& \frac{-6\dot{a}\ddot{a}}{(3a^2H^2+3k)^2} \nn
                                                             &\propto& -\ddot{a} < 0\,, \label{flatsol}
\en
\noi This means that the universe evolves towards flatness, curing the unnatural fine tuning required by the Hot Big Bang model.

\subsubsection{Solution to horizon problem}
\label{ssec:shp}

\noi The horizon problem arises from the fact that the size of causally connected regions increases during all eras of the Hot Big Bang model.\\
During inflation, however, the size of the comoving horizon decreases by definition, as shown in Eq.~(\ref{infdef2}). It follows that, provided inflation lasts long enough, the horizon problem can be solved in the context of inflationary cosmology.\\

\noi Thus, the Hot Big Bang model is completed by having inflation as an add-on.

\subsection{Realising inflation}
\label{sec:ri}

\noi Combining the Friedmann equations (\ref{FRW1}) and (\ref{FRW2}), one finds

\be
\ddot{a}>0 \Leftrightarrow \rho + 3p <0.
\label{infdef3}
\en

\noi Since the energy density $\rho$ is always positive, the condition (\ref{infdef3}) can only be satisfied if the pressure $p$ is negative. The easiest way to achieve this, is to consider a universe whose energy and pressure are dominated by a scalar field $\phi$, called the inflaton. If the scalar field is homogeneous ($\phi\equiv\phi(t)$) and sees a potential $V(\phi)$, its energy density and pressure are given by

\be
\rho_\phi &=& \frac{1}{2}\dot{\phi}^2+V(\phi),\label{eqi:phien}\\
p_\phi &=& \frac{1}{2}\dot{\phi}^2-V(\phi).\label{eqi:phip}
\en

\noi Depending on how the energy density is distributed between the potential and kinetic terms, it is possible for a scalar field to have negative pressure and therefore to source an era of inflation. Indeed, Eq.~(\ref{infdef3}) is satisfied provided that $\dot{\phi}^2 < V(\phi)$.\\

\noi Substituting Eqs.~(\ref{eqi:phien}) and (\ref{eqi:phip}) into the Einstein Eqs.~(\ref{FRW1}) and (\ref{FRW2}), one finds the equations of motion for the scalar field $\phi$. Assuming a spatially flat universe ($k=0$), one finds

\be 
&&H^2=\frac{1}{3\mpl^2}\left[\frac{1}{2}\dot{\phi}^2 + V(\phi)\right],\\
&&\ddot{\phi} + 3H\dot{\phi} = -\frac{dV}{d\phi}.
\label{eqi:EOM}
\en

\noi A very useful simplification can be done in the regime where $V(\phi) \gg \dot{\phi}\mpl^2$ and $|\ddot{\phi
}|\ll 3H|\dot{\phi}|$. It is called the slow-roll approximation and it allows to reduce the second order differential equation (\ref{eqi:EOM}) to a first order one:

\be 
&&H^2 \simeq \frac{V(\phi)}{3\mpl^2},\\
&&3H\dot{\phi} \simeq -\frac{dV}{d\phi}.
\label{eqi:SR}
\en

\noi It is useful at this point to introduce the slow-roll parameters:

\be
\epsilon &=& \frac{\mpl^2}{2}\left(\frac{V'}{V}\right)^2,\label{eqi:eps}\\
\eta &=& \mpl^2\frac{V''}{V}.\label{eqi:eta}
\en

\noi Here and in what follows, the prime $'$ denotes derivation with respect to $\phi$. It is necessary, for the slow-roll approximation to be valid, that

\be 
\epsilon\ll1, \;\;\;\;\;\; |\eta|\ll1.\label{eqi:srconds}
\en

\noi The conditions (\ref{eqi:srconds}) are necessary but not sufficient for Eqs.~(\ref{eqi:SR}) to be valid: they only restrict the shape of the potential, leaving $\dot{\phi}$ as a free parameter. However, it can be proven that when the slow-roll conditions (\ref{eqi:srconds}) are met, even if $\dot{\phi}$ is chosen so as to violate the slow-roll approximation, Eqs.~(\ref{eqi:SR}) are an attractor solution: given enough time, the system tends to follow them. 

\subsection{Amount of inflation and evolution of scales}
\label{ssec:amofiandevoscales}

\noi The ratio between the scale factor at the end of inflation to its value at the beginning quantifies the amount of inflation a spacetime region underwent. For convenience, the logarithm of such quantity is taken, defining the e-foldings $N$:

\be N(t)\equiv\ln \frac{a(t_{\rm end})}{a(t_{\rm in})}.\label{eqi:efolds}\en

\noi This measures the amount of expansion from time $t_{\rm in}$ to the end of inflation at time $t_{\rm end}$.\\
\noi If the slow-roll approximation is valid, we can use the inflaton as a clock, and consider $N$ as a function of $\phi$:

\be N\equiv\ln \frac{a(t_{\rm end})}{a(t_{\rm in})}=\int_{t_{\rm in}}^{t_{\rm end}}Hdt \simeq -\int_{\phi_{\rm in}}^{\phi_{\rm end}}\frac{V}{V'}d\phi\,.
\label{eqi:Nphi}
\en
Now consider a comoving length scale $k^{-1}$ (where $k$ is a comoving wavenumber of a Fourier decomposition of a given function, as introduced at the beginning of Section~\ref{sec:qfandp}). An important question concerning its evolution is whether $k$ is larger or smaller than $k_H=aH$, the wavenumber corresponding to the Hubble radius.\\
\noi Assume $k$ starts its evolution inside the horizon, $k>k_H$. During inflation the comoving Hubble radius decreases, therefore the scale $k$ may exit the horizon at some time during the inflationary phase. Once outside the horizon, fluctuations corresponding to the scale $k^{-1}$ are conserved. When inflation comes to an end, the Hubble radius starts increasing again, and the scale $k^{-1}$ may eventually reenter the horizon, with properties determined by the primordial fluctuations that were frozen during the superhorizon evolution. A scale of particular interest is the scale that today equals the Hubble radius $a_0H_0$, which we will call $k_0$. As we will see, the properties of the CMB anisotropies are directly related to the evolution of such scale.\\

\noi In order to study the evolution of the scale $k_0$, we need to know when, during inflation, it crossed the horizon. The answer depends on the evolution of the universe from the end of inflation to the present, given that the scale $k_0$ reenters the horizon in present times. Since we do not know exactly the history of the universe throughout that period, some assumptions have to be made. According to the simplest scenario, after inflation the universe reheats and behaves as if matter dominated. Once reheating is completed, the universe is radiation dominated and the Hot Big Bang is restored. Assuming instantaneous transitions between the different regimes, one finds that $k_0$ exits the horizon approximately $60$ e-folds before the end of inflation. The uncertainty is due to our lack of knowledge of the energy scales associated with inflation.\\

\subsection{Chaotic inflation}
\label{ssec:ci}

The chaotic inflation scenario describes a primordial universe whose energy density is dominated by the inflationary degrees of freedom \cite{Linde:1983gd}. Its simplest realisation consists of a single scalar field minimally coupled to gravity with potential of the form 

\be 
V=\frac{\lambda \phi^n}{n}\,,
\en

\noi where $\lambda\ll1$ and $n>0$ are dimensionless coupling constants.\\
\noi The slow-roll parameters (\ref{eqi:eps}) and (\ref{eqi:eta}) are particularly simple and given by:

\be
\epsilon = \frac{n^2}{2\phi^2}\;\;\;\;\; {\rm and} \;\;\;\;\;\eta=\frac{n(n-1)}{\phi^2}
\label{eqi:cisr}
\en

\noi Provided that $\phi>n/\sqrt{2}$, the slow-roll conditions are met. If these conditions hold long enough, the system will approach the attractor solution given in Eqs.~(\ref{eqi:SR}), which in this case read

\be
3H^2   &\simeq& \frac{\lambda \phi^n}{n},\nn
3H\dot{\phi} &\simeq& -\lambda\phi^{n-1}.
\label{eqi:cieom}
\en

\noi The equations can be solved exactly and the solution is:

\be
\phi=\sqrt{\phi_{\rm i}^2 - 2nN},\label{eqi:ciphi}
\en

\noi where $\phi_{\rm i}$ is the value of the inflaton some time early in the inflationary phase and $N$ the e-folds of expansion. Taking $\phi=n/\sqrt{2}$ we find that $60$ e-folds before the end of inflation $\phi_{\rm i} \simeq \sqrt{120n}$. Assuming chaotic inflation is the right model, this is the value of $\phi$ when the mode $k_0$, corresponding to the size of the Hubble radius today, crossed the horizon. Hereafter, the value of a quantity at the time $k_0$ crosses the horizon will be labeled by a $*$ (that is $\phi_*\simeq\sqrt{120n}$).

\subsection{Eternal inflation}
\label{ssec:ei}

\noi Imagine a primordial universe whose energy density is dominated by the scalar field $\phi$, as in chaotic inflation. Quantum fluctuations ensure that the field's configuration is not homogeneous in space. As a consequence, different regions of the universe undergo different expansion histories, with inflation ending in some regions while it still continues in other. Given that inflating regions grow in size at a rate much greater than non-inflating regions, spacetime rapidly comes to be dominated by the former, with sparse oases where inflation ended and the energy density has been successfully transferred to radiation. Such a scenario is called eternal inflation \cite{eternal}.\\

\noi An inflating region of spacetime is said to be in the eternal inflation regime if quantum fluctuations dominate over the classical displacement of the inflationary degrees of freedom:

\be 
\Delta\phi_{\rm cl} = \dot{\phi} \lesssim \Delta\phi_{\rm qu} \simeq \frac{H^2}{2 \pi}\,.
\label{eqei:ei}
\en
If this is the case, when inflation ends in the region, quantum fluctuations will drive parts of it back into the inflationary phase. These smaller regions will expand exponentially faster than the rest, thus dominating that region of spacetime.\\

\noi It is worth noting that the eternal inflation scenario generates a universe that is inhomogeneous on the largest scales. Indeed, in this scenario, the  homogeneity is local and an exclusive consequence of inflation. The universe is divided into inflating and non-inflating regions, the non-inflating regions forming a network of mini-universes where the energy has been succesfully tranferred to non-inflationary degrees of freedom.

\chapter{Quantum Fluctuations and Perturbations}
\label{sec:qfandp}

\noi In Section~\ref{sec:SMc}, the metric Eq.~(\ref{FRW}) was introduced and we argued that it correctly describes large scales of our region of the universe. However, on any scale we are able to observe, there are inhomogeneities. Inflation provides a mechanism for these inhomogeneities to arise: they are related to the quantum fluctuations of the inflationary degrees of freedom. These fluctuations backreact with spacetime, generating the structures we observe today. Therefore quantum fluctuations in the matter content of the universe source perturbations of the geometry of spacetime. In this Chapter we will compute the power spectrum of a scalar field in an inflating spacetime and derive the curvature perturbation, a suitable quantity for understanding the relics of inflation in the CMB.

\subsection{Fourier Transform and Random Fields}
\label{FTandRF}

\subsubsection{Fourier transform}

In what follows, we will extensively use the properties of the Fourier transform, which we briefly introduce here.\\
\noi Consider a generic function $g(t,\textbf{x})$. Inside a cubic box of comoving size $L$, the Fourier forward and backward transforms can be defined as \cite{Lyth:2009zz}

\be 
g(t,\textbf{x}) = \frac{1}{L^3}\sum_n \tilde{g}_n(t) e^{i\textbf{k}_n\cdot \textbf{x}}\,,\;\;\;\;\;\;\;\;\;\; \tilde{g}_n(t) = \int g(t,\textbf{x}) e^{-i\textbf{k}_n\cdot \textbf{x}}d^3x\,,\label{eq:fourierdis}
\en
where the wavevectors $\textbf{k}_n$ form a cubic lattice with spacing $2\pi/L$ and can be labelled without loss of generality by a number $n$.\\
\noi Requiring the coefficients $\tilde{g}_n(t)$ to be real implies

\be 
\int e^{i(\textbf{k}_n - \textbf{k}_m)\cdot \textbf{x}}d^3x = L^3\delta_{nm}\,,
\label{realdis}
\en
where the Kronecker delta $\delta_{nm}=1$ if $n=m$ and zero otherwise.\\
\noi When $L\rightarrow\infty$, the sum becomes an integral and we have

\be
g(t,\textbf{x}) = \frac{1}{(2\pi)^3}\int \tilde{g}(t,\textbf{k})e^{i\textbf{k}\cdot \textbf{x}}d^3k\,,\;\;\;\;\;\tilde{g}(t,\mathbf{k}) = \int g(t,\textbf{x}) e^{-i\textbf{k}\cdot \textbf{x}}d^3x\,,\label{eq:fourier}
\en
Requiring $\tilde{g}(t,\mathbf{k})$ to be real now gives

\be 
\int e^{i(\textbf{k} - \textbf{q})\cdot \textbf{x}}d^3x = (2\pi)^3\delta^3(\textbf{k} - \textbf{q})\,,
\en
where $\delta^3$ is the $3$-dimensional Dirac delta function.\\ 

\noi Some comments regarding the comoving size $L$ used here are in order. In cosmology, we tipically consider functions defined over regions of finite size. For such functions, the discrete Fourier transform, Eq.~(\ref{eq:fourierdis}) is the appropriate expansions. However, from a practical point of view we can consider a box whose size is much larger than today's observable universe and approximate the wavevectors $\textbf{k}$ as continuous. This allows to use the transformations Eqs.~(\ref{eq:fourier}), which give neater equations with respect to the ones derived using the discrete Fourier transform.\\

\subsubsection{Random fields}

\noi In the context of cosmology, the function $g(t,\textbf{x})$ is usually associated with a random field. This means that it is extracted from a set of functions $g_\alpha(t,\textbf{x})$ called ensemble, with the property that every member of the set is extracted with a probability $P_\alpha$. The function $g(t,\textbf{x})$ is then referred to as a realization of the ensemble. For our purposes, the set of functions can be taken to be continuous and the probabilities $P_\alpha$ are replaced by a probability density function $\textbf{P}(g)$, or PDF.\\
\noi Looked at from this perspective, the anisotropies of the CMB can be understood as a realization of the ensemble of perturbation functions. If such perturbations arise from inflation, their PDF can be derived from the study of inflationary dynamics. In practice, working out the exact form of the PDF is a daunting task. Usually, cosmologists work with correlators. For a discrete ensemble of random variables, the real space two-point correlator (or function) of $g(t_0,\textbf{x})\equiv g(\textbf{x})$ evaluated at time $t_0$ is \cite{Lyth:2009zz}

\be
\langle g(\textbf{x}) g(\textbf{y}) \rangle = \sum_\alpha P_\alpha g_\alpha(t_0,\textbf{x}) g_\alpha(t_0,\textbf{y})\,,
\label{dis2pc}
\en
where $\textbf{x}$ and $\textbf{y}$ are two points of the constant time hypersurface. The n-point correlator is defined analogously.\\
\noi The statistical $n^{\rm th}$-moment in terms of the PDF $\textbf{P}(g)$ is defined as

\be
\langle g^n(\textbf{x})  \rangle = \int \textbf{P}(g) g^{n} dg\,.
\label{moment}
\en 

\section{Vacuum Fluctuations during Inflation}
\label{ssec:vacFRW}

\noi An important, if not the most important, feature of inflation is that it magnifies quantum fluctuations from sub-horizon scales to super-horizon scales, providing a mechanism for the origin of large scale structures in the universe. Given the fact that, once stretched outside the horizon, fluctuations are frozen until they reenter it, a natural question to ask is how these perturbations look like at horizon exit. To answer this question, we need to understand the properties of vacuum fluctuations in an inflating universe. \\
\noi We start by considering an inflationary phase driven by a scalar field $\phi(t,\textbf{x})$. If the metric describing the spacetime on which this quantum field lives is not globally invariant under time transformations (and this is the case of the FRW metric Eq.~(\ref{FRW})), it is impossible to define a quantum vacuum state using the canonical formalism, which relies on the decomposition of the Fourier modes of the field into positive and negative frequency modes.\\
\noi For a massive scalar field, this issue can be overcome by noting that, analogously to a black hole, the inflating spacetime has an event horizon. It has therefore a Hawking temperature. This can be seen by considering the metric (\ref{FRW}) with an exponentially growing scale factor $a=e^{Ht}$ or, equivalently, with a constant Hubble rate $H$, as is approximately the case during inflation. Such a spacetime is called de Sitter space and subsituting $t\rightarrow i\lambda$ we can write its metric in Euclidean form as \cite{Linde:2005ht}
\be
ds^2 = d\lambda^2 + \frac{\cos^2 H\lambda}{H^2}d\Sigma^2\,.
\label{dSmet}
\en
This metric corresponds to a four-sphere $S^4$ with radius $H/\pi$. Scalar fields will have fluctuations which are periodic in $\lambda$ with period $2\pi/H$, which is equivalent to consider quantum statistics at a temperature $T = H/(2\pi)$ \cite{Gibbons:1976ue}.
Physically, the appearence of this temperature is due to the inaccessibility to an observer of information coming from states beyond the event horizon and therefore to the necessity of averaging over such states \cite{Gibbons:1977mu}.\\
\noi This 'cosmological' temperature has one peculiarity: it arises from a periodicity in all four spacetime directions and therefore leads to vacuum fluctuations qualitatively different from the usual spectrum of thermal fluctuations. For a massive field $\phi$ with mass $m$, the second moment at a given time is \cite{Linde:2005ht}

\be 
\langle \phi(\textbf{x})^2 \rangle = \frac{3H^4}{8\pi^2m^2}\,.
\label{bd}
\en
This state was first derived in \cite{Bunch:1978yq}. It tells us that the fluctuations of a massive field are suppressed by its mass. In the limit that $m^2\gg H^2$, the field practically has no fluctuations. On the other hand, Eq.~(\ref{bd}) diverges as $m\rightarrow 0$. This is problematic since in the early stages of inflation, the effective mass of a scalar field is negligibly small. To see this, consider the equation of motion of a massive field $\phi(t,\textbf{x})$ in an expanding spacetime with metric (\ref{FRW}):

 \be
  \ddot{\phi}(t,\textbf{x})+3H\dot{\phi}(t,\textbf{x})-(\nabla^2+m^2)\phi(t,\textbf{x}) = 0\,,
  \label{eq2:26}
 \en
 where $m^2=V''(\phi)$ is the mass and the gradient is defined as
 \be
  \nabla^2 = \frac{1}{a^2}\Sigma\frac{\partial^2}{\partial x_i^2}\,.
  \label{eq2:27}
 \en
For a Fourier mode $\phi(t,\mathbf{k})$, we find

 \be
  \ddot{\phi}(t,\textbf{k})+3H\dot{\phi}(t,\textbf{k})+(\frac{k^2}{a^2} + m^2)\phi(t,\textbf{x})= 0\,.
  \label{eq2:four}
 \en
 \noi During inflation, the scale factor undergoes an accelerated growth, $\ddot{a}>0$. This implies that, provided inflation lasts long enough, there exists a time in the past when the $k^2/a^2$ term in the equation of motion dominates over the mass term, that is $ k^2\gg m^2a^2$. Therefore, at such time the scalar field is effectively massless and the two-point function Eq.~(\ref{FRW}) diverges. To understand better this peculiar behaviour, in what follows we derive the vacuum state in a different way.\\

\noi We start by noting that on scales smaller than the Hubble radius $(aH)^{-1}$, the universe (\ref{FRW}) is well approximated by the Minkowski metric and gravity is negligible. This allows to unambiguously define the vacuum state of a field $\phi(t,\textbf{x})$ on such scales. For a massless field, at tree level it is  fully described by the two-point function of the field and its conjugate momentum $\pi(t,\textbf{x})=\dot{\phi}(t,\textbf{x})$ (note that we are using the same notation for the functions and the corresponding operators). They are given by \cite{Peskin:1995ev}

\be \langle \phi^*(\mathbf{k})\phi(\mathbf{q})\rangle &=& \frac{1}{2k}(2\pi)^3\delta^3(\mathbf{k}-\mathbf{q}),\nn
    \langle \pi^*(\mathbf{k})\pi(\mathbf{q})\rangle &=& \frac{k}{2}(2\pi)^3\delta^3(\mathbf{k}-\mathbf{q}).
    \label{eqi:vac}
\en
 \noi Assuming a scalar field $\phi(t,\textbf{x})$ has an initial state well described on subhorizon scales by Eqs.~(\ref{eqi:vac}), we want to understand how its fluctuations look like at the time of horizon exit.\\
 \noi To this end, we expand the field inside a comoving box of size $L\gg (aH)^{-1}$ into a homogeneous part and a perturbation as
  
 \be
  \phi(t,\textbf{x})=\phi(t) + \delta\phi(t,\textbf{x})\,. 
  \label{eq2:28}
 \en
  
 \noi The linearized equation of motion for the perturbation $\delta\phi(t,\textbf{x})$ is given by
  
 \be
  \delta\ddot{\phi} + 3H\delta\dot{\phi}+\left(-\nabla^2 + m^2\right)\delta\phi=0\,.
  \label{eq2:29}
 \en
  
 \noi In terms of the Fourier modes $\delta\phi(t,\textbf{k})\equiv\delta\phi_k$, this becomes
  
 \be
  \delta\ddot{\phi_k} + 3H\delta\dot{\phi_k} + \left(\frac{\textbf{k}^2}{a^2}+ m^2\right)\delta\phi_k=0\,.
  \label{eq2:30}
 \en
 Note that the perturbation $\delta\phi_{k}$ does not depend on the direction of the wavevector $\textbf{k}$, but only on its magnitude $|\textbf{k}|\equiv k$.\\
 \noi Physical scales exit the comoving horizon when the expansion rate accelerates. Therefore, at a sufficiently early stage of inflation, the scales which correspond to the observable universe were well inside the causal  
 horizon. As argued above, on such scales, the quantum vacuum can be unambiguously defined and is given by Eqs.~(\ref{eqi:vac}).\\

\noi In order to understand how the perturbation $\delta\phi_k$ evolves from subhorizon to superhorizon scales, we should solve Eq.~(\ref{eq2:30}) with initial conditions given by Eqs.~(\ref{eqi:vac}). It cannot be solved analytically in the general case. However, assuming the scale $k$ crosses the horizon when its mass is still negligible and ignoring the variation of $H=H_*$, we have
  
 \be
  \delta\ddot{\phi_k} + 3H_*\delta\dot{\phi_k} + \left(\frac{k}{a}\right)^2\delta\phi_k = 0\,.
  \label{eq2:32}
 \en
 This differential equation can be solved analytically. The solution such that in the limit $t\rightarrow-\infty$ we recover the Minkowski vacuum is \cite{liddlelyth}
  
 \be
  \delta\phi_k = \sqrt{\frac{1}{2L^3k^3}}H_*\left(i+\frac{k}{aH_*}\right)\exp\left(\frac{ik}{aH_*}\right)\,,
  \label{eq2:34}
 \en
\noi Finally we find that the two-point function is given by \cite{liddlelyth}
  
 \begin{eqnarray}
  \langle \delta\phi_k\delta\phi^*_{q}\rangle &\equiv& (2\pi)^3\delta^3(\textbf{k} - \textbf{q})L^3|\delta\phi_k|^2\, \nn
   &=& (2\pi)^3\delta^3(\textbf{k} - \textbf{q})\left( \frac{1}{2ak}+ \frac{H_*^2}{2k^3}\right)\,.
   \label{eq2:35}
 \end{eqnarray}
 \noi As expected, when $H=0$ and $a=1$ we recover the Minkowski vacuum Eqs.~(\ref{eqi:vac}). The second term derives from the nature of the inflating spacetime. In order to get an intuition of what it means, we look at it from the viewpoint of canonical quantization, although we already pointed out that such formalism does not work in an expanding spacetime. From such perspective, this term arises because the inflating spacetime, in addition to the Minkowski quantum fluctuations, contains $\phi$ particles with occupation numbers $n_k = H^2/(2k^2)$. Well inside the comoving horizon $k\gg H$ and these perturbations are highly suppressed. However, on superhorizon scales they are dominant and characterize the inflationary universe.\\
	\noi The real space two-point function is obtained by integrating Eq.~(\ref{eq2:35}) over $\int d^3k$. The first term can be renormalized as in flat spacetime quantum field theory. On the other hand, the second term gives a logarithmic contribution and diverges in the long wavelength limit. Therefore we recover the peculiar behaviour of Eq.~(\ref{bd}) in the limit $m\rightarrow 0$. In an inflating spacetime the boundary conditions solve this issue: the inflationary phase begins and finishes at well defined times, providing physical cutoffs for the real space two-point function.\\ 
 \noi A quantity that we will need in what follows is the power spectrum of $\phi$, which we formally define in subsection~\ref{ssec:gauss}. It is given by 
 
 \be
 \mathcal{P}_{\phi}(k,t)\equiv\frac{L^3k^3}{2\pi^2}|\delta\phi_k|^2\,.\label{eqi:pphi}
 \en
 
 \noi At horizon crossing, it becomes
 
 \be 
 \mathcal{P}_{\phi}(k,t_*) = \left.\left(\frac{H}{2\pi}\right)^2\right|_{k=aH}\,.\label{eqi:pphihc}
 \en

 \noi Since in the vacuum quantum fluctuations of $\phi$ are Gaussian, the power spectrum and the mean value $\langle \phi \rangle$ are the only quantities needed to describe it (see subsection~\ref{ssec:gauss}).\\

 \section{Curvature Perturbation}
 
\noi The metric (\ref{FRW}) is homogeneneous and isotropic and correctly describes the causal structure of the universe on large scales. However, as we argued before, on the scales that we experience the universe is inhomogeneous and full of structures. Thus, (\ref{FRW}) cannot be an exact description of spacetime. A more realistic description is that of a metric that can be described in the form of \emph{homogenous and isotropic + perturbations}. In what follows we introduce first order perturbations of (\ref{FRW}) and of the energy momentum tensor $T^{\mu}_{\nu}$ and derive mathematical expressions for the curvature perturbation, a quantity central to modern theoretical cosmology.\\ 

\subsection{First order metric perturbations}

\noi The metric (\ref{FRW}) plus small perturbations can be written as 
 
 \be
 ds^2 = \left[g_{\mu\nu} + \delta g_{\mu\nu}\right]dx^{\mu}dx^{\nu}\,,
\label{deltag}
\en
with $|g_{\mu\nu}| \gg |\delta g_{\mu\nu}|$. In terms of conformal time $d\tau = dt/a$, the background FRW metric (\ref{FRW}) becomes

\be
g_{\mu\nu}dx^{\mu\nu}  = a(\tau)^2 \left(-d\tau^2 + d\Sigma^2 \right)\,.
 \label{eqi:FRWct}
 \en
To first order, the most general perturbation one can write down is \cite{Mukhanov:1990me}

 \be
 \delta g_{00}&=&-2a(\tau)^2A(\tau,\mathbf{x})\,,\nn
 \delta g_{0i}&=&a^2(\tau)\left[\partial_i B(\tau,\mathbf{x})+S_i(\tau,\mathbf{x})\right]\,,\nn
 \delta g_{ij}&=&a^2(\tau)\left[(-2\psi(\tau,\mathbf{x}))\delta_{ij}+2\partial_i\partial_jE(\tau,\mathbf{x})\right.\nn
              && \left.+\partial_i F_j(\tau,\mathbf{x})+\partial_j F_i(\tau,\mathbf{x}) + h_{ij}\right]\,.
\label{eqi:gij}
 \en
 
 \noi The indices $i$,$j$ run from $1$ to $3$, representing the spatial dimensions.\\
\noi Scalar perturbations are given by the functions $A$, $B$, $\psi$ and $E$. They are at the origin of structures in the universe \cite{Mukhanov:1990me}. \\
\noi Vector perturbations are described by the functions $S_i$ and $F_i$. They decay very quickly and usually do not lead to interesting cosmological signatures \cite{Mukhanov:1990me}.\\
\noi Tensor perturbations $h_{ij}$ correspond to gravitational waves. Although they are generated during inflation, they typically have a very low amplitude and do not have important effects on structure formation \cite{Mukhanov:1990me}.\\
\noi Scalar, vector and tensor perturbations are decoupled and can therefore be studied separately. 

\subsection{First order energy-momentum tensor perturbations}

The rotational and translational invariance of the unperturbed energy-momentum tensor imply that it takes the perfect fluid form \cite{Weinberg:2008zzc}

\be
T^{\mu}_{\nu}=(\rho+p)u^{\mu}u_{\nu} + p\delta^{\mu}_{\nu}\,,
\label{stress}
\en
\noi where $u^{\mu}$ is the $4$-velocity of the fluid.\\
\noi Perturbing Eqs.~(\ref{stress}) as $T^{\mu}_\nu + \delta T^\mu_\nu$ gives \cite{Weinberg:2008zzc}

\be
\delta T^0_0 &=& -\delta\rho\,,\nn
\delta T^0_i &=& (\rho+p)(\partial_i\delta u + \delta u_i^V)\,,\nn
\delta T^i_0 &=& \frac{\rho+p}{a(\tau)^2}\left(a(\tau)\partial_i B(\tau,\mathbf{x})+a(\tau)S_i(\tau,\mathbf{x})-\partial_i\delta u-\partial u_i^V\right)\,,\nn
\delta T^i_j &=& \delta^i_j\delta p + \partial^i\partial_j\pi^S+\partial^i\pi_j^V+\pi^{i\;T}_{j}\,,\nn
\delta T^\lambda_\lambda &=& 3\delta p - \delta\rho + \nabla^2\pi^S\,,
\label{stressper}
\en
where $i,j$ run over the spatial degrees of freedom whereas $\lambda$ runs over all spacetime degrees of freedom and we used the summation convention.\\ 
\noi The perturbations (\ref{stressper}) are related to perturbations of the Einstein tensor by

\be
G^\mu_\nu + \delta G^\mu_\nu = T^\mu_\nu + \delta T^\mu_\nu\,.
\label{pertein}
\en
The unperturbed quantities balance each other leaving us with the equations

\be
\delta G^\mu_\nu = \delta T^\mu_\nu\,.
\label{pertein2}
\en
Using this relation it is possible to derive evolution equations for the perturbations (see \cite{Weinberg:2008zzc} for a complete treatment).

\subsection{Gauge transformations}

\noi An interesting question is how the perturbations are affected by coordinate transformations such as

\be
x^{\mu} \rightarrow \tilde{x}^{\mu} = x^{\mu} + \xi^{\mu}\,,
\label{coord}
\en
where $\xi^{\mu}$ is an infinitesimal function of spacetime. At a particular point of spacetime, the metric tensor after the change of coordinates (\ref{coord}) becomes \cite{Mukhanov:1990me}

\be
\tilde{g}_{\mu\nu}(\tilde{x}^{\gamma})&=&\frac{\partial x^{\alpha}}{\partial \tilde{x}^{\mu}}\frac{\partial x^{\beta}}{\partial \tilde{x}^{\nu}}g_{\alpha\beta}(x^{\gamma})\nn
                                   &\approx& g_{\mu\nu}(x^{\gamma}) + \delta g_{\mu\nu}(x^{\gamma}) -  g_{\mu\beta}\xi^{\beta}_{\nu} -  g_{\alpha\nu}\xi^{\alpha}_{\mu} \,,
	\label{metrictrsf}
	\en
	where we neglected non-linear terms in $\delta g$ and $\xi$.\\
	\noi Neglecting vector and tensor perturbations, the metric perturbations after the change of variables (\ref{coord}) become \cite{Mukhanov:1990me}
	
	\be
  A &\rightarrow& \tilde{A} = A - \frac{1}{a}\left(a\xi^{0}\right)'\,,\nn
	B &\rightarrow& \tilde{B} = B+\kappa' - \xi^{0} \,,\nn
	\psi &\rightarrow& \tilde{\psi} = \psi + \frac{a'}{a}\xi^0\,, \nn
	E&\rightarrow&\tilde{E} = E+\kappa\,,
	\label{scgauge}
	\en
	where we decomposed $\xi^i=\xi^i_\bot+\partial\kappa/(\partial x^i)$ into a $3-$vector with zero divergence $\xi^i_\bot$ and a scalar function $\kappa$ ($i$ spans the spatial dimensions). The prime $'$ denotes the derivate with respect to conformal time $\tau$. It is not difficult to construct combinations of the four functions (\ref{scgauge}) which do not depend on the coordinate transformation (\ref{coord}), or in other words that are gauge invariant. The simplest gauge invariant linear combinations are \cite{Mukhanov:1990me}
	
	\be 
	\Phi \equiv A - \frac{1}{a}\left[a(B-E')\right]'\,\;\;\;\;\;\;\Psi\equiv-\psi+\frac{a'}{a}\left(B-E'\right)\,.
	\label{gaugeinv}
	\en
	The two quantities are suitable observables, since they do not depend on the observer that is measuring them.\\
	\noi To derive mathematical expressions for them, we have the freedom to choose particular values for the parameters $\xi^0$ and $\kappa$. Fixing them to given values amounts to choosing a gauge.
	
	\subsection{Newtonian gauge and the curvature perturbation}
	
\noi A convenient gauge consists in choosing  $\xi^0$ and $\kappa$ so that $B=0$ and $E=0$. Such choice is called the Newtonian gauge and it reduces the scalar perturbations of the metric to
	
	\be 
	\delta g_{00} = -2a(\tau)^2A(\tau,\textbf{x})\,,\;\;\;\;\;\;\delta g_{ij} = -a(\tau)^2\psi(\tau,\textbf{x})\delta_{ij}\,,
	\label{newtper}
	\en

\noi The function $\psi$ (equivalent to $\Psi$ for this choice of gauge) is of particular interest here as it describes scalar perturbations on constant time hypersurfaces. Its evolution equation derived from Eqs.~(\ref{pertein2}) in the Newtonian gauge is \cite{Weinberg:2008zzc}

\be 
3(\rho+p)\dot{\psi} = \delta\dot{\rho} + 3\frac{\dot{a}}{a}(\delta\rho+\delta p) +\nabla^2\left[\frac{\rho+p}{a^2}\delta u + \frac{\dot{a}}{a}\pi^S\right]\,.
\label{psievo}
\en

\noi Now, consider $\psi$ evaluated on a constant energy density hypersurface \cite{Mukhanov:1990me}:    
 
 \be
 \zeta = -\psi|_{\delta\rho=0}\,.
 \label{eqi:zeta}
 \en
 \noi $\zeta$ is called the curvature perturbation and it has an important feature: it is conserved on superhorizon scales provided that the universe is dominated by a single fluid with a unique equation of state \cite{zetaConv}. Indeed, on superhorizon scales, the operator $\nabla^2$ is negligible, as can be seen by using the properties of the Fourier transform, and Eq.~(\ref{psievo}) evaluated on a constant energy density hypersurface gives

\be
\dot{\zeta} = -\dot{\psi}|_{\delta\rho=0} = \frac{H}{\rho+p}\delta p\,.
\label{zetaevo}
\en
Therefore if $\delta p$ vanishes, $\zeta$ is conserved on superhorizon scales. This is the case if the universe is dominated by a single fluid.\\

\noi Recall that, at the end of inflation, the universe is dominated by the inflationary degrees of freedom, which in general do not have a unique equation of state. Therefore, $\zeta$ is not conserved yet at that time. It is only when the universe reheats and becomes radiation dominated that $\zeta$ becomes time-independent, making it a suitable observable for understanding the signatures of inflation and reheating. 

\subsection{The power spectrum of \texorpdfstring{$\zeta$}{TEXT}}

\noi During slow-roll inflation, the curvature perturbations can be expressed as \cite{liddlelyth}

\be 
\zeta = \left(\frac{H}{\dot{\phi}}\delta\phi_{\mathbf{k}}\right)_{t_*}\,.\label{eqi:srzeta}
\en

\noi If there is a single dynamically relevant scalar field during inflation, the power spectrum (which is formally defined in Eq.~(\ref{g2})) of $\zeta$ is \cite{liddlelyth}

\be 
\mathcal{P}_{\zeta}(k) = \left[\left(\frac{H}{\dot{\phi}}\right)^2\mathcal{P}_\phi (k)\right]_{t_*}\,.\label{eqi:pzeta}
\en

\noi The quantity $\mathcal{P}_\phi (k)$ is given in Eq.~(\ref{eqi:pphihc}). Replacing yealds

\be 
\mathcal{P}_{\zeta}(k) = \left[\left(\frac{H}{\dot{\phi}}\right)^2\left(\frac{H}{2\pi^2}\right)\right]_{t_*}\,.\label{eqi:pzeta2}
\en

\noi The spectrum (\ref{eqi:pzeta2}) follows the power law

\be
\mathcal{P}_{\zeta}(k) \propto k^{\ns-1}\,,
\label{eqi:spectralindex}
\en

\noi where $\ns$ is the spectral index. If $\ns=1$ the spectrum is scale invariant. It is highly constrained by observations: according to the PLANCK data, $\ns= 0.9603\pm0.0073$ at $68\%$ confidence level \cite{planck}.\\
\noi In reality, there is no reason to expect $\mathcal{P}_\zeta$ to follow exactly the power law in Eq.~(\ref{eqi:spectralindex}). We can nevertheless define an effective spectral index:

\be
\ns(k) - 1 = \frac{d\ln\mathcal{P}_\zeta}{d\ln k}\,.
\label{eqi:effspi}
\en

\noi In the limit in which $\ns$ is constant with respect to $k$, Eq.~(\ref{eqi:spectralindex}) and Eq.~(\ref{eqi:effspi}) are equivalent.\\

\noi In order to find an expression for $\ns$ as a function of the inflationary degrees of freedom, we substitute the slow-roll expressions Eqs.(\ref{eqi:SR}) into Eq.(\ref{eqi:pzeta2}). Find

\be 
\mathcal{P}_\zeta = \frac{1}{12\mpl^2}\frac{V^3}{V'^2} = \frac{1}{24\pi^2\mpl^2}\frac{V}{\epsilon}\,.
\label{eqi:szetaSR}
\en

\noi Using again the slow-roll equations and the fact that derivatives are evaluated at horizon crossing, gives \cite{liddlelyth}

\be 
\ns-1=-6\epsilon_*+2\eta_*\,.
\label{eqi:spiSR}
\en

\noi Provided that the slow-roll conditions (\ref{eqi:srconds}) are satisfied, we obtain a nearly scale-invariant spectral index, as observed. Wether $\ns$ is greater or smaller than one depends on the inflationary model and, since $\ns<1$ has been observed, constitutes a viability condition.

\chapter{Observables and Separate Universes Approximation}
\label{ch:sepuni}

In this Chapter, we will present a formalism allowing to derive in a simple way observables associated with the anisotropies of the CMB (\cite{Starobinsky:1986fxa, Sasaki:1995aw, Lyth:2005fi, Seery:2008qj}, for other methods see \cite{otherMethods}). The starting idea is that at the time the CMB photons were emitted $t_{rec}$, the Hubble distance was much smaller than at present time $t_0$: $(aH)^{-1}|_{t_{rec}} \ll (aH)^{-1}|_{t_0} $. Indeed, regions of the sky separated by more than $2^{\circ}$ have not been in causal contact since inflation.\\
\noi Consider two spatially separated regions at $t_{rec}$, such that the distance between them is much greater than $(aH)^{-1}|_{t_{rec}}$. To a good approximation, we can assume that these regions evolve as independent spacetimes, each with its own FRW metric. Indeed, because the spatial gradients of super-horizon modes are negligibly small, the separate universes approximation is nearly exact. 

\section{The \texorpdfstring{$\delta N$}{TEXT} formalism}
\label{sec:dNf}

\noi The separate universes approximation allows to understand the curvature perturbation $\zeta$ in a novel way. First, we rewrite the spatial part of the perturbed FRW metric (\ref{eqi:gij}) keeping only the scalar perturbation $\psi$ and in terms of normal time:

\be 
g_{ij}=a^2(t)e^{-2\psi(t,\mathbf{x})}\delta_{ij}\,.\label{eqdn:gij}
\en

\noi We can define an effective scale factor $\tilde{a} = ae^{-\psi}$. Then, the number of e-folds between times $t$ and $t'$ is:

\be
N(t,t',\mathbf{x}) = \log\frac{a^2(t)e^{-2\psi(t,\mathbf{x})}}{a^2(t')e^{-2\psi(t',\mathbf{x})}}= \psi(t',\mathbf{x}) - \psi(t,\mathbf{x}) + \log\frac{a(t)}{a(t')}\,.
\label{eqdn:Nttpr}
\en

\noi Choosing the initial hypersurface to be flat ($\psi(t',\mathbf{x}) = 0$) and evaluating at a final hypersurface of constant energy density $\rho_f$, we use the definition (\ref{eqi:zeta}) to identify

\be
\zeta = -\psi(t,\mathbf{x})|_\rho = N(t,t',\mathbf{x}) - \log\frac{a(t)}{a(t')}\,.\label{eqdn:zeta}
\en

\noi Thus, we can write

\be 
\zeta = \delta N_*^f\,, \label{eqdn:zeta2}
\en
where $\delta N$ represents the difference in expansion between causally disconnected spacetime regions, $*$ represents the initial flat hypersurface and $f$ the final hypersurface of energy density $\rho_f$. Note that for $\zeta$ to be a conserved quantity, the final energy density $\rho_f$ must be chosen some time after the end of inflation when the universe is dominated by a single fluid, generally taken to be radiation. Eq.~(\ref{eqdn:zeta2}) is the central result of the $\delta N$ formalism and we will use it extensively throughout this thesis. It tells us that if we know how the scale factor depends on the initial conditions, we can determine the curvature perturbation $\zeta$ on super-horizon scales.\\
\noi During slow-roll inflation, the dynamically relevant degrees of freedom are represented by the background value of some scalars $\varphi^{\alpha}$, where $\alpha$ emphasises the fact that there might be more than one. Then we can write 

\be \zeta = \delta N_*^f(\delta\varphi^{\alpha})\,. \en

\noi If the amplitude of the perturbations $\delta\varphi^{\alpha}$ is sufficiently small, the statistics of the curvature perturbations can be determined by Taylor expanding

\be
\delta N^f_* \approx \frac{\partial N^f_*}{\partial \varphi^{\alpha}_*} \delta\varphi^{\alpha}_* + \frac{1}{2}\frac{\partial^2 N^f_*}{\partial \varphi^{\alpha}_*\varphi^{\beta}_*} \delta\varphi^{\alpha}_*\delta\varphi^{\beta}_*\,,
\label{eqdn:taylordn}
\en
where we used the summation convention. In what follows, we will use the common notation $N_{\alpha}={\partial N^f_*}/{\partial \varphi^{\alpha}_*}$.\\
\noi It is straightforward to derive the amplitude of the curvature perturbation $A_\zeta$ using Eq.~(\ref{eqdn:taylordn}):

\be
A_\zeta^2 = N_{\alpha}N_{\alpha}\frac{H_*^2}{4\pi^2}\,.
\label{eqdn:ampl}
\en
The PLANCK results fix the amplitude to $10^9\times A_\zeta^2 = 2.18863^{+0.05317}_{-0.05831}$ at $68\%$ confidence level \cite{planck}.\\
\noi Somewhat more involved is the expression for the spectral index $\ns$ \cite{Sasaki:1995aw,largeNG1b} : 

\be
\ns =1   - 2\epsilon_* + \frac{2}{H_*}\frac{\dot{\varphi}^{\alpha}_*N_{\beta}N_{\alpha\beta}}{N_{\alpha}N_{\alpha}} \,.
\label{eqdn:spectrum}
\en
The PLANCK results give $\ns= 0.9603\pm 0.0073$ at $68\%$ confidence level \cite{planck}.

\section{Non-Gaussianity}
\label{sec:ng}

In the previous section \ref{sec:dNf}, we derived the $\delta N$ formalism expressions for the amplitude $A_\zeta$ and the spectral index $\ns$ of the curvature perturbation. There are other observables that we have the tools to calculate: quantities such as the two-point function and three-point function of $\zeta$. These are measurable quantities, and inflationary models predict a variety of different magnitudes and shapes for them. They constitute therefore an observational test which is very sensitive to the details of the dynamics of inflation.\\
\noi In this section we will introduce the concept of gaussianity and deviations from it, construct an observable to measure the degree of non-gaussianity of the CMB anisotropies and derive an expression to compute this quantity from theoretical inflationary models.

\subsection{Gaussian statistics}  
\label{ssec:gauss}

Let $g(\mathbf{x})$ be a random field defined in a region of volume $L^3$ of the universe. Following Eqs.~(\ref{eq:fourierdis}), it can be expressed in terms of Fourier modes as

\be 
g(\textbf{x}) = \frac{1}{L^3}\sum_n \tilde{g}_n e^{i\textbf{k}_n\cdot \textbf{x}}\,.
\label{eqdn:gauss}
\en
The simplest random field is a Gaussian random field. It is statistically homogeneous and isotropic. It can be defined as one whose Fourier modes $\tilde{g}_n$ have no other correlation than the reality condition Eq.~(\ref{realdis}) \cite{Lyth:2009zz}:

\be
\langle \tilde{g}_n\tilde{g}^*_m\rangle = \delta_{nm} P_{g_n}\,,
\label{grf}
\en
with $P_{g_n}=\langle |\tilde{g}_n|^2\rangle$. For $n\neq 0$, $\langle g_n \rangle=0$ because of homogeneity. For $n=0$, $\langle g_0 \rangle \equiv \langle g\rangle$ can be interpreted as the mean value of the field over the volume $L^3$. The correlators between odd numbers of points vanish because they cannot be expressed in terms of Eq.~(\ref{grf}). Even correlators, on the other hand, can be expanded in terms of the two-point correlator. For example, the $4$-point function can be expressed as

\be
\langle \tilde{g}_1\tilde{g}_2\tilde{g}_3\tilde{g}_4\rangle = \langle \tilde{g}_1\tilde{g}_2\rangle \langle\tilde{g}_3\tilde{g}_4\rangle +\langle \tilde{g}_1\tilde{g}_3\rangle\langle\tilde{g}_2\tilde{g}_4\rangle +\langle \tilde{g}_1\tilde{g}_4\rangle \langle\tilde{g}_2\tilde{g}_3\rangle \,.
\label{4poc}
\en
Higher order even correlators can be expanded similarly.\\
\noi In the limit that $L\rightarrow\infty$, $g(\mathbf{x})$ becomes

\be
g(\textbf{x}) = \frac{1}{(2\pi)^3}\int \tilde{g}(\textbf{k})e^{i\textbf{k}\cdot \textbf{x}}d^3k\,.
\label{contg}
\en
Up to third order, the correlators are given by (with $g_k\equiv \tilde{g}(\textbf{k})$)
\be
\langle g_k\rangle &=& 0 \;\;\;\;\;\;\; {\rm for}\;\textbf{k}\neq 0 \label{g1}\,,\\
\langle g_k g_q \rangle &=&  (2\pi)^3 \delta^3(\textbf{k} + \textbf{q}) P_g\,,\label{g2}\\
\langle g_{k_1} g_{k_2} g_{k_3} \rangle &=& 0\,, \label{g3}
\en
where we defined the power spectrum $P_g\equiv\frac{2\pi^2}{k^3}\mathcal{P}_g(k)$ (this definition is used for non-gaussian fields as well). Note that $P_g$ and $\mathcal{P}_g$ are both referred to as power spectra. The delta function in the two-point correlator, analogously to the Kronecker delta in Eq.~(\ref{eqdn:gauss}), is a consequence of statistical homogeneity.\\
\noi As in the discrete case, $g_0 = \langle g \rangle$ is the mean value.\\
\noi The $4$-point correlator can be expanded as in Eq.~(\ref{4poc}). It can also be written as

\be 
\langle g_{\textbf{k}_1}g_{\textbf{k}_2}g_{\textbf{k}_3}g_{\textbf{k}_4}\rangle =& (2\pi)^6 \delta^3(\textbf{k}_1 + \textbf{k}_2) \delta^3(\textbf{k}_3 + \textbf{k}_4)P_g(k_1)P_g(k_3)\nn
                                                                                 &+{\rm permutations}\,.
\label{4pcspec}
\en

\noi In position space, the second moment $\langle g^2(\textbf{x}) \rangle$, or mean-square $\sigma^2_g (\textbf{x})$, are related to the power spectra by 

\be 
\sigma^2_g(\textbf{x}) \equiv \langle g^2(\textbf{x}) \rangle = \frac{1}{(2\pi)^3}\int_0^\infty P_g(k)d^3k = \int_0^\infty \mathcal{P}_g(k)\frac{d^k}{k}\,.
\label{sigmasq}
\en
The delta function in Eq.~(\ref{g2}) implies that $\sigma^2_g$ is independent of position.\\
\noi As in Fourier space, all odd correlators vanish in position space, as can be shown using Eqs.(\ref{g1}),~(\ref{g2}) and~(\ref{g3}), implying that if $\tilde{g}(\textbf{k})$ is gaussian, so is $g(\textbf{x})$. For example, the real space $2$-point correlator is given by

\be 
\langle g(\textbf{y})g(\textbf{x}+\textbf{y}) \rangle &=& \frac{1}{(2\pi)^3}\int P_g(k)e^{i\textbf{k}\cdot \textbf{x}} \nn
                                                      &=& \int_0^\infty \mathcal{P}_g(k)\frac{\sin(kx)}{kx}\frac{dk}{k}\,,
\label{2pr}
\en
In the second line we used spherical symmetry with $x=|\textbf{x}|$.\\
\noi Summarising, gaussianity means that the statistics can be fully described by the two quantities $\langle g \rangle$ and $\sigma^2_g$. The correlators between different points can be derived using the reality condition Eq.~(\ref{g2}).\\
\noi The PDF $\textbf{P}(g)$ associated with Gaussian statistics can be derived from the above properties and is given by

\be
\textbf{P}(g)=\frac{1}{\sqrt{2\pi}\sigma_g}\exp\left[-\frac{(g-\langle g \rangle)^2}{2\sigma^2_g}\right]\,.
\label{gaussPDF}
\en

\subsection{Parametrisation of non-gaussianity}
\label{ssec:fnldef}

The Gaussian distribution (\ref{gaussPDF}) is a very particular and simple case of statistical distribution. In fact, the Central Limit Theorem of probability theory states that a set of independent random variables with finite variance and mean tends to follow a Gaussian distribution. Reversing the logic and applying it to physics, if perturbations in a quantity follow a Gaussian distribution, it means they arise from independent random variables. If these variables are described by one or more fields, the fields must be non-interacting and in their vacuum state. However, if perturbations are non-Gaussian, we can measure quantities such as the three-point correlator and higher order statistical momenta to learn about the interacting theory giving rise to the pattern of perturbations.\\

\noi In the context of cosmology, learning about the statistics of the curvature perturbation $\zeta$ might teach us something about inflation dynamics. Generalising Eq.~(\ref{g2}), the momentum space $n$-point correlator of $\zeta$ can be written as

\be
\langle \zeta_{\mathbf{k_1}}...\zeta_{\mathbf{k_n}} \rangle = (2\pi)^3\delta(\mathbf{k_1}+...+\mathbf{k_n})P_n(\mathbf{k_1},...,\mathbf{k_n})\,,\label{eqng:ncorr}
\en
where $P_n$ is the $(n-1)$-spectrum. For example, $P_3$ is called the bispectrum and it vanishes if the curvature perturbation is Gaussian. Therefore, deviations from non-gaussianity can be determined by measuring $P_3$, which is usually parametrised in a conveninent way using a non-linearity parameter $\fnl$. Different definitions of such parameter have been considered in the literature. A possibility is to define it as the second order contribution to perturbations \cite{Komatsu:2000vy,Maldacena:2002vr}

\be 
\zeta = \zeta_g^2 + \theta\fnl\zeta_g^2\,,
\label{fnl1}
\en
where $\zeta_g$ is Gaussian and $\theta$ is a generic numerical factor. Using the properties of the Gaussian statistics, it is straightforward to show that a non-zero $\fnl$ sources a non-zero bispectrum. Indeed

\be
\langle \zeta\zeta\zeta \rangle &=& \theta\fnl \langle \zeta_g\zeta_g\zeta^2_g \rangle + {\rm permutations}\nn
                                 && \theta^3\fnl^3  \langle \zeta^2_g\zeta^2_g\zeta^2_g\rangle + {\rm permutations}\,,
\label{ng3p}
\en
where, for simplicity, we omitted to explicitly write the $k$-dependence. Eq.~(\ref{ng3p}) highlights how the third order correlation is sourced by the second order in perturbation.\\
\noi An alternative definition, and the one we will use in this thesis,  uses the ratio of the bispectrum $P_3$ to the spectrum $P_2$:

\be 
\fnl({\mathbf k_1},{\mathbf k_2},{\mathbf k_3}) = \frac{6}{5}\frac{P_3({\mathbf k_1}, {\mathbf k_2},
{\mathbf k_3})}{P_2({\mathbf k_1}, {\mathbf k_2})P_2({\mathbf k_2}, {\mathbf k_3})+{\rm permutations}}\,.\label{eqng:fnldef}
\en
A so defined $\fnl$ is also referred to as the reduced bispectrum. Note that it is a function of the scales $\mathbf{k}_1$,$\mathbf{k}_2$, and $\mathbf{k}_3$. In truth, $\fnl$ only depends on two of the three vectors, since the third momentum is fixed by the $\delta$-function in Eq.~(\ref{eqng:ncorr}) that constrains us to triangular shapes. Experimentally, measurements focus on particular shapes. The most studied shapes are the local form \cite{Komatsu:2000vy}, or squeezed form, corresponding to triangles such that $|\mathbf{k}_1|\simeq|\mathbf{k}_2|\gg|\mathbf{k}_3|$; the equilateral form \cite{Creminelli:2006rz}, corresponding to $|\mathbf{k}_1|\simeq|\mathbf{k}_2|\simeq|\mathbf{k}_3|$; the orthogonal form \cite{Senatore:2009gt} which roughly consists of shapes that are orthogonal to both the local and the equilateral forms.\\
\noi In this thesis we will focus on the local form. In an inflating universe, using the $\delta N$ formalism result Eq.~(\ref{eqdn:taylordn}) and assuming slow-roll is a good approximation at the time of horizon crossing, we find \cite{Lyth:2005fi}

\be 
 \fnl^{\rm local} = \frac{5}{6}\frac{N_{\alpha}N_{\beta}N_{\alpha\beta}}{(N_{\alpha}N_{\alpha})^2} +    
        \ln(kL)\mathcal{P}_\zeta\frac{N_{\alpha\beta}N_{\beta\gamma}N_{\gamma\alpha}}{(N_{\alpha}N_{\alpha})^3}\,.
\label{fnldn1}
\en
The scale dependence is exclusively in the second term, it is logarithmic and small in magnitude compared to the scale independent term and is usually thought of as a loop correction. Therefore, we can express $\fnl^{\rm local}$ in terms of a constant contribution, which may be interpreted as an average, and a shape-dependent one. From here on, we will focus on the shape-independent term of $\fnl^{\rm local}$ and refer to it as $\fnl$:

\be 
 \fnl = \frac{5}{6}\frac{N_{\alpha}N_{\beta}N_{\alpha\beta}}{(N_{\alpha}N_{\alpha})^2}\,.
 \label{eqng:fnldn}
\en 
\noi

\noi In deriving Eq.~(\ref{fnldn1}), it was assumed that the only source of non-gaussianity comes from the higher order terms in the $\delta N$ expansion Eq.~(\ref{eqdn:taylordn}). In general, however, the field perturbations $\delta\varphi^{\alpha}$ will also generate some non-gaussianity. The assumption that the non-gaussianity of the field perturbations is negligible is justified by the slow-roll approximation \cite{liddlelyth}.\\

\noi Of course, Eq.~(\ref{eqng:fnldef}) is not the only way to measure non-gaussianity. In fact, $\fnl$ might be zero and the curvature perturbation still be non-Gaussian. In general, a non-Gaussian distribution requires an infinite number of parameters to be described: we must know all the correlation functions, for $n$ going all the way up to infinity! However, we must start somewhere and given that obervations give us reasonably refined constraints on $\fnl$, it seems as a good starting point. Higher order spectra are much less constrained by present observations because of the increasing practical challenges.\\

\subsection{Experimental measurements of non-gaussianity}
\label{ssec:fnlOBS}

The most precise measurements of the properties of the CMB are due to the PLANCK collaboration \cite{planck}. The constraint on the local reduced bispectrum $\fnl$ is

\be
\fnl = 2.7 \pm 5.8\,\,\,\,(68\%{\rm C.L.})\,.
\label{fnlOBS}
\en
On the equilateral parameter $\fnl^{\rm equil}$

\be
\fnl^{\rm equil} = -42 \pm 74 \,\,\,\,(68\%{\rm C.L.})\,.
\label{fnleq}
\en
On the orthogonal parameter $\fnl^{\rm orth}$

\be
\fnl^{\rm orth} <-25 \pm 39\,\,\,\,(68\%{\rm C.L.})\,.
\label{fnlorth}
\en

\noi PLANCK considerably increased the precision of the constraints, although it did not rule out a vanishing $\fnl$. 

\subsection{Numerical computation of observables}
\label{ssec:numDeltaN}

In the following chapters, we will investigate the super-horizon signatures of a series of inflationary models. It is not always possible to compute the quantity $\delta N_*^f$ and its derivatives analytically, and we will often use numerical techniques.\\
\noi Consider an inflationary model with $\varphi^{\alpha}$ degrees of freedom. Then, the algorithm proceeds as follows. The homogeneous equations of motion of the system are solved numerically  

\be
\ddot{\varphi}^\alpha + 3 H \dot{\varphi}^\alpha + \frac{\partial V}{\partial \varphi^\alpha} &=& -\Gamma_\alpha \dot{\varphi}_\alpha\,,\nonumber\\
\dot{\rho}_{\rm rad} + 4 H \rho &=& \Gamma_\alpha \dot{\varphi}_\alpha^2\,,
\label{eqnumdN:eoms}
\en
where
\be
H^2 = \frac{1}{3 \mpl^2}\left(\frac{1}{2}\dot{\varphi}^\alpha \dot{\varphi}^\alpha +V +\rho_{\rm rad}\right)\,.
\label{endennum}
\en
The $\Gamma$'s in the equations of motion have been introduced to simulate the decay of the inflationary degrees of freedom $\varphi^{\alpha}$ to radiation. They are crucial since it is only when all the energy has been transferred to radiation that the curvature perturbation $\zeta$ and its primordial statistical properties are conserved. They must be subdominant compared to the energy scale of the system until the very end of inflation, that is $\Gamma \ll H_*$.\\
\noi In order to compute $\delta N_*^f(\varphi^{\alpha})$, the system is solved starting from initial conditions $\varphi^{\alpha}_{\rm i}$ such that $60$ e-folds or more of expansion are realised along the trajectory. Ideally, the integration should stop when all the energy has been transferred to $\rho_{\rm rad}$, however, in practice, this takes an infinite amount of time. Therefore, the evolution is followed until most of the energy has decayed into radiation, say $99.9\%$ and the Hubble rate at this time, $H_f$ is used to define the final hypersurface. The values of the fields $\varphi^{\alpha}_*$ $60$ e-folds before the end of inflation is also recorded. The numerical system is then re-run after varying the initial conditions at horizon crossing $\varphi^{\alpha}_*$ and the number of e-folds to the same final Hubble rate $H_f$ is recorded. The operation is repeated until we have enough points of the function to build up a finite difference approximation to the derivatives of the function $N_*^f(\varphi^{\alpha})$ required to compute the observables (\ref{eqdn:ampl}), (\ref{eqdn:spectrum}) and (\ref{eqng:fnldn}).\\

\subsubsection{A more detailed description of the numerical method}

\noi For a given set of parameters, the initial conditions are chosen as follows. First of all, since the integration starts in the slow-roll regime, Eqs.~(\ref{eqi:SR}) can be used to determine the time derivates of the fields $\dot{\varphi}^{\alpha}$.  Note that the integrated trajectory must be in the classical regime as opposed to the eternal inflation regime described in subsection~\ref{ssec:ei}. In practice this means checking that in Eq.~(\ref{eqei:ei}) the right hand side is larger than the left hand side. Furthermore, the trajectory has to lead to at least $60$ e-folds of inflation, so that the field values at horizon crossing $\varphi^{\alpha}_*$ can be inferred. The trajectory can be identified using anlytical approximations such as slow-roll, which should always be valid at horizon crossing, and other suitable simplifications of the equations of motion or by numerically spanning the space of initial field values. The initial $\rho_{\rm rad}$ is set to zero and the $\Gamma$s must be chosen to be at least a few order of magnitude smaller than the initial Hubble rate $H_{\rm i}$. Given that $H$ varies very slowly during inflation, the hierarchy between the two quantities need not be extremely large, but one should always check that $\Gamma<H$ during inflation and for more than a few oscillations after. The effect of varying the $\Gamma$s on the observables is the subject of recent papers \cite{Elliston:2011dr,Leung:2013rza}. The authors found that although the spectral index $\ns$ is quite insensitive to the change of the decay rates, $\fnl$ depends more significantly on them. Therefore, one should either check that varying the decay rates does not have a significant effect on the final values of the observables, or have physical arguments to justify a particular choice of $\Gamma$s.\\

\noi To compute the observables (\ref{eqdn:ampl}), (\ref{eqdn:spectrum}) and (\ref{eqng:fnldn}), the first and second derivatives of $ N_*^f(\varphi^{\alpha})$ need to be approximated. We used the finite difference formulas

\be
\frac{\partial N_*^f}{\partial \varphi^\alpha} &\simeq& \frac{N_*^f(\varphi^\alpha+2\Delta\varphi^{\alpha})-N_*^f(\varphi^\alpha)}{\Delta\varphi^{\alpha}}\,,\nn
\frac{\partial^2 N_*^f}{\partial (\varphi^\alpha)^2}&\simeq& \frac{N_*^f(\varphi^\alpha+2\Delta\varphi^{\alpha}) - 2N_*^f(\varphi^\alpha+\Delta\varphi^{\alpha})+N_*^f(\varphi^\alpha)}{(\Delta\varphi^{\alpha})^2}\,,\nn
\frac{\partial^2 N_*^f}{\partial \varphi^\alpha\partial\varphi^{\beta}}&\simeq&  \left[N_*^f(\varphi^\alpha+2\Delta\varphi^{\alpha},\varphi^\beta+2\Delta\varphi^{\beta}) - N_*^f(\varphi^\alpha+\Delta\varphi^{\alpha},\varphi^{\beta})\right.\nn
                                                                            && \left.-N_*^f(\varphi^\alpha,\varphi^\beta+\Delta\varphi^{\beta})+N_*^f(\varphi^\alpha,\varphi^{\beta})\right]\frac{1}{4\Delta\varphi^{\alpha}\Delta\varphi^{\beta}}\,,
\label{finitediff}
\en
where $\alpha\neq\beta$. The total number of trajectories to be integrated is $9$. It is possible to reduce them to $6$, but this comes at the expense of precision. The differences $\Delta\varphi^{\alpha}$ must be chosen carefully for Eqs.~(\ref{finitediff}) to provide a good approximation fo the derivatives. They must lead to $\delta N_*^f/N_*^f\lesssim 10^{-2}$, otherwise the discretization error becomes dangerously large. Also, the numerical precision of the machine imposes a lower bound on $\Delta\varphi^{\alpha}$. In practice one must check how the finite difference approximation depends on $\Delta\varphi^{a}$ by choosing increasingly smaller values, until the round-off error becomes too large. For the systems we integrated, we found that choosing $\Delta\varphi^{\alpha}=10^{-3}\varphi^{\alpha}$ is a good compromise.\\

\noi To integrate the system, we used MATLAB $7.10$ and subsequent versions. The most efficient algorithm to integrate Eqs.~(\ref{eqnumdN:eoms}) is the differential equation solver ode$113$, which interpolates between different methods depending on the behaviour of the solutions. Given that the integration starts during inflation, when the fields are slowly rolling along the potential, and it finishes when they are oscillating about their minima, such a flexible solver is very convenient. The absolute tolerance enforced was at least $10^{-20}$ for the final integrations. The relative tolerance was fixed to $5\times10^{-14}$ for all the integrations.\\

\noi For our purposes, a last generation desktop or personal computer was sufficient. The integration times were of the order of minutes to build up a finite difference approximation.

\section{Observables from Chaotic Inflation}
\label{sec:ciNG}

In this section, we will apply the $\delta N$ formalism results to chaotic inflation. In particular, we are interested in the non-gaussianity generated on super-horizon scales.\\
\noi In the previous section, we derived Eqs.~(\ref{eqdn:ampl}),~(\ref{eqdn:spectrum}) and (\ref{eqng:fnldn}) for, respectively, the amplitude of the curvature perturbation $A_\zeta$, the spectral index $\ns$ and the non-gaussianity parameter $\fnl$. The only quantity required to evaluate these observables is $N^f_*(\varphi)$, where $\varphi$ represents the inflationary degrees of freedom.\\
\noi The potential we considered in \ref{ssec:ci} is

\be 
V=\frac{\lambda \phi^n}{n}\,,
\en 
and, during the slow-roll phase, the dynamics follow the attractor solution

\be
3H^2   \simeq \frac{\lambda \phi^n}{n}\,\,\,\,\, {\rm and} \,\,\,\,\,   3H\dot{\phi} &\simeq& -\lambda\phi^{n-1}\,.
\label{eqobs:chSR}
\en
The field $\phi$ being the only degree of freedom in this model, we need to find $N_*^f(\phi)$. Using Eq.~(\ref{eqi:Nphi}), we have

\be
N_*^f(\phi) &\simeq& \int_{\phi_f}^{\phi_*}\frac{V}{V'}d\phi\nn
            &=&      \frac{1}{n}\int_{\phi_f}^{\phi_*} \phi d\phi\nn
            &=&      \frac{1}{2n}\left(\phi_*^2-\phi_f^2\right)\,.
\label{eqobs:Nch}
\en
The derivatives are now straightforward to compute:

\be
\frac{\partial N_*^f}{\partial\phi_*} = \frac{\phi_*}{n} \,\,\,\,\, {\rm and} \,\,\,\,\, \frac{\partial^2 N_*^f}{\partial\phi_*^2} = \frac{1}{n}\,.
\label{eqobs:dNch}
\en
In the subsection \ref{ssec:ci}, we found $\phi_*\simeq 15.5$. We need three more quantities to compute the observables: $H_*$, $\dot{\phi_*}$ and $\epsilon_*$. Using Eqs.~(\ref{eqobs:chSR}), we find:

\be
H_*  \simeq \sqrt{\frac{\lambda\phi_*^n}{3n}}\,,\,\,\,\,\,
\dot{\phi}_* \simeq -\sqrt{\frac{n}{3}\lambda\phi_*^{n-2}}\,,\,\,\,\,\,
\epsilon_*   \simeq\frac{n^2}{2\phi_*^2}\,.
\label{eqobs:chiSR2}
\en
Substituting the relevant quantities into Eqs.~(\ref{eqdn:ampl}),~(\ref{eqdn:spectrum}) and (\ref{eqng:fnldn}) and using $\phi^2_*\simeq 120n$ gives:

\be
A^2_\zeta = \frac{\lambda\phi_*^{n+2}}{12n^3\pi^2}
\label{eqobs:champl}
\en
for the amplitude,

\be
\ns = 1  - \frac{n(n+2)}{\phi_*^2} = 0.9833 - 0.0083n
\label{eqobs:chns}
\en
for the spectral index and

\be
\fnl = \frac{5}{3\phi_*^2} \simeq 0.007
\label{eqobs:chfnl}
\en
for $\fnl$.\\
\noi The amplitude is fixed by observations to be $10^9\times A_\zeta^2 = 2.18863^{+0.05317}_{-0.05831}$ at $68\%$ confidence level \cite{planck} and we can use it to evaluate the free parameter of the model. For $n=2$, find $m\equiv\lambda\simeq4\times10^{-6}$. Requiring the spectral index $\ns$ to be inside the $68\%$ C.L. range $\ns=0.9603\pm0.0073$ of PLANCK, implies the constraint $2\leq n <4$ . On the other hand, the non-gaussianity parameter $\fnl$ is very small and undetectable from a practical point of view. \noi Therefore, the simplest inflationary scenarios can lead to predictions for the spectral index $\ns$ and non-gaussianity parameter $\fnl$ in good agreement with PLANCK data.\\

\noi In fact, it has been rigorously proven that for inflationary models with a single relevant degree of freedom with canonical kinetic term, the Fourier space $3$-point function of the curvature perturbation is related to the spectrum by the following consistency relation \cite{consist}  

\be
\lim_{k_1\rightarrow0} \langle \zeta_{\mathbf{k}_1}\zeta_{\mathbf{k}_2}\zeta_{\mathbf{k}_3} \rangle &=& -(2\pi)^3\delta^3\left(\sum_i\textbf{k}_i\right)P_\zeta(\mathbf{k}_1)P_\zeta(\mathbf{k}_3)\frac{d\log k_3^3P_\zeta(\mathbf{k}_3)}{d\log k_3}\nn
                                                                                                    &\simeq&  \delta^3\left(\sum_i\textbf{k}_i\right)P_\zeta({\mathbf{k}_1})P_\zeta(\mathbf{k}_3)(1-\ns)\,,
\label{consist}
\en
where we neglected numerical factors and $k\equiv|\textbf{k}|$.\\
\noi This relation tells us that, in the squeezed limit, the $3$-point function is suppressed by the factor $(1-\ns)\simeq 0.01$. Using the definition of $\fnl$ given in Eq.~(\ref{eqng:fnldef}), we find $\fnl\simeq(1-\ns)$. Therefore, the local non-gaussianity produced by canonical single field models is always very small and undetectable with the current technology. If slow-roll is a good approximation, it can be shown that a similar consistency relation applies to the shape dependent $\fnl$, implying that the non-gaussianity is negligibly small for all triangular shapes \cite{Maldacena:2002vr}.

\chapter{Hybrid Inflation}
\label{ch:hi}

In the previous Section \ref{sec:ciNG}, we argued that canonical single scalar field models of inflation always generate unobservable non-gaussianity and red spectral index. Given the observational situation outlined in the subsection \ref{ssec:fnlOBS}, this is compatible with current observations. Nevertheless, it is important to understand how more-complex, and perhaps realistic, models of inflation change these results.\\  

\noi A multi-field model of particular interest is known under the name of hybrid inflation and was first introduced by Andrei Linde in 1993 \cite{Linde:1993cn}. It is given by the two real scalar fields potential

\be 
V(\phi,\chi) = \frac{1}{2}m^2\phi^2 + \frac{1}{2}g^2\phi^2\chi^2 + \frac{\lambda}{4}\left(\chi^2-v^2\right)^2\,,
\label{eqhi:hi}
\en
where $m$, $v$ are dimensionfull and  $\lambda$, $g$ are dimensionless parameters. The potential (\ref{eqhi:hi}) combines two successful ideas, one coming from cosmology, that is chaotic inflation, and one coming from particle physics, that is spontaneous symmetry breaking. For this reason it is called hybrid inflation. Potentials of this type are easily embedded in high-energy particle theories such as supersymmetry or its local version supergravity \cite{SUSY}, as we will see in the next chapter, allowing for intriguing connections to be made between particle physics and cosmology.\\

\noi The phenomenology of (\ref{eqhi:hi}) as an inflationary model is very rich and depends sensitively on which term of the potential dominates the inflationary phase. Below, we will describe the features that are common to all the inflationary scenarios based on (\ref{eqhi:hi}).\\

\noi The mass of the field $\chi$ when it sits at $\chi=0$ is given by:

\be 
m^2_\chi= -\lambda v^2 + g^2\phi^2 \equiv g^2(\phi^2-\phic^2),
\label{eqhi:msqchi}
\en
where $\phic=\sqrt{\lambda}v/g$ is the critical value of $\phi$, that at which $m^2_\chi=0$. For $\phi>\phic$ the mass (\ref{eqhi:msqchi}) is positive and $\chi=0$ is stable. However, for $\phi<\phic$, $\chi$ becomes tachyonic at zero and the symmetry breaking phase transition to the global minimum $(\phi=0,\chi=\pm v)$ of the potential (\ref{eqhi:hi}) begins.\\
\noi Inflation typically starts at large field values $\phi>\phic$, with $\langle\chi\rangle=0$. For this reason, we will call $\phi$ the inflaton. In this phase, the inflaton rolls down its effective quadratic potential

\be
V_{\rm eff}(\phi)= \frac{1}{2}m^2\phi^2 + \frac{\lambda v^4}{4}\,.
\label{eqhi:veffphi}
\en
This phase can be either vacuum dominated or $\phi$ dominated, depending on the hierarchy between the two terms in (\ref{eqhi:veffphi}).\\
\noi Eventually, $\phi$ reaches the critical value $\phic$ and the symmetry breaking phase transition to the global minimum starts. Two limiting regimes can be identified: either $\chi$ rapidly acquires a heavy mass and the transition is fast and highly non-linear, or $\chi$ remains light for several Hubble times and a second phase of inflation happens.\\

\noi In the following sections we will study three different regimes of hybrid inflation. In section \ref{sec:vdhchi} we consider vacuum dominated inflation with a fast phase transition. This is the regime in which hybrid inflation was first considered. In section \ref{sec:vdlchi}, inflation is still vacuum dominated but the phase transition is slow. Finally, in section \ref{sec:philchi}, we study the regime in which inflation is $\phi$ dominated and the phase transition is slow.

\section{Vacuum Dominated with Heavy \texorpdfstring{$\chi$}{TEXT}}
\label{sec:vdhchi}

\noi In this regime, the inflationary scenario is as follows: initially $1>\phi>\phic$, $\langle \chi \rangle =0$ and the potential (\ref{eqhi:hi}) is vacuum dominated, $V_{\rm eff}=\lambda v^4/4$. When $\phi$ reaches the critical value, $\chi$ becomes tachyonic and the symmetry breaking phase transition starts, promptly ending inflation. This phase is highly non-linear and it is frequently referred to as the waterfall phase.\\
\noi Of course, if inflation starts at $\phi$ values which are large enough, the potential will initially be dominated by the inflaton's mass term. However, only the last $60$ e-folds of inflation are relevant for observations. What we really mean, is that at least during those last $60$ e-folds of inflation, the potential is dominated by the constant term. 

\subsection{Constraints on parameter space}
\label{ssec:cps}

The small field value assumption $1>\phi>\phic$ translates to $\lambda v^2 \ll g^2$. This condition, together with the requirement that the potential (\ref{eqhi:hi}) be vacuum dominated, implies $\lambda v^4 > m^2$. Another feature of this regime is that $\chi$ is heavy, meaning that once $\phi$ reaches $\phic$, the symmetry breaking phase transition is fast and happens in less than $1$ e-fold, thus ending inflation. This requires the potential (\ref{eqhi:hi}) to be steep enough in the $\chi$ direction, leading to the constraint $v<1$. An additional necessary condition is that $\phi$ rolls fast through $\phic$ so that $\chi$ rapidly acquires a heavy mass. We can estimate the constraint on the parameters by assuming the slow-roll approximation to be still valid when $\phi$ rolls through $\phic$. For $\phi>\phic$ and $\langle \chi \rangle =0$, the slow-roll equation of motion of the inflaton is
\be
\frac{\partial \phi}{\partial N} \simeq - \frac{4m^2\phi}{\lambda v^4}\,,
\label{eqvdh:eom1}
\en
where we used the relation

\be
\frac{\partial}{\partial t} &=& \frac{\partial N}{\partial t} \frac{\partial}{\partial N} = \frac{\partial\ln a}{\partial t}\frac{\partial}{\partial N}\nn
                            &=& H\frac{\partial}{\partial N}\,.
\en
Linearising about $\phic =\phi - \Delta \phi$ gives
\be
 \Delta N \simeq \Delta\phi\phic\lambda v^4/(4m^2)\,.
\label{dNlinn}
\en
Furthermore, replacing the linearised $\phi$ into Eq.~(\ref{eqhi:msqchi}) gives 
\be
m^2_\chi = g^2(\phi^2-\phic^2) \simeq 2g^2\phic\Delta\phi + \mathcal{O}(\Delta\phi^2)\,.
\label{dphilinn}
\en 
Combining Eqs~(\ref{dNlinn}) and~(\ref{dphilinn} and neglecting second order in $\Delta\phi$, we find that $|m^2_\chi| < H^2 = \lambda v^4/12$ for $\Delta N \approx 0.01\lambda v^6/m^2$, a combination of parameters that we require to be less than unity. Summarising, this regime requires the parameter choices:

\be
\lambda v^2 < g^2\,,\;\;\;\;&&\;\;\;\;\lambda v^6 < 100 m^2\,, \nn
\lambda v^4 > m^2\,, \;\;\;\;&&\;\;\;\;  v<1 \,.
\label{eqvdh:parameters}
\en

\subsection{Observables}
\label{ssec:obsvdhc}

Let's now explore the dynamics and observational signatures of this scenario. \\
\noi Initially, $\mpl>\phi>\phic$, $\langle \chi \rangle = 0$ and the inflaton evolves according to Eq.~(\ref{eqvdh:eom1}), which has solution

\be
\phi=\phi_*\exp\left[-\frac{4m^2}{\lambda v^4}N\right]\,.
\label{eqvdh:phi}
\en
The amount of inflation between the horizon crossing value $\phi_*$ and $\phic$ is

\be
N_*^{\rm crit} = \frac{\lambda v^4}{4m^2}\ln\frac{\phi_*}{\phic}\,
\label{eqvdh:Nc}
\en
and

\be 
\frac{\partial N_*^{\rm crit}}{\partial \phi_*}= \frac{\lambda v^4}{4m^2}\frac{1}{\phi_*}\,,\;\;\;\;\;\;\;\;\frac{\partial^2 N_*^{\rm crit}}{\partial \phi_*^2}  = -\frac{\lambda v^4}{4m^2}\frac{1}{\phi_*^2}\,.
\label{eqvdh:dNc}
\en
Therefore, the non-gaussianity parameter $\fnl$ evaluated at $\phic$ is:

\be 
\fnl^{\rm crit} = -\frac{10}{3}\frac{m^2}{\lambda v^4}\,,
\label{eqvdh:fnlc}
\en
which, given the parameter choices in Eq.~(\ref{eqvdh:parameters}), is very small and undetectable.\\
\noi The spectral index evaluated at $\phic$ is:

\be 
\ns^{\rm crit} &=& 1 + \frac{m^2}{\lambda v^4}\left(12 - 16\frac{m^2\phi_*^2}{\lambda v^4}\right)\nn
               &\simeq& 1 + \frac{12m^2}{\lambda v^4} > 1\,,
\label{eqvdh:nsc}
\en
which is highly disfavored by current observations.\\

\noi However, when $\phi$ reaches $\phic$ the curvature perturbation is not yet conserved and the observables (\ref{eqvdh:fnlc}) and (\ref{eqvdh:nsc}) might change by the end of the waterfall phase. Since we are interested in observables associated with the CMB, we need to understand how the phase transition affects long wavelength modes. To that end, we present the following argument, due to Mulryne et al \cite{dmarg}. Assuming that the mass of $\chi$ changes instantaneously from $0$ to $-\lambda v^2$ when $\phi$ reaches $\phic$, we can write down the potential of $\chi$ during the phase transition:

\be
V(\chi) = \frac{\lambda}{4}\left(\chi^2-v^2\right) = V_0 - \frac{\lambda v^2}{2} \chi^2 + \frac{\lambda\chi^4}{4}\,.
\label{eqvdh:vchi}
\en
Given that once $\phi$ rolls past $\phic$ the system reaches the global minimum in less than $1$ e-fold, we can neglect the expansion and drop the first time derivative in the equations of motion. We have
\be 
\ddot{\chi} - \nabla^2\chi - \lambda v^2 \chi^2 + \lambda \chi^3 = 0\,.
\label{eqvdh:eomchi}
\en
We are interested in the evolution of the long wavelength modes of $\chi$, or in other words, of $\chi$ smoothed over a large length scale $L\gg H^{-1}$. Let us refer to the smoothing operator as $s_L$ and to the smoothed $\chi$ as $s_L[\chi]$.  The smoothing can be defined in a series of ways, but for our purposes all we need is that the smoothing is a linear projection, namely

\be
s_L[c_1\chi_1 + c_2\chi_2] = c_1s_L[\chi_1]+c_2s_L[\chi_2]
\label{eqvdh:linsmooth}
\en
for any constant $c_1$ and $c_2$, and

\be
s_L\left[s_L[\chi]\right] = s_L[\chi]\,.
\label{eqvdh:smoothproj}
\en
Given these properties, we can use the smoothing to decompose $\chi$ into a short-wavelength part $\chi_S$ and a long-wavelength part $\chi_L$ as:

\be
\chi = \chi_S + \chi_L\,.
\label{eqvdh:chils}
\en
The equation of motion for the long-wavelength part $\chi_L$ is obtained by applying the operator $s_L$ to the equation of motion (\ref{eqvdh:eomchi}):

\be
\ddot{\chi}_L -\lambda v^2 \chi_L + \lambda \chi_L^3 + \lambda\Delta =0\,,
\label{eqvdh:eomchil}
\en
where $\Delta = s_L[\chi^3] - \chi^3_L$. Crucially, the $\chi$ field being heavy, the gradient term is strongly suppressed on super-horizon scales and can be safely dropped. As can be seen from Eq.~(\ref{eqvdh:eomchil}), $\chi_L$ has the same equation of motion as the unsmoothed $\chi$ apart from the additional effective term $\lambda\Delta$. In order to estimate this term, we use the decomposition (\ref{eqvdh:chils}) and the properties (\ref{eqvdh:linsmooth}) and (\ref{eqvdh:smoothproj}) of the operator $s_L$:

\be
\Delta = 3\chi_L^2s_L[\chi_S] + 3\chi_Ls_L[\chi_S^2]+s_L[\chi_S^3]\,.
\label{eqvdh:delta1}
\en
The first term vanishes because of the definition (\ref{eqvdh:chils}). The last term vanishes in the limit in which the symmetry $\chi\rightarrow-\chi$ is exact. Although during the transition this symmetry is broken, we assume that the last term is negligible. There remains the middle term, which, once the short-wavelength modes reach the minimum of the potential, $\chi_S^2=v^2$, gives

\be
\Delta = 3v^2\chi_L\,.
\label{eqvdh:delta2}
\en
The equation of motion for $\chi_L$ becomes:

\be
\ddot{\chi_L} + 2\lambda v^2\chi_L + \lambda \chi^3_L = 0\,.
\label{eqvdh:eomchil2}
\en
When the short-wavelength modes reach the minimum, they kill the tachyonic instability of the long-wavelength modes, stabilizing them. For the parameter choices (\ref{eqvdh:parameters}), the short wavelength modes reach the minimum nearly instantaneously when $\phi$ rolls past $\phic$ and no perturbations are generated on long wavelengths modes. Therefore,  the observational predictions on super-horizon scales are not affected by the tachyonic phase transition, and the prediction for $\fnl$ and $\ns$ remain those given in Eqs.~(\ref{eqvdh:fnlc}) and (\ref{eqvdh:nsc}):

\be 
\fnl &=& -\frac{10}{3}\frac{m^2}{\lambda v^4}\,, \nn
\ns  &\simeq& 1 + \frac{12m^2}{\lambda v^4} > 1     \,.
\label{eqvdh:obs}
\en
Thus this scenario generates a power spectrum that on CMB scales has properties that are at odds with current observational bounds \cite{planck,observations}.

\subsection{Non-equilibrium aspects of the waterfall phase}
\label{ssec:neqhi}

Although, as we argued above, the waterfall phase does not affect observables on super-horizon scales, on scales smaller than the horizon its signatures are highly non-trivial. The dynamics of this phase are characterized by non-equilibrium effects and the generation of bound structures that drastically affect the evolution of causal regions. It is referred to as tachyonic preheating and it has an important role in reheating the universe after inflation (see \cite{Copeland:2002ku}). Below, we describe the main aspects of this phase.\\

\noi If $\chi$ is a real scalar, the symmetry breaking generates a network of domains where different choices of vacuum have been made, either $\chi=+v$ or $\chi=-v$, with properties determined by causality \cite{Kibble:1976sj}. These domains are separated by domain walls, regions of space where $\chi$ interpolates between $+v$ and $-v$, going through $\chi=0$. They are characterized by a non-zero energy density, with a surface tension roughly given by $\sigma_{\rm DW}\simeq V(\phi=0,\chi=0)$. Assuming the area of a wall inside a Hubble region  of volume $\mathcal{V} \simeq H^{-3}$ is approximately $\mathcal{A} \simeq H^{-2}$, the energy density of the wall is $\rho_{\rm DW} = \sigma_{\rm DW}\mathcal{A}/\mathcal{V}\propto H$. Given the first Friedmann equation (\ref{FRW1}), we can estimate the wall's contribution to the total energy density as

\be
\Omega_{\rm DW} = \frac{\rho_{\rm DW}}{\rho_{\rm tot}} \propto \frac{1}{H} \propto t.
\label{eqneqhi:dwen}
\en
It means that the fraction of energy contained in the domain wall grows with time and, eventually, it comes to dominate the evolution of the causal region containing it. This is in contrast with observation and severely constrains the surface tension $\sigma_{\rm DW}$, requiring it to be of order of a few $MeV$, incompatibly with the higher energy scale required for inflation \cite{topDef}.\\
\noi This issue can be solved by requiring $\chi$ to be  a multi-component field. In fact, in SUSY derived hybrid models $\chi$ generally transforms as a non-trivial representation of a group $\mathcal{G}$ \cite{minSUSY}. Depending on the dimensions of $\mathcal{G}$, the symmetry breaking phase transition can generate different types of defects, these being cosmic strings if $\mathcal{G}=O(2)$, monopoles if $\mathcal{G}=O(3)$ and textures if  $\mathcal{G}=O(4)$ \cite{topDef}. For higher dimensional symmetry groups, no topological defects are created \cite{topDef}. If the broken continuous symmetry is global, the transition generates massless scalars, known as Goldstone bosons, which do not decay and lead to considerable issues in cosmology. If the symmetry is not exact, however, the Goldstone bosons acquire a small mass and might lead to interesting phenomenologies. Gauging the symmetry kills the Goldstone bosons, however this comes at the expense of considerable complications in the model.\\
\noi Of all the defects mentioned above, cosmic strings are those with the most interesting cosmological effects \cite{topDef}. They can possibly play an important role in structure formation following the end of inflation \cite{Contaldi:1998qs}.\\

\noi A less catastrophic aspect, but yet constrained by observations, is the creation of primordial black holes during the waterfall phase \cite{GarciaBellido:1996qt} . Intuitively, they originate from the large density fluctuations characterising the symmetry breaking transition: overdense regions, such that $\delta\rho/\rho \approx \mathcal{O}(1)$, can potentially form black holes. These overdense regions are not an exclusive feature of hybrid inflation \cite{bholes}. In order for a model to be viable, the black holes created cannot be too big and too many, providing a constraint on the scale at which they form.
For the hybrid potential, the constraint is satisfied in the limit that $\lambda v^2 \gg H^2$ \cite{lythBH}.\\

\noi Other objects that are created during the phase transition are known as oscillons \cite{Copeland:2002ku,Amin:2011hj}. They are localised configurations in which the inflaton is oscillating about the minimum $\phi=0$ with large amplitude and high frequency. Perhaps the most surprising feature of these objects is that they can have a long lifetime, and might have an important role in reheating the universe after inflation. It is interesting to note that, if the inflaton were a complex field, these objects would be Q-balls \cite{Broadhead:2003ni} and conservation of charge would prevent them from decaying, making them potential candidates for dark matter.   

\section{Vacuum Dominated with Light \texorpdfstring{$\chi$}{TEXT}}
\label{sec:vdlchi}

Considering all the problematic issues arising from the non-equilibrium dynamics of tachyonic preheating, it is tempting to consider regimes in which inflation continues after the instability. The velocity of the symmetry breaking phase transition depends on the mass of the field $\chi$: the smaller the mass, the slower the rolling along the trajectory to the vacuum. Depending on the speed and the displacement from $\chi=0$ of the trajectory when it goes through the instability, inflation after the critical point can last much more than $60$ e-folds of expansion \cite{Clesse:2010iz}. In such a scenario, topological defects and every other feature of the spontaneous symmetry breaking are stretched outside the horizon and do not affect observations. The intermediate regime in which inflation after the instability lasts a non negligible number of e-folds but not enough to erase signals of the transition from the sky, generates an amplitude of density perturbations on scales associated with the instability in tension with primordial black hole constraints.\\

\noi In this Section we will consider a regime in which inflation after the critical point lasts for more than $60$ e-folds and the potential (\ref{eqhi:hi}) is vacuum dominated.\\

\subsection{Constraints on parameter space}
\label{ssec:cps2}

In this regime, $\chi$ remains light for at least $60$ e-folds after the instability, that is $|m^2_\chi| < H^2$ has to be satisfied for at least $60$ e-folds. Since the mass of $\chi$ is $m^2_\chi \simeq g^2(\phi^2 - \phic^2)$, requiring it to remain light for a long time after the transition is equivalent to require that $\phi$ rolls very slowly through the instability. This gives a condition on the inflaton's mass, which can naively be estimated using analogous approximations to the ones used in subsection~\ref{ssec:cps}. Vacuum domination at the critical point implies that $3H^2=\lambda v^4/4$. Solving for $|m^2_\chi| < H^2$ and neglecting second order in $\delta\phi$, gives $\lambda v^6 \gtrsim 5\times10^3 m^2$.\\
\noi Of course, the length of the transition depends also on the value $\chi_{\rm i}$ at the critical point: the larger it is, the faster the transition. Requiring that $\phi$ rolls faster than $\chi$ for at least $60$ e-folds after the instability, again using the linear approximation for $\phi$ during the slow-roll regime as in subsection~\ref{ssec:cps}, gives $\chi_{\rm i}\lesssim8\times10^{-3}v^2\phic$.\\
\noi Furthemore, if $\chi$ is too close to the origin, quantum fluctuations can severely affect the dynamics after the instability by driving $\chi$ over the potential barrier at $\chi=0$. These quantum jumps can potentially generate domain walls inside the horizon, affecting subsequent dynamics in a non-trivial way. Therefore, we require $\chi$ to be larger than the quantum fluctuations scale $\chi >\Delta \chi_{\rm qu} \approx \frac{H}{2\pi}$. Demanding this condition to be satisfied at the critical point $(\phic,\chi_{\rm i})$ is sufficient for ensuring the subsequent evolution remains classical as long as $\chi$ is light. One finds $\chi_{\rm i} > 0.23 \sqrt{\lambda} v^2/\mpl$, where $\lambda$ is constrained to be considerably less than unity by the amplitude of primordial perturbations, as we will see in the following Sections. The constraints on parameter space therefore are

\be
v<1\,,\;\;\;\;       &&  \;\;\;\; \lambda v^6  >  5000 m^2 \,,\nn
\phi_{\rm i} = \phic\,,\;\;\;\; &&\;\;\;\; 8\times10^{-3} v^2\phic > \chi_{\rm i} 1 >  0.23 \sqrt{\lambda} v^2\,.
\label{eqvdl:parameters}
\en

\subsection{Inflation dynamics}
\label{ssec:obsvdlchi}

\begin{figure}[t]
\centering
\includegraphics[width= 10cm]{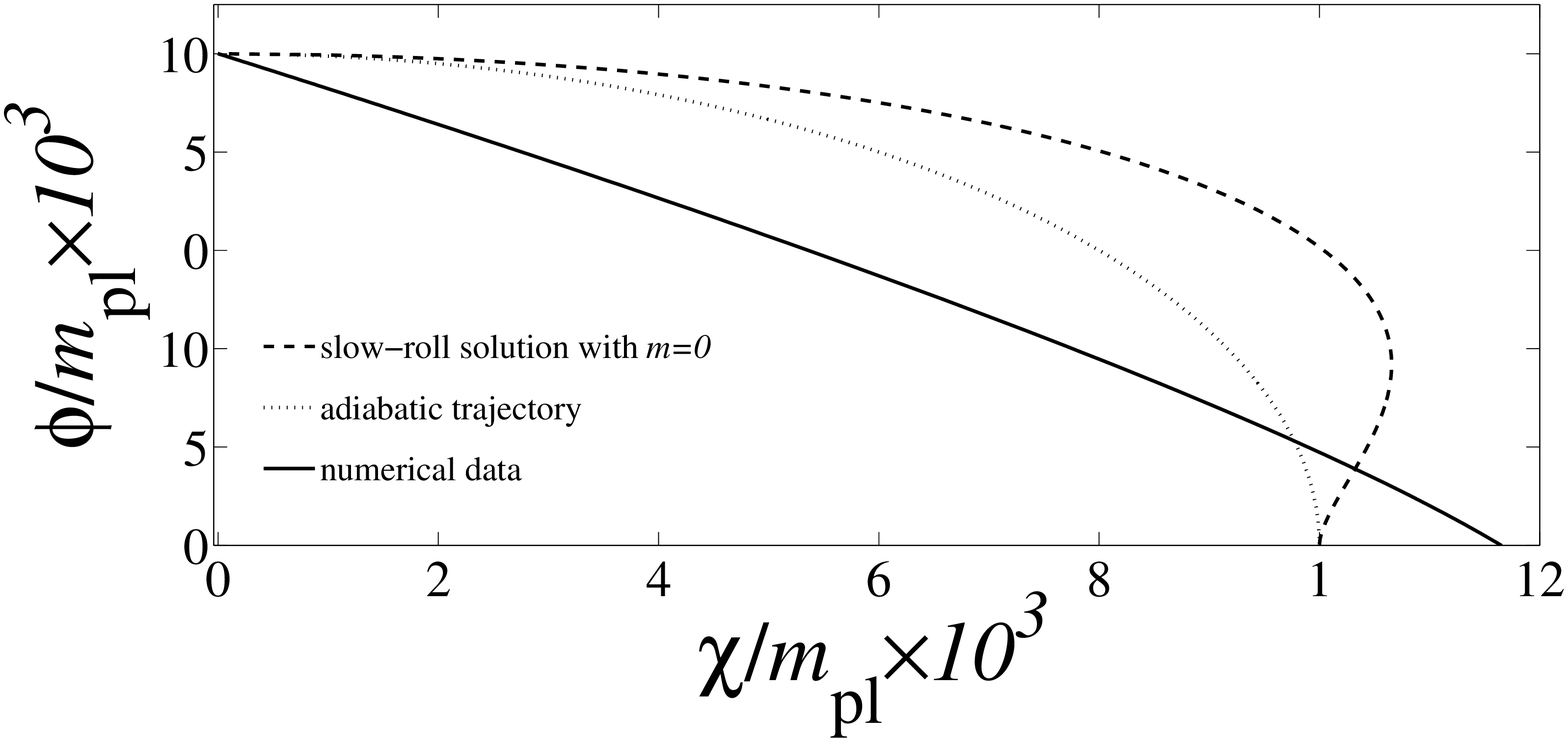}
\caption{Comparison between a numerically integrated trajectory for which $96$ e-folds of inflation are realised (continuous line), the adiabatic trajectory Eq.~(\ref{eqvdlc:adchi}) (dotted line in the middle) and the slow-roll trajectory with $m=0$ Eq.~(\ref{eqvdlc:chiphi}) (dashed line in the top). The parameters are fixed as $v=\phic=0.01$, $m=1.47\times10^{-20}$ and $\lambda=1.29\times10^{-19}$. The initial conditions for the numerical trajectory are $\phi_{\rm i}=\phic$, $\chi_{\rm i} = 3\times10^{-9}$, with the first derivatives fixed by the slow-roll equations (\ref{eqvdlc:srEOMs}). The slow-roll trajectory has the same initial field values and the adiabatic solutions starts at $\phi_{\rm i}=\phic$, $\chi_{\rm i} = 0$. The numerical trajectory satisfies the PLANCK constraint on the amplitude $A_\zeta$ and gives $\ns=0.93$ and $\fnl=0.029$. The spectral index is outside the $95\%$ confidence limit bounds from PLANCK. This figure highlights that neither the slow-roll nor the adiabatic solutions are good descriptions of the motion.} 
\label{vdlfields}
\end{figure}

\begin{figure}
\centering
\includegraphics[width= 10cm]{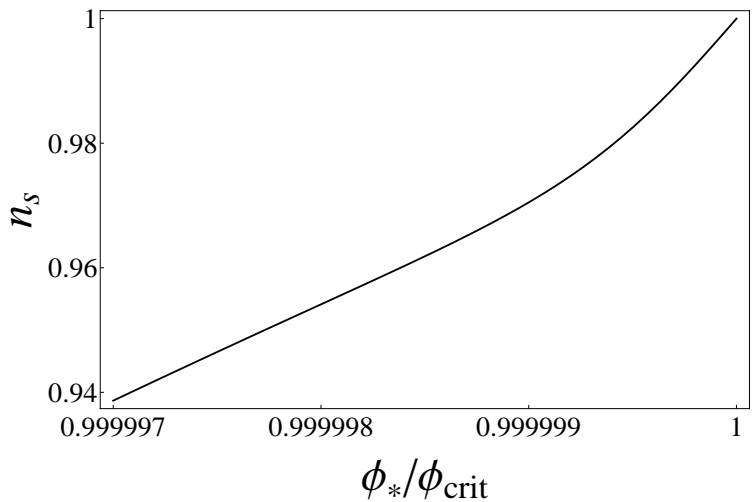}
\caption{The spectral index $\ns=1-6\epsilon_*+2\eta_*$ as a function of $\phi_*/\phic$ for fixed $\chi_*=10^{-7}\mpl$. The other parameters are fixed as $v=0.03\mpl$, $\phic=0.08\mpl$, $m=3.5\times10^{-8}$ and $\lambda=1$. These parameters don't satisfy the observational constraint on the amplitude of perturbations and the plot is just meant to illustrate the facts that $\ns<1$ and the smaller $\phi_*$, the smaller $\ns$.}
\label{vdlnsvsf}
\end{figure}

As we argued above, inflation before the instability is irrelevant to observations for parameters satisfying (\ref{eqvdl:parameters}). After the instability, trajectories tend towards the valley of minima given by

\be 
\frac{\partial V}{\partial\chi} = 0\,\, \Rightarrow\,\, \chi^2 = v^2\left(1 - \frac{\phi^2}{\phic^2}\right)\,.
\label{eqvdlc:adchi}
\en
Of course, Eq.~(\ref{eqvdlc:adchi}) represents an ideal trajectory. More realistically, the system during the inflationary phase evolves according to the slow-roll equations of motion:

\be
\frac{\partial \phi}{\partial N} &= -\frac{m^2+g^2\chi^2}{3H^2}\phi\,,\nn
\frac{\partial \chi}{\partial N} &= -\frac{g^2\phi^2+\lambda(\chi^2-v^2)}{3H^2}\chi\,.
\label{eqvdlc:srEOMs}
\en
The equations (\ref{eqvdlc:srEOMs}) cannot be solved analytically in the general case. However, the constraints (\ref{eqvdl:parameters}) imply that, in this regime, $m$ is very small and we can solve for $\chi$ as a function of $\phi$ in the limit in which $m=0$:

\be 
\chi(\phi) =\sqrt{\frac{\left(\frac{\phi}{\phic}\right)^{2\phic^2/v^2}\left(\chi_{\rm i}^2(\phic^2-v^2)+v^4\right)+v^2(\phic^2-\phi^2-v^2)}{\phic^2-v^2}} \,.
\label{eqvdlc:chiphi}
\en
The trajectories (\ref{eqvdlc:adchi}), (\ref{eqvdlc:chiphi}) and a numerically integrated solution are compared in Fig.~\ref{vdlfields}. The difference between the three trajectories is striking, highlighting the fact that neither the slow-roll nor the adiabatic solutions are good descriptions of the motion.\\ 

\noi We showed in the previous Section \ref{sec:vdhchi} that the spectral index generated when vacuum dominated inflation ends abruptly at the instability is greater than one, see Eqs.~(\ref{eqvdh:obs}), in conflict with observations. How does a phase of inflation after the instability change this result?\\
\noi In Fig.~\ref{vdlnsvsf}, the slow-roll expression for the spectral index $\ns=1-6\epsilon_*+2\eta_*$ is plotted against $\phi_*/\phic$ for representative parameters. We find that $\ns<1$ and it tends to decrease as $\phi_*$ decreases, spanning the observationally compatible range (in agreement with the results of Clesse \cite{Clesse:2010iz}). Therefore, a phase of inflation after the instability cures the blue spectral index $\ns>1$ predicted when inflation ends abruptly at the phase transition.\\
\noi In a recently appeared paper \cite{Clesse:2013jra}, the authors derived analytical formulas for the spectral index $\ns$ and the reduced bispectrum $\fnl$ in this regime, and confirmed their validity with numerical simulations. They arrived at the conclusion that inflation along waterfall trajectories can't satisfy the observational constraints on the amplitude of the perturbations $A_\zeta$ and the spectral index $\ns$ simultaneously.

\section{\texorpdfstring{$\phi$}{TEXT} Dominated with Light \texorpdfstring{$\chi$}{TEXT}}
\label{sec:philchi}

So far we only considered inflationary regimes in which the potential (\ref{eqhi:hi}) is vacuum dominated. Another interesting possibility is that the potential be dominated by the inflaton's mass term at horizon crossing \cite{mpHP}. In this scenario, the exponential expansion is driven by $\phi$ rolling down its quadratic potential, just as in chaotic inflation. The trajectory rolls through the instability slowly enough for the scalar field $\chi$ to  remain dynamically irrelevant for at $60$ more efolds of expansion, with a mass $m^2_\chi \ll H^2$. As the universe's energy density drops, eventually $m^2_\chi \approx H^2$ and $\chi$'s motion becomes important, potentially sourcing different super-horizon signatures than chaotic inflation (see Section \ref{sec:ciNG}).     

\subsection{Constraints on parameter space}
\label{ssec:plcparch}

For the potential (\ref{eqhi:hi}) to be dominated by the inflaton's mass term, the relation $m^2\phi_*^2 \gg 2V_0\equiv \lambda v^4/2$ has to be satisfied. Also, we wil require that at least $60$ e-folds of expansion happen after the instability. As shown in \ref{ssec:ci}, this implies that $\phic \gtrsim 15.5$. This condition is equivalent to $\lambda v^2 \gg g^2$. Furthermore, we will only consider sub-planckian vacuum expectation values: $v<1$. Summarising, the requirements on the parameters are 

\be
\lambda v^2 \gg g^2\,,\;\;\;\;&&,\;\;\;\;m^2 \phi^{*2} \gg \lambda v^4\,,\nonumber\\ 
  v<1\,,\;\;\;\;&&\;\;\;\;\phic \gtrsim 16\,,
\label{eqplc:param}
\en

\subsection{Analytical estimates}
\label{ssec:anes}

For parameters satisfying the constraints (\ref{eqplc:param}), during the inflationary phase the interaction term $g^2\phi^2 \chi^2 $ is sub-dominant to the other terms and the potential (\ref{eqhi:hi}) is approximately of the separable form: 

\be
V(\phi,\chi) = V_\phi(\phi) + V_\chi(\chi)\,.
\en
We will therefore use this approximation for analytic calculations, and later check its validity using full numerical simulations.\\

\noi For separable potentials, there exist analytic expressions 
for the derivatives of $N$ \cite{Vernizzi:2006ve, GarciaBellido:1995qq}. 
The starting point for these formulae are the slow-roll equations 
of motion

\be
\dot{\phi} = -\frac{V_{,\phi}}{3H}\,,\nn
\dot{\chi} = -\frac{V_{,\chi}}{3H}\,.
\label{eqplc:sr}
\en
It follows that 

\be
N = \int^f_* d \phi \frac{-V(\phi, \chi) }{  V_{,\phi} } =  \int^f_*d \phi  \frac{-V_\phi}{ V_{,\phi} }+ \int^f_* d \chi \frac{-V_\chi }{  V_{,\chi} }\,.
\en
Therefore, for example, 
\be
N_{,\chi} =\left. \frac{V_\chi}{ \mpl^2 V_{,\chi} }\right |_* - \left. \frac{V_\chi}{  V_{, \chi} }\right |_f \frac{\partial \phi^f}{\partial \phi_*}\,.
\en
The complexity of the calculation then reduces to determining 
the partial derivative of the final field 
value on a constant energy density hypersurface with respect 
to the initial field value. However, when the dynamics become 
adiabatic during inflation, this derivative tends to zero. Therefore, 
one usually finds that\footnote{It is possible that this 
is not the case if $V_\chi/V_{,\chi}|_f$ diverges as the derivative 
tends to zero (see \cite{Elliston:2011dr} for a full discussion).}
 $N _{\chi}= V_\chi/( V_{,\chi}|_*)$.
The second derivatives 
of $N$ then follow by simple differentiation.\\

\noi We can now proceed to compute the derivatives. For $\chi_*\ll v$, we find

\be 
N _{\phi}    &\approx& \frac{\phi_*}{2}\,,                 \label{eqplc:nphi}\\
N_{\chi}     &\approx& -\frac{V_0}{m^2}\frac{1}{\chi_*}\,, \label{eqplc:nchi}\\
N_{\phi\phi} &\approx& \frac{1}{2}\,,                      \label{eqplc:n2phi}\\
N_{\phi\phi} &\approx& \frac{V_0}{m^2}\frac{1}{\chi_*^2}\,.\label{eqplc:n2chi}
\en
In the limit that $\chi_*\rightarrow0$ the derivatives with respect to $\chi$ tend to infinity and dominate over the derivatives with respect to $\phi$. This regime corresponds to $\chi$ being very close to the top of its self-interactive potential at the time of horizon crossing. In such a scenario, it is known that $\chi$'s descent from the ridge can generate a large non-gaussianity via the hilltop mechanism \cite{Kim:2010ud,Elliston:2011dr} if its tachyonic mass squared is much greater than the vacuum energy at the top of the hilltop. We find that $\fnl$ in this regime is given by

\be
\fnl \approx \frac{10}{3v^2 }\,,
\label{eqvdlc:anpred}
\en
Thus, depending on the value of $v$, $\fnl$ can be large provided that $\chi_*$ is not too large.\\

\noi The condition for $\chi$ to source the dominant contribution to $\delta N$ is $|N_{\chi}| > |N_{\phi}|$, which translates to
$\chi_* \simeq 0.03v^2$. We arrive at the interesting situation in which the $\phi$ field sources inflation, while the contribution due to the $\chi$ field dominates $\zeta$ and can be extremely non-Gaussian. On the other hand, consideration of Eq.~(\ref{eqdn:spectrum}) implies that the 
spectral index is generically red tilted and close to scale 
invariant, as will be confirmed by the full numerical 
simulations which follow.\\

\noi Guidance from analytic arguments, therefore, appears to lead to a 
large non-gaussianity in the following scenario. 
Long before the last $60$ e-folds of inflation, $\phi$ begins its evolution 
with $\phi>\phi_{\rm crit}$, 
and some value of $\chi$. 
Since the mass of $\chi$ is initially positive it will evolve 
towards its minimum at $\chi=0$ as $\phi$ evolves towards $\phi_{\rm crit}$. 
Assuming the classical evolution of $\chi$ reaches $\chi=0$ before $\phi$ 
reaches $\phi_{\rm crit}$, then at this time the value of $\chi$ 
will be dominated by its quantum diffusion during the period 
before the transition for which it was light.  
After the transition for the parameter 
choices discussed, more than $60$ e-folds of 
inflation occur and $\chi$ is still light when 
$\phi =16$ roughly $60$ e-folds 
before the end of inflation. Its
vev at this time will then likely still be close to $\chi=0$. 
The $\chi$ field will only roll significantly when 
$\chi$ becomes heavy. If this occurs 
before $\phi$ evolves to its minimum, it will roll during 
the slow-roll inflationary phase, otherwise $\phi$ will begin 
oscillating and begin to decay into radiation before $\chi$ 
begins to roll. In the later case $\chi$ will 
perhaps source an extremely short secondary inflationary 
phase as it rolls. It is clear, however, that Eq.~(\ref{eqvdlc:anpred})
will remain 
at best a \emph{rough estimate}. This is why we 
employ numerical simulations in our study. \\

\noi Using Eqs.~(\ref{eqplc:nphi}-\ref{eqplc:n2chi}) we can write analytic expression for the asymptotic value of $\fnl$ even when $\chi_*$ becomes larger and Eq.~(\ref{eqvdlc:anpred}) becomes obsolete.
At $\chi_*\sim v^2$, we reach a regime where $|N_\phi|>|N_\chi|$. If the numerator of Eq.~(\ref{eqng:fnldn}) is still dominated by $\chi$, we find
\be
\fnl=\frac{5}{6}\frac{N_\chi^2N_{\chi\chi}}{N_\phi^4}\approx \frac{v^6}{\chi_*^4},
\label{equ:fnltrans}
\en
indicating that $\fnl$ drops sharply. At even larger values of  $\chi_*$, we find
$N_{\phi\phi}>N_{\chi\chi}$, which implies  $\fnl = 5N_{\phi\phi}/(6N_\phi^2)$.
This gives
\be
\fnl = \frac{5}{3\phi_*^2}\approx 0.007.
\label{anpredphi}
\en
Hence, we should find a smooth but clear transition from the significant non-gaussianity predicted by Eq.~(\ref{eqvdlc:anpred}) to practically Gaussian single-field behavior (\ref{anpredphi}) at around $\chi_*\approx 0.03v^2$.

\subsection{Initial conditions}
\label{sec:ics}

The scenario discussed above requires a small initial value for 
the waterfall field, $\chi_*\ll v^2$, at $\phi=\phi_{\rm crit}$.
In this subsection we discuss the naturalness of this condition.
This is an instance of the `measure problem' which is unresolved in general. 
However, we can apply some heuristic arguments. \\
\noi In \ref{ssec:anes}, we outlined a plausible sequence of events 
in which, long before the hybrid transition, both $\phi$ and 
$\chi$ took large values, and the classical evolution 
of $\chi$ tended to zero before $\phi$ reached $\phi_{\rm crit}$. 
The initial values of $\phi$ and $\chi$ can be motivated by 
assuming a phase of eternal inflation during which the path the 
fields follow is dominated by quantum fluctuations, rather than 
classical rolling. The condition for such behaviour is that the 
distance moved in field space in a Hubble time, is less than  
a typical quantum fluctuation \cite{eternal}
\be
\sqrt{\dot{\phi}^2 +\dot{\chi}^2}> \frac{H^2}{2 \pi}\,.
\label{eq:eternal}
\en
Using the slow roll equations of motion (\ref{eqplc:sr}) this becomes a constraint 
on the field values, and defines a one dimensional surface in field space at 
which the dynamics becomes predominantly classical.\\

\noi Beginning at this surface, therefore, 
one might wonder whether $\chi$ will indeed reach $\chi=0$ (according to its classical 
rolling) before  $\phi$ reaches $\phi_{\rm crit}$. The answer to this question 
will be extremely parameter dependent, and will likely also depend on the 
position on the surface from which the fields originate. It is easy, however,  
to build up a rough picture of the expected behaviour. 
Considering cases where both $\phi$ and 
$\chi$ are significantly displaced from zero initially, $\chi$ will likely 
initially be the more massive field for the parameter choices we are focusing on (\ref{eqplc:param}). 
This is because $\chi$ feels a mass of $1/2 \lambda \chi^2 +g^2 \phi^2$ in comparison with 
$m^2 + g^2 \chi^2$, the mass of $\phi$. Moreover, 
we expect $\lambda>g^2$, unless $v$ is much less than $1$. 
In the following we will assume this condition on $\lambda$ and $g$.   
The mass of 
$\phi$ would only be greater initially, therefore, 
if $m^2 \gtrsim \lambda \chi_{\rm e}^2 + g^2 \phi_{\rm e}^2$ 
(where subscript $\rm e$ represents values on the eternal inflation surface). \\

\noi Assuming that $\chi$ is indeed more massive initially, it will evolve rapidly towards 
$\chi=0$, while $\phi$ will remain nearly frozen. 
As $\chi$ decreases, there will come a time when $\lambda \chi^2 < m^2$. At this point 
if the condition $m^2 \gtrsim g^2 \phi^2_{\rm e}$ is met then the 
$\chi$ field will cease to be the more massive, and 
its evolution will slow significantly, and it will likely not 
reach $\chi=0$. On the other hand, if this condition is not met we 
expect that $\chi$ does reach zero.  The necessary condition for 
$\chi$ to reach zero, therefore, is that $m^2 < g^2 \phi^2_{\rm e}$, which ensures 
that $\chi$ is always the more massive field. \\
\noi This condition is likely only very approximate, but from some limited 
numerical probing it appears to capture the rough behaviour of the fields.  
In Section~\ref{sec:workExample} we will look carefully 
at a particular parameter choice and probe the surface defined by Eq.~(\ref{eq:eternal}). We find 
a complicated picture, however this rough analysis is a useful aid.

\begin{figure}[t]
\centering
\includegraphics[width= 10cm]{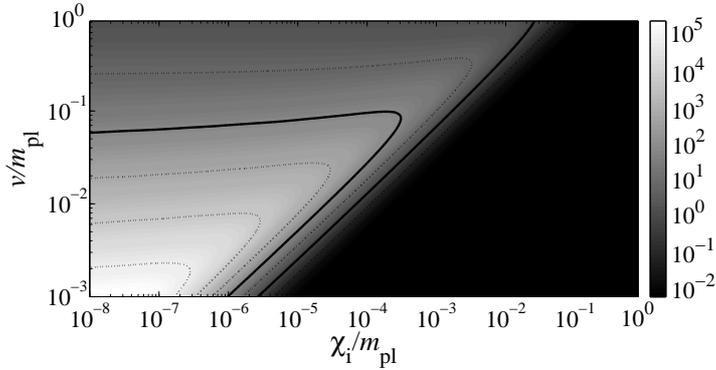}
\caption{The non-gaussianity parameter $\fnl$ as a function of $\chi_i$ and $v$, with other parameters 
given by Eq.~(\ref{eqplc:param}). Lighter shades indicate higher $\fnl$. The black contour lines correspond to, from left to right, $\fnl=100$ and $\fnl=1$. Values $\fnl\gtrsim 10$ are ruled out by PLANCK.}
\label{confnl}
\end{figure}
\begin{figure}
\centering
\includegraphics[width= 10cm]{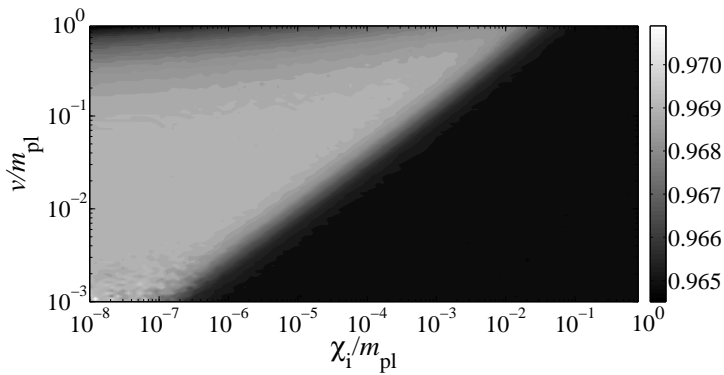}
\caption{The spectral index $n_s$ 
as a function of $\chi_i$ and $v$, with other parameters given by Eq.~(\ref{eqplc:param}). Lighter shades indicate higher $n_s$. All values obtained are compatible with PLANCK observations.}
\label{conns}
\end{figure}
\begin{figure}[t]
\centering
\includegraphics[width= 10cm]{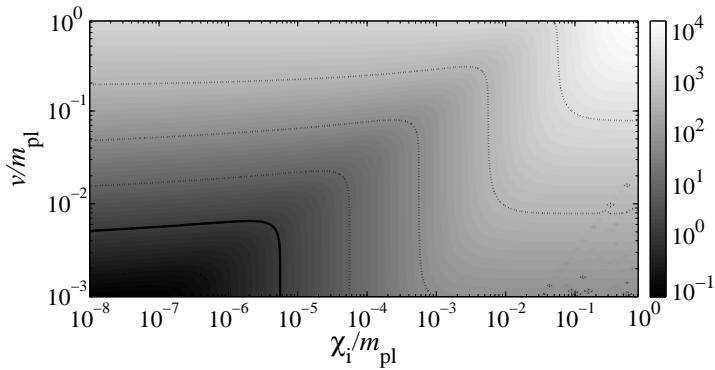}
\caption{The ratio $\chi_*/H_*$ 
as a function of $\chi_i$ and $v$, with other parameters given by Eq.~(\ref{eqplc:param}). Lighter shades indicate higher $\chi_*/H_*$. The thicker contour line corresponds to $\chi_*/H_*$=1. Values below this are unnatural because quantum fluctuations would generally give $\chi_*\gtrsim H_*$}
\label{conDH}
\end{figure}
\begin{figure}
\centering
\includegraphics[width= 10cm]{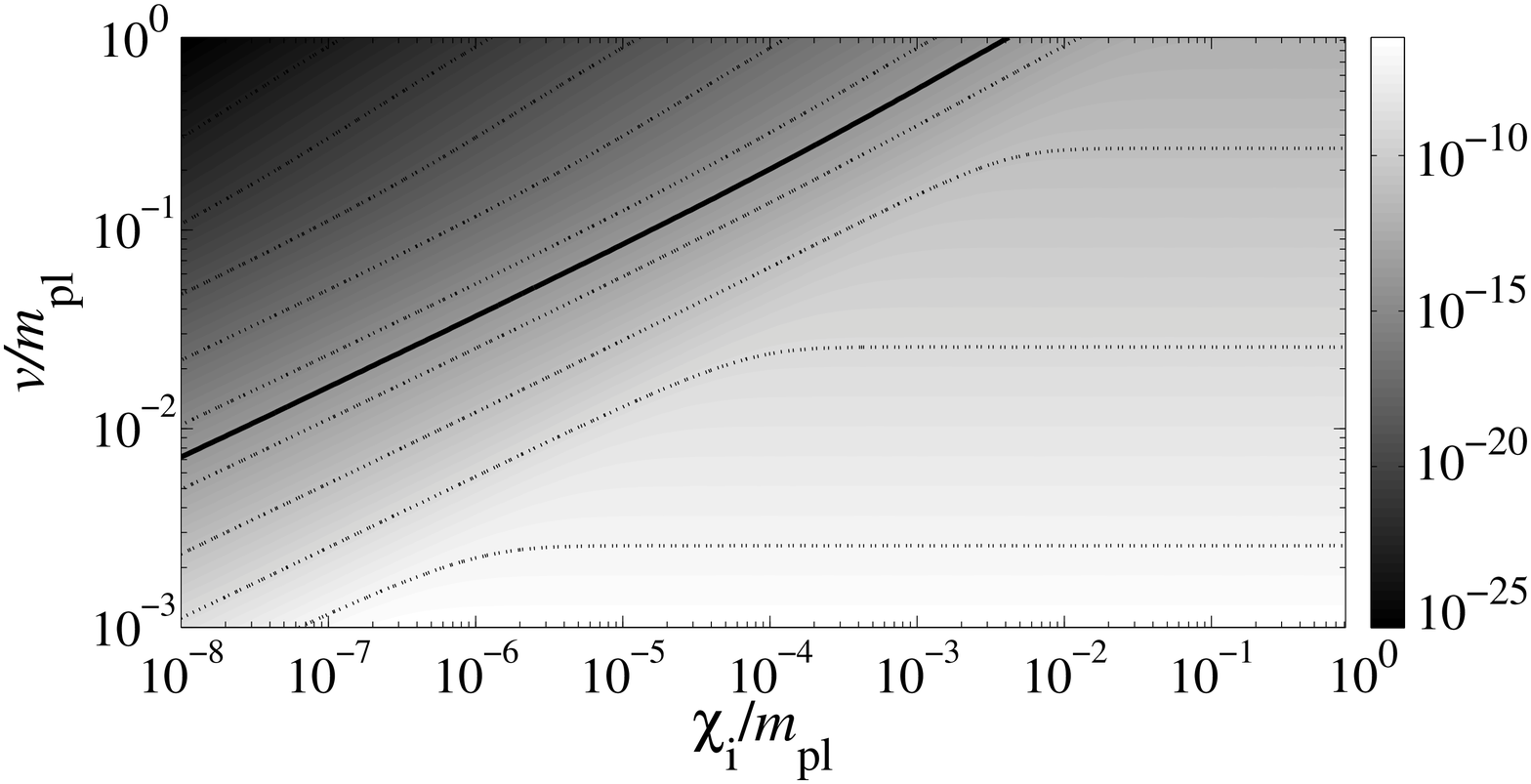}
\caption{The dimensionless parameters $\lambda$ 
as a function of $\chi_i$ and $v$, with other parameters given by Eq.~(\ref{eqplc:param}). Lighter shades indicate higher $\lambda$. The thicker contour line corresponds to $\lambda=10^{-14}$.}
\label{conLambda}
\end{figure}

\subsection{Numerical analysis}
\label{sec:numerical}

Having identified the interesting parameter range, we now compute the
statistics of the perturbations numerically.
We evolve the full non-slow-roll equations 
of motion Eqs.~(\ref{eqnumdN:eoms}) and (\ref{endennum}) following the method described in subsection~\ref{ssec:numDeltaN}.\\
\noi For simplicity we reduce the number of parameters which 
can be varied, by fixing
\be
\phi_{\rm crit} &=& \sqrt{1000},\nonumber\\
m&=&\sqrt{\lambda}v,\nonumber\\
g&=& \sqrt{\lambda} v/ \phi_{\rm crit},\nonumber\\
\Gamma&=& 10^{-1} \sqrt{\lambda} v,
\label{equ:paramchoices}
\en
and we explore the ranges
\be
10^{-3} <&v&<1,\nonumber\\ 
10^{-6}<&\chi_{\rm i}/v&<1.
\en
Our parameter choice effectively means that the 
magnitude of $H_*$ and of $\Gamma$ is proportional to $\sqrt{\lambda}$, 
and so can be scaled simply by changing the 
value of $\lambda$. On the other hand, since the derivatives of $N^f_*$, which 
enter formulae for observational quantities, are unaltered by such a scaling, changing 
the value of $\lambda$ leaves these derivatives unaltered. 
For each 
choice of $\chi_{\rm i}$ and $v$, we fix $\lambda$ such that the 
resulting amplitude $A_\zeta$ is in agreement with the observational 
requirement. The rescaling properties just discussed, however, 
mean that this can be done retrospectively once the derivatives of $N$ have been  
calculated (see below).\\

\noi As discussed in Section~\ref{sec:dNf}, we need to calculate the amount of 
expansion between a flat and an equal-energy density hypersurface, $N_*^f(\phi_*,\chi_*)$, 
which we obtain by solving Eqs.~(\ref{eqnumdN:eoms}) for different initial values $\phi_*$ and $\chi_*$. 
To that end, we follow the algorithm describe in subsection \ref{ssec:numDeltaN}. We follow the evolution until $99.9 \%$ of the energy density is in the radiation component.\\ 
\noi Once we have obtained $N_*^f(\phi_*,\chi_*)$, we calculate the spectral 
index $\ns$ and $\fnl$ using Eqs.~(\ref{eqdn:spectrum}) and (\ref{eqng:fnldn}), respectively.
The results in Fig.~\ref{conns} show that for all the parameters we considered, the 
spectral index is contained in the region 
$0.9649 \le \ns \le 0.9701$. Therefore they are all compatible with the observations.\\

\noi On the other hand, Fig.~\ref{confnl} shows that a 
wide range of values for $\fnl$ are obtained,  
$7\times 10^{-3} \lesssim \fnl  \lesssim 5 \times 10^5$. 
All parameters giving $\fnl \gtrsim 10$ are excluded by 
PLANCK measurements \cite{planck}, constraining the parameter space.
For $\chi_i\lesssim 2v^2/\mpl$, values $v\lesssim 0.07\mpl$ are ruled out,
and larger values of $\chi_i$ are not constrained.\\

\noi Furthermore, Fig.~\ref{confnl} confirms the validity  
of the analytical arguments and estimates in Section~\ref{ssec:anes} . To a good approximation, we can identify $\chi_*\approx \chi_i$. 
Our results show that as long as $\chi_i$ is small enough, $\fnl$ is independent of it and decreases by one 
order of magnitude for a two orders of magnitude increase in 
$v/\mpl$, as predicted by Eq.~(\ref{eqvdlc:anpred}).
Towards the right, at $\chi_i\sim 0.1v^2$, $\fnl$ starts to fall in accordance with Eq.~(\ref{equ:fnltrans}), and reaches a constant value $\fnl \approx 0.007$ in good agreement with Eq.~(\ref{anpredphi}). \\
\noi As mentioned above in this section, with the parametrisation given in Eq.~(\ref{equ:paramchoices}), we can fix the parameters of the model retrospectively, constrained by the amplitude of the power spectrum. As can be seen in Fig.~\ref{conLambda}, $10^{-6}\lesssim\lambda\lesssim10^{-25}$. \\

\noi An issue of importance is whether quantum fluctuations will 
affect which side of the potential the waterfall field is rolling along. 
This happens if the range of quantum fluctuations is greater 
than the distance of 
$\chi$ to its hilltop. It is problematic because such a scenario leads to 
a configuration with domain walls, which we want to avoid. 
For a massless scalar field, the range is given by $H_*/(2\pi)$. For a massive field, 
$H_*/(2\pi)$ is an upper bound. Thus, excluding parameters such that $\chi_*/H_* \le 1$ 
settles the issue. As Fig.~\ref{conDH} shows, the excluded values are approximately
$v \le 5 \times 10^{-3}$ and $\chi_i< v^2 $. 
\begin{figure}
\centering
\includegraphics[width= 10cm]{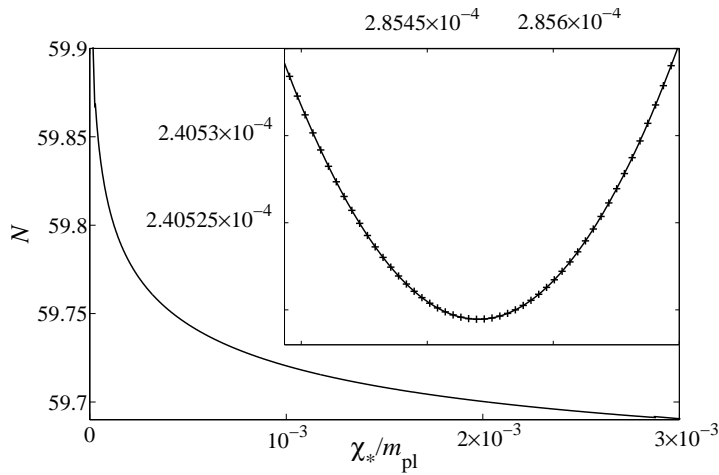}
\caption{$N$ as a function of $\chi_*$ for the parameters shown in Eq.~(\ref{equ:paramchoices}) with $v=0.2\mpl$. The results in Section~\ref{sec:dNf} assume that the quadratic Taylor expansion in Eq.~(\ref{eqdn:taylordn}) is a good approximation. This is clearly not the case for the whole range of $\chi_*$, but for a range that corresponds to the values present in one comoving volume representing the currently observable universe it is. This can be seen in the inset, where the data for such a range with the linear term subtracted is plotted, together with a quadratic fit. The plot is indeed quadratic in $\chi_*$.}
\label{fit}
\end{figure}
\begin{figure}
\centering
\includegraphics[width= 10cm]{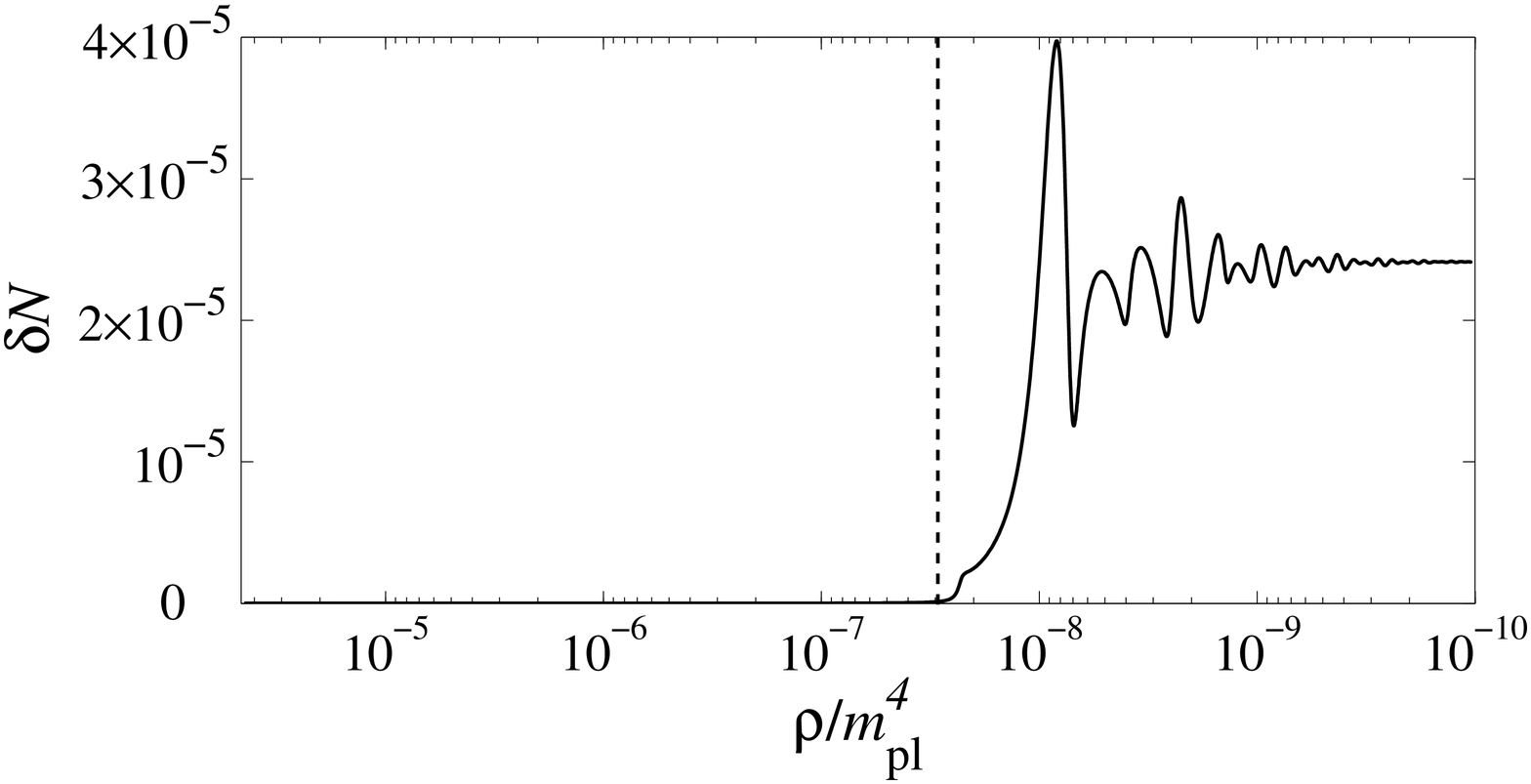}
\caption{Comparison of two initial conditions $\chi_i$ and $\chi_i+\delta\chi_i$ with parameters in Eq.~(\ref{equ:paramchoices}) and $\delta\chi_i=1.4\times 10^{-7}\mpl$. The plotted function is the difference in $N$ as a function of energy density $\rho$, and corresponds to the curvature perturbation between points with the two different initial conditions.
The vertical dashed line  indicates when the waterfall field $\chi$ starts rolling down the 
ridge.}
\label{dNevo}
\end{figure}
\begin{figure}
\centering
\includegraphics[width= 10cm]{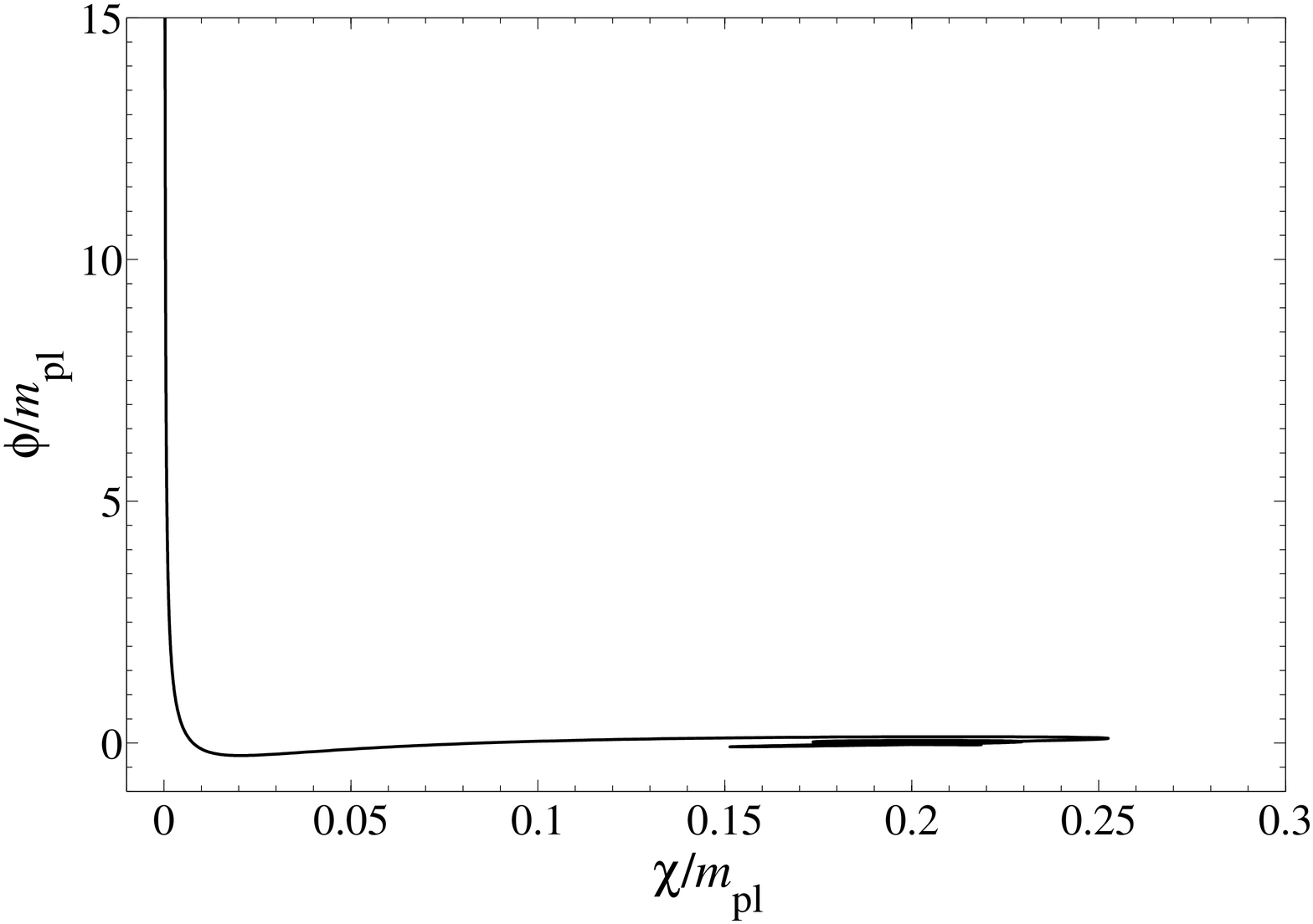}
\caption{Trajectory in $\chi$-$\phi$ field space for the working example, with parameters given in Eq.~(\ref{eq:example}). The plot starts at horizon crossing, with $\phi_*=15.02\mpl$.}
\label{fieldplot}
\end{figure}
\subsection{A working example}
\label{sec:workExample}
Our results indicate that there is a fairly large window of parameters that are compatible with 
observations but lead to observable non-gaussianity. We pick one representative example
for more detailed study:
\begin{eqnarray}
\label{eq:example}
v&=&0.2\,,\nonumber\\
\lambda &=& 5.322\times 10^{-15}\,,\nonumber\\
g^2 &=& 2.128\times 10^{-19}\,,\nonumber\\
\phi_i &=& \phi_{\rm crit}\,,\nonumber\\
\chi_i &=& 10^{-3}v\,.
\end{eqnarray}
These parameters give
\begin{eqnarray}
\fnl&=&24.14\,,\nonumber\\
n_s&=&0.968\,.\nonumber\\
\end{eqnarray}
Besides a value of the spectral index in the centre of the 
permitted observational range, these parameters give us a large value of $\fnl$, close to the 
favoured WMAP value, $\fnl \approx 20 \pm 10 $. Note that the PLANCK results \cite{planck}, released subsequently to the making of this study, rule out this value. Nevertheless, the following analysis remains qualitatively interesting and applies to values within the new observational bounds (\ref{fnlOBS}).\\ 

\noi In Fig.~\ref{fit}, we show $N$ for a wide range of field values $\chi_*$. 
The expressions (\ref{eqng:fnldn}) and (\ref{eqdn:spectrum}) for $\fnl$ and $n_s$ assume quadratic Taylor expansion (\ref{eqdn:taylordn}) of this function. 
It is clear from the figure that this assumption is not valid over the whole range shown. However, the actual range of $\chi_i$ present in a comoving volume corresponding to the universe observable today is smaller, roughly $\delta\chi_i\approx \sqrt{N/2\pi}H_*$, where $N\approx 60$. 
The inset shows this range of $\chi_*$, with the linear term subtracted. The remaining contribution is small and almost exactly quadratic, demonstrating that the Taylor expansion (\ref{eqdn:taylordn}) is valid.\\

\noi It is interesting to confirm our expectation that the 
dominant contribution to $\delta N$ comes from the 
rolling of the waterfall field from its hilltop. 
This can be seen clearly in Fig.\ref{dNevo} which plots the evolution 
of $\delta N$ together with the point at which the waterfall field's 
evolution becomes significant.\\

\begin{figure}
\centering
\includegraphics[width= 10cm]{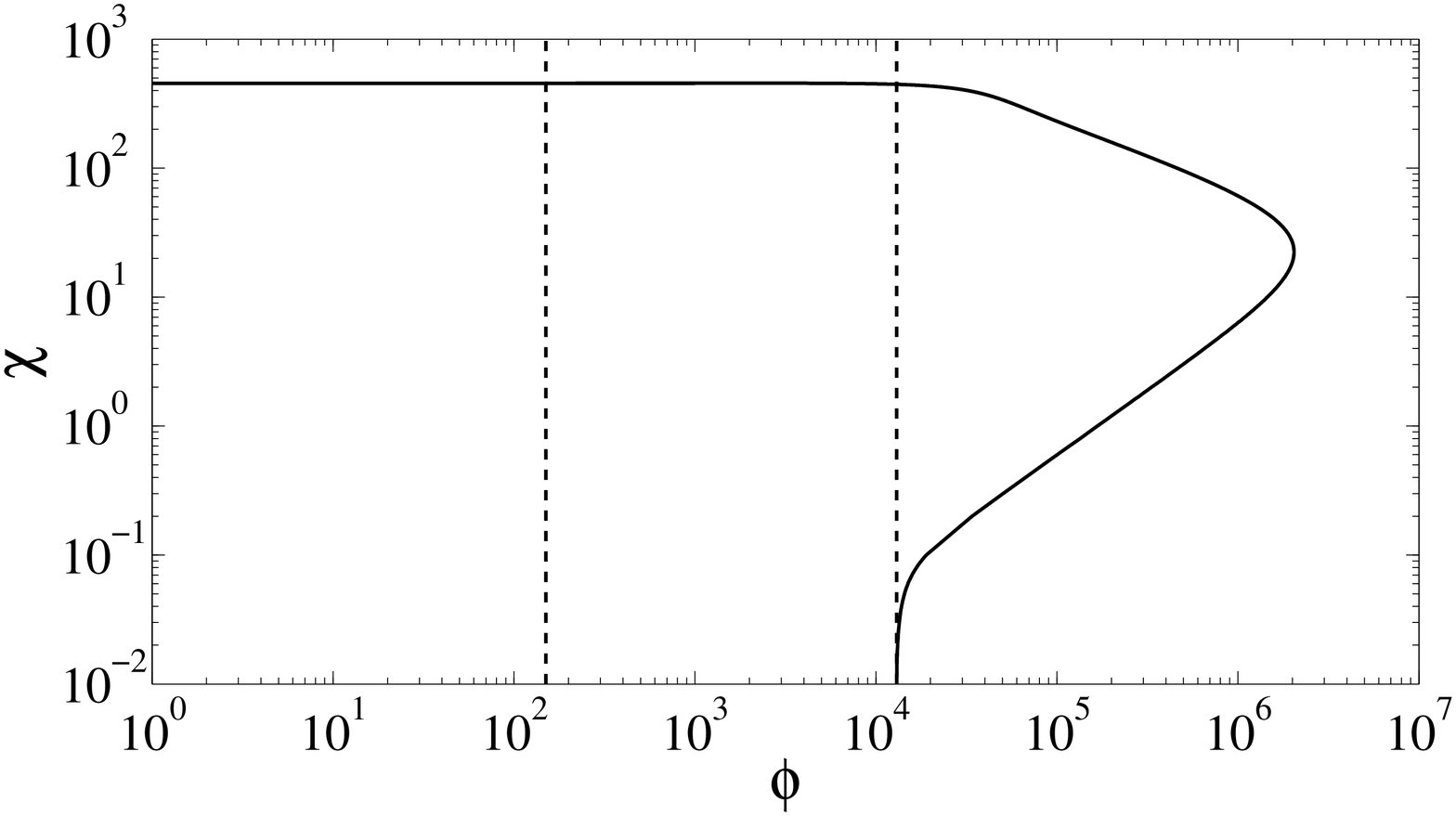}
\caption{A plot of the boundary in field space on which quantum fluctuations become greater than the classical 
rolling of the 
fields. The three regions, described in the text, demarcate initial conditions on the surface for which $\chi$ does not reach $\chi=0$ before 
$\phi$ reaches $\phi_{\rm crit}$ (far left), initial conditions which do (middle), and initial conditions for which eternal inflation does not end 
(far right). Because of the symmetry of the potential, only positive values of $\phi$ and $\chi$ are shown.}
\label{fig:Eternal}
\end{figure}

\noi Finally, we ask whether 
the initial conditions we have assumed in this Section are reasonable. 
Importantly, for the parameter values given in example (\ref{eq:example}), 
we find that the condition, given in Section~\ref{sec:ics}, for mass structure to be correct
for $\chi$ to evolve classically to $\chi=0$, $\phi_{\rm e}^2>m^2/g^2$, is met if $\phi_{\rm e} > 32\mpl$. 
This is necessarily satisfied when $\phi$ is above $\phi_{\rm crit}$ in this 
case. Of course, in practice we will need $\phi$ to be significantly larger than 
this value in order that 
$\chi$ has time to evolve to zero. Numerically we find that $\phi_{\rm e}>150\mpl$ is acceptable 
for $\chi\ll v^2$ at $\phi_{\rm crit}$, which 
is satisfied on a significant proportion of  the surface defined by Eq.~(\ref{eq:eternal}). This can be seen 
by considering Fig.~\ref{fig:Eternal}, which plots the surface on which eternal 
inflation ends for the parameter values at hand. For values of $\phi\gg150$ the evolution 
is predominantly in the $\chi$ direction. Since the surface is complicated, an evolution which leaves the eternal regime may 
re-enter it, as can be seen from the figure. The dashed lines demarcate three regimes, trajectories 
originating from the surface in the region on the far left do not evolve to $\chi=0$ 
before $\phi$ reaches $\phi_{\rm crit}$, while trajectories in the middle region do.  Trajectories in the region on the 
far right originating from the upper eternal inflation boundary evolve classically towards $\chi=0$, but 
reenter an eternal inflationary regime. The vast majority of trajectories which remain classical, however, 
evolve to $\chi \ll v^2$ when $\phi = \phi_{\rm crit}$.

\section{Discussion}

In this Chapter, we studied three different regimes of inflation driven by the potential (\ref{eqhi:hi}). First, we considered vacuum dominated inflation with a sharp tachyonic phase transition ending inflation, see Section \ref{sec:vdhchi}. It leads to a blue spectral index ($\ns>1$) on scales that exited the horizon $60$ e-folds before the end of inflation, corresponding to scales observed today in the CMB, in disagreement with observational constraints, and a small and negative non-gaussianity parameter $|\fnl|<1$, which is undetectable by ongoing and foreseeable experiments. Furthermore, we argued that the non-equilibrium dynamics of the tachyonic preheating generate objects such as domain walls which are highly constrained by observations and incompatible with the energy scale at which the transition is expected to happen in an inflationary scenario \cite{GarciaBellido:1996qt}. We reach the conclusion that this regime is, to put it mildly, disfavoured by experimental data.\\

\noi In Section \ref{sec:vdlchi}, we relaxed the requirement that inflation ends with a sharp phase transition and considered a regime in which inflation is still vacuum dominated, but the universe undergoes considerable expansion even after the phase transition, as first considered in \cite{Clesse:2010iz}. The model was found to be in tension with observations in recent work \cite{Clesse:2013jra} .\\

\noi Last, but not least, in Section \ref{sec:philchi} we investigated the scenario in which inflation is driven by the inflaton's mass term, as in chaotic inflation. Provided that the critical value $\phic>\phi_*\approx15.5\mpl$, inflation lasts for more than $60$ e-folds after the transition and we need not to worry about relics from the symmetry breaking. In this scenario, the symmetry breaking field remains light until inflation ends, frozen at the hilltop of its self-interacting potential. As the universe's energy density drops, it eventually falls down the hilltop and potentially sources a highly non-Gaussian curvature perturbation, analogously to the curvaton mechanism \cite{curv}. We found that this scenario very naturally predicts the observed spectral index $\ns$ and can generate $\fnl$ ranging from $10^{-2}$ to $10^{5}$, leading to strong constraints on parameter space from the PLANCK results \cite{planck}. \\

\noi While earlier work has considered  
the hybrid model with a light waterfall field and the resulting non-gaussianity 
\cite{barnaby, Abolhasani:2011yp} (and hybrid inflation 
with two 
light inflaton fields \cite{largeNG1,inhom2})
the analysis presented in Section~\ref{sec:philchi}
is the only studies of non-gaussianity in the case in which 
more than $60$ e-folds of evolution occur after the hybrid transition.\\  
 
\noi Finally we note that as discussed in \ref{sec:ics} an interesting 
feature of the hybrid model is that it offers an 
explanation for the initial conditions at horizon crossing which lead to 
a large non-gaussianity. While parameter choices are necessarily fine tuned, 
the initial conditions need not be, at least for some subset of the parameters 
which can produce a large non-gaussianity. This is in contrast to other two field 
models which produce a large non-gaussianity due to inflationary dynamics \cite{largeNG1, largeNG2,Elliston:2011dr}, which 
require both carefully chosen model parameters, and offer no explanation for the origin of the
finely tuned horizon crossing conditions.\\

\noi Assuming inflation happened in the early universe, and assuming it was driven by the hybrid potential (\ref{eqhi:hi}), we are left with the question of which of the three regimes described in this chapter is more likely to occur. Lacking an understanding of the fundamental physics that might give rise to the hybrid potential, it is impossible to discrimate between the three regimes, apart through comparison to observations. However, if embedded in a more fundamental theory, the relations between the different parameters of the potential (\ref{eqhi:hi}) might be fixed by higher energy dynamics, leading to a natural choice of the possible phenomelogies. With this issue in mind, in the next Chapter we will explore the link between hybrid inflation and supersymmetry.

\chapter{Supersymmetric Hybrid Inflation}
\label{ch:susyhi}

\noi Inflationary models embedded in high energy particle physics theories are of great importance. They allow to put to the test intriguing connections between cosmology and high energy physics.
In Chapter~\ref{ch:hi} we studied in detail three limiting regimes of hybrid inflation, mentioning that it is a well motivated model from a particle physics perspective. In particular, we argued that potentials with the same features as (\ref{eqhi:hi}) are easily embedded in the supersymmetric framework and its local version supergravity. In this Chapter we justify our claim by describing how hybrid potentials are derived from supersymmetric models.\\
\noi A lot of work has been invested in deriving hybrid potentials directly from high energy frameworks, such as F-term versions in N=1 SUSY, D-term versions in SUGRA, P-term versions in N=2 SUSY, D-brane versions and many more (see \cite{minSUSY,SUSY} for a few representative examples).\\

\noi As we showed in Section~\ref{sec:philchi}, inflationary models with light scalar fields, besides the inflaton field, can have interesting and distinctive observable signatures. If the light scalar is coupled to the inflaton and reheating at the end of inflation involves non-equilibrium processes the observational signatures are richer, including anisotropies in the gravitational wave background~\cite{Bethke:2013aba} and highly non-Gaussian contributions to the curvature perturbation~\cite{preh}. If the perturbations are subdominant, they would generally lead to a small non-gaussianity parameter $f_{\rm NL}$ on large scales~\cite{Suyama:2013dqa} and would therefore be compatible with the PLANCK data.\\

\noi The effects of non-equilibrium dynamics of light scalars have been studied mostly in the context of preheating with a parametric resonance, but analogous observable signatures should also be produced during tachyonic preheating~\cite{Copeland:2002ku} in hybrid inflation models. As discussed in subsection~\ref{ssec:neqhi}, tachyonic preheating can involve highly non-trivial non-equilibrium phenomena such as  formation of topological defects, Q-balls or oscillons, which can all be influenced by the light scalar field and therefore lead to observable signatures.\\

\noi The simplest hybrid inflation model studied in Chapter~\ref{ch:hi} consists of two fields, the inflaton $\phi$ and the waterfall field $\chi$. When inflation ends rapidly, which is required for non-equilibrium processes, the waterfall field has to be heavy, and therefore it would have no effect on cosmological scales.
In bosonic models it is possible to add another scalar field by hand, but if the new field is coupled to the two other fields, it is generally not light. It is therefore interesting to consider the scenario in the context of supersymmetric hybrid inflation in which case the lightness of the field is protected by supersymmetry.\\

\noi In Section~\ref{sec:SUSYhi} we briefly introduce the minimal supersymmetric hybrid inflation model.
In Section~\ref{sec:3HI}, we derive an F-term version in N=1 SUSY with three dynamically relevant scalar fields \cite{mpHPS}.  We analyze an inflationary scenario in which the waterfall field is heavy and the symmetry breaking phase transition is fast, as in Section~\ref{sec:vdhchi}. We study the primordial perturbations generated on super-horizon scales and find that the model can produce the observed amplitude and spectral index. 

\section{Minimal Supersymmetric Hybrid Inflation}
\label{sec:SUSYhi}

\noi Hybrid inflation models derived from SUSY can be of two types, F-term and D-term hybrid inflation (for a review of inflationary model building from SUSY see \cite{SUSYinf}). Of the two, the F-term type attracts more attention because it naturally fits with the Higgs mechanism. In this section, we will describe how the minimal SUSY F-term hybrid inflation potential is derived.\\

\noi The starting point is the superpotential \cite{minSUSY}

\be 
W=\alpha\Phi\left(X\overline{X}-\frac{v^2}{2}\right)
\label{susyhi}
\en

\noi where $\Phi$ is a gauge singlet containing the inflaton and $X$, $\overline{X}$ is a conjugate pair of superfields transforming as non-trivial representations of a gauge group. It is the most general form of superpotential consistent with the R-symmetry, under which $W\rightarrow e^{i\gamma}W$, $\Phi\rightarrow e^{i\gamma}S$ and $X\overline{X}$ is invariant \cite{minSUSY}. In this sense, the superpotential (\ref{susyhi}) is natural.\\

\noi The scalar potential is given by  \cite{SUSYinf}

\be
V=\sum_\alpha\left|\frac{\partial W(S^{\alpha})}{\partial S^{\alpha}}\right|^2,
\label{susyvdef}
\en
where the sum is taken over all the superfields $S^{\alpha}$.\\
\noi For the superpotential (\ref{susyhi}), we find

\be
V=\alpha^2|\Phi|^2\left(|X|^2+|\overline{X}|^2\right)+\alpha^2\left|X\overline{X} - \frac{v^2}{2}\right|^2.
\label{Vhi}
\en

\noi There is a single global minimum and it preserves SUSY \cite{minSUSY}:

\be 
X = \overline{X} = \pm \frac{v}{\sqrt{2}} \;\;\;\;\;\;\;\;\; \Phi=0.
\en

\noi However, in the tree level potential (\ref{Vhi}) there is no term that drives $\Phi$ to zero. It is no longer the case when radiative corrections are taken into account \cite{minSUSY}. For $|\Phi|>0$ SUSY is broken, meaning that one-loop corrections to the potential do not cancel. They are given by \cite{minSUSY}

\be 
\Delta V = \sum_{i}\frac{(-1)^F}{64\pi^2}M_i^4\ln\frac{M_i^2}{\Lambda^2}
\label{1loop}
\en

\noi where $F=0,1$ respectively for a boson or a fermion and $M_i$ is the effective mass measured at a scale $\Lambda$. These corrections lift the potential in the $\Phi$ direction, driving the system towards the global minimun \cite{minSUSY}.\\

\noi This model is appealing for a series of reasons, perhaps the main one being the possibility to embed it in the particle physics framework. Furthermore, the potential (\ref{Vhi}) has only two parameters, $\alpha$ and $v$. The relations to the parameters of the hybrid potential (\ref{eqhi:hi}) are  $\alpha^2=g^2=\lambda$ and requiring the fields to travel sub-Planckian distances implies $v<1$.\\ 
\noi The dynamics of the phase transition depend on the hierarchy between the masses of the two fields $\Phi$ and $\sqrt{X\overline{X}}$, with behaviour interpolating from that described in Section~\ref{sec:vdhchi} to that described in Section \ref{sec:vdlchi}. It follows that the model can predict a red spectral index $\ns<1$, as shown in Fig.~\ref{vdlnsvsf} provided that inflation lasts for more than $60$ e-folds after the phase transition. However, a recently appeared paper \cite{Clesse:2013jra} claims that this regime does not lead to a viable cosmology.

\section{Supersymmetric Hybrid Inflation with a Light Scalar Field}
\label{sec:3HI}

\noi In this section we derive a SUSY hybrid model with a $3^{\rm rd}$ scalar field which is light at horizon crossing.\\ 

\noi The superpotential is

\be W=\alpha\Phi\left(X\overline{X}-\frac{v^2}{2}\right)+\frac{m}{2}\Phi^2, \label{w}\en

\noi where $\Phi$ is a gauge singlet and $X$, $\overline{X}$ is a conjugate pair of superfields in $N=1$ superspace. The Kahler potential is supposed to give the canonical kinetic terms plus negligible terms. Note that in (\ref{w}) $\Phi$ breaks the R-symmetry. This can be justified by assuming that the R-symmetry is not exact, with $m\ll\alpha v$. Otherwise, since no observation tells us that this is a fundamental symmetry of nature, we can take the view that (\ref{w}) is just a phenomenological model.\\

\noi The scalar sector of the potential is given by
\be V &=& \alpha^2\left(X\overline{X}-\frac{v^2}{2}\right)^2 + m^2|\Phi|^2 +2\alpha^2|\Phi|^2|X|^2\nn 
         &&  + \alpha m (\Phi + \Phi^*)\left(X\overline{X}-\frac{v^2}{2}\right)\,. 
\label{vscalar}
\en
Let us expand the field $\Phi$ into its real and imaginary components:
$\Phi = \frac{1}{\sqrt{2}}(\phi + i\sigma)$.
The potential (\ref{vscalar}) becomes
\be 
V &=& \frac{\alpha^2}{4}\left(\chi^2 - v^2\right)^2 + \frac{m^2}{2}(\phi^2+\sigma^2)\nn
     && +  \frac{\alpha^2}{2}\chi^2(\phi^2+\sigma^2) + \frac{\alpha m}{\sqrt{2}}\phi\left(\chi^2-v^2\right)\,,
 \label{vs}
\en
where, without loss of generality, we constrained $X=\overline{X}=\chi/\sqrt{2}$, $\chi$ being a real scalar field. Note that the fields $\phi$ and $\sigma$ share the same mass, whereas $\chi$ does not.
\\

The potential (\ref{vscalar}) has three degenerate global minima with $V=0$:
a symmetry-preserving vacuum (SPV) at 
\begin{equation}
\phi = w, \quad \chi=\sigma=0,
\end{equation}
with $w=\alpha v^2/(\sqrt{2}m)$, and
two symmetry-breaking vacua (SBV) at 
\begin{equation}
\phi=\sigma=0,\quad \chi=\pm v.
\end{equation}
We are interested in inflationary trajectories that end up at the SBV, with a fast phase transition driven by the waterfall field $\chi$. Its mass is given by $m^2_{\chir} = -\alpha^2v^2+\alpha^2(\phi^2+\sigma^2)+\sqrt{2}\alpha m\phi$ and it becomes tachyonic inside the circle in the $\phi-\sigma$ plane defined by
\be 
\left(\phi+\frac{m}{\sqrt{2}\alpha}\right)^2+\sigma^2=v^2+\frac{m^2}{2\alpha^2}\,.
\label{circle}
\en

\begin{figure}
\centering
\includegraphics[width= 8cm]{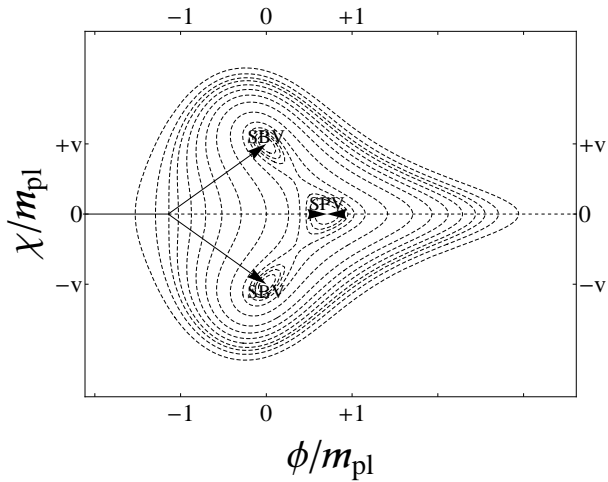}
\caption{Contour plot of the potential~(\ref{vs}) with $\sigma=0$, $V(\phi,\chi,0)$. We choose $m=v=\mpl$ and $\alpha=1$ for drawing purposes, although these parameters are unrealistic. The arrows represent possible inflationary trajectories. If inflation starts at some $\phi_{\rm i} > 0$, then the trajectory hits the SPV. On the other hand, if $\phi_{\rm i} < 0$, the trajectory may end up at the SBV.} 
\label{Vphichi}
\end{figure}

\begin{figure}
\centering
\includegraphics[width= 8cm]{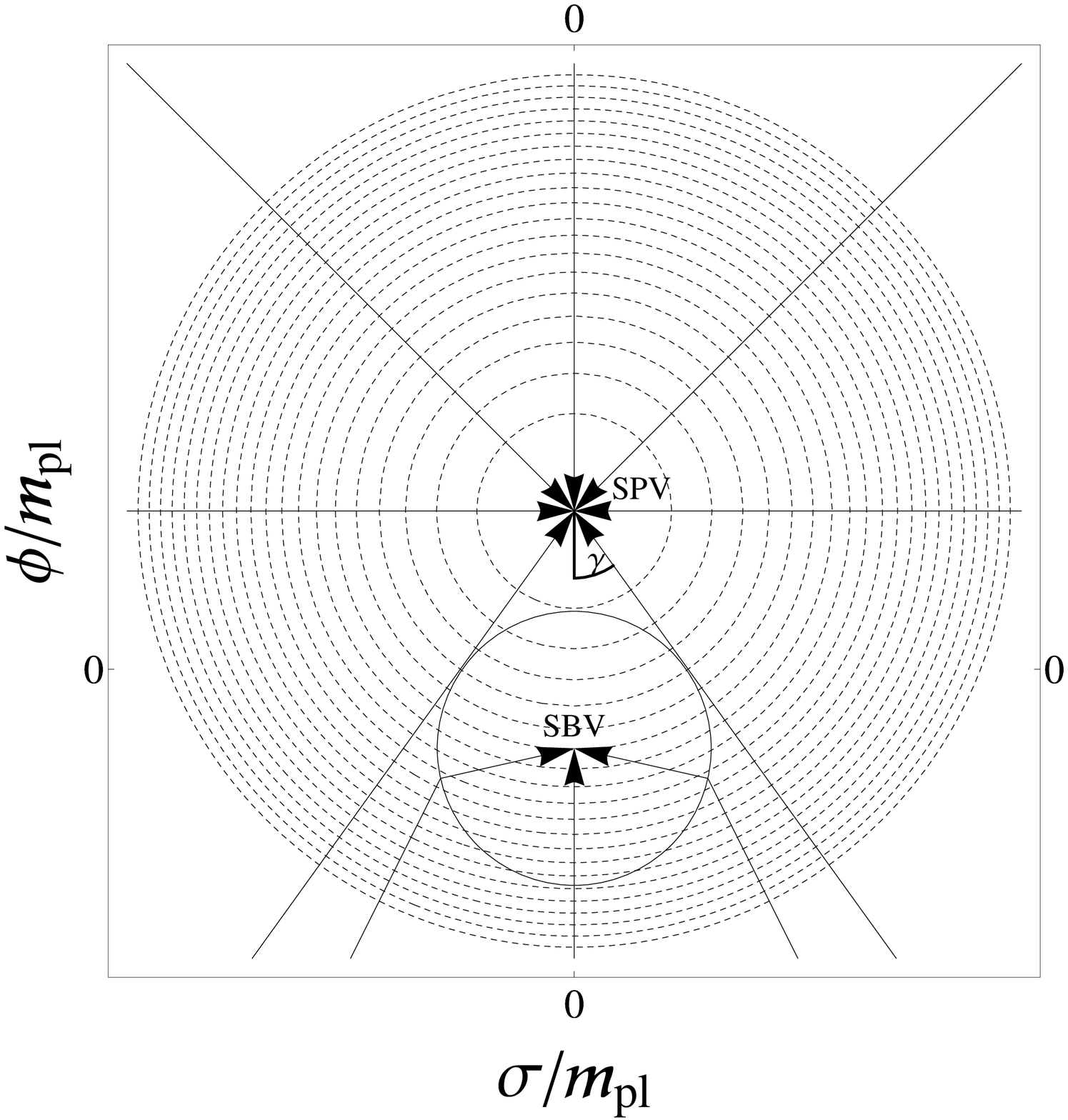}
\caption{Contour plot of the potential~(\ref{vs}) with $\chi=0$, $V(\phi,0,\sigma)$. We choose $m=v=\mpl$ and $\alpha=1$ for drawing purposes, although these parameters are unrealistic. The arrows represent possible inflationary trajectories and the circle corresponds to $m^2_\chi = 0$. Trajectories at an angle greater than $\gamma$ to the $\sigma=0$ line don't hit the $m^2_\chi<0$ region and the SBV.} 
\label{Vphisigma}
\end{figure}

\noi The possible inflationary trajectories are shown in Figs.~\ref{Vphichi} and \ref{Vphisigma}. The key fact is that, at large field values, the global attractor of the system is the SPV. Only trajectories that, on their way to the SPV, hit the tachyonic region $m^2_\chi<0$, can potentially end up at the SBV, as can be seen in Fig.~\ref{Vphichi}. However, hitting the instability region is not sufficient to guarantee that the system ends up at the SBV: the waterfall field $\chi$ still has to sufficiently roll down the ridge while in the tachyonic region, otherwise the trajectory crosses the instability region and reaches the SPV. This means the waterfall field has to be heavy: $\alpha v > m$ or, equivalently $r\equiv m/(\alpha v) \ll 1$. For such parameters, the waterfall is fast, analogously to the regime discussed in Section \ref{sec:vdhchi}.\\

\noi Fig.~\ref{Vphichi} highlights the fact that trajectories rolling from positive $\phi$ end up at the SPV, constraining the interesting trajectories to those with initial condition $\phi<0$. Fig.~\ref{Vphisigma} strengthens the constraint: only trajectories within an angle $\gamma$ or $(2\pi-\gamma)$ from the $\sigma=0$ trajectory in the $\phi-\sigma$ plane hit the instability region. At large field values $\phi\gg v$ and $\sigma\gg v$, $\phi$ and $\sigma$ are related by an approximate $U(1)$ symmetry, as can be seen in Fig.~\ref{Vphisigma}. This suggests that at such large field values, trajectories will come from random directions with probability uniformly distributed along a circle centred at the SPV. Under these assumptions, the probability of hitting the instability region (\ref{circle}) is given by the ratio of the arc corresponding to trajectories entering the tachyonic region to the circumference of the circle, or, in terms of the angle $\gamma$ defined in Fig.~\ref{Vphisigma}, $P_{\rm SBV}=\gamma/\pi$. Doing some trigonometry, we find that the likelihood is given by

\begin{equation}
\label{gamma}
P_{\rm SBV}=\frac{\gamma}{\pi}=\frac{1}{\pi}\arcsin\frac{c\sqrt{c^2+2}}{c^2+1}
\approx\frac{\sqrt{2}}{\pi}c. 
\end{equation}
In the limit $r \ll 1$, we find $\sin(\gamma) \approx \sqrt{2} r$, which is small. Therefore we can expand Eq.~(\ref{gamma}) around $\gamma=0$ to find $\gamma \approx \frac{m}{\alpha v}$ and the likelihood of hitting the SBV is $P_{\rm SBV} = \frac{m}{\pi\alpha v}$, a very small probability. In order to avoid fine-tuning issues as much as possible, we will therefore focus on the largest possible ratio $r$ that leads to the desired phenomenology. As shown in what follows, this ratio turns out to be $r\approx10^{-3}$, which gives a likelihood $P_{\rm SBV} \approx 10^{-4}$, clearly not enough to avoid fine-tuned initial conditions. Thus, for most of the trajectories, inflation sourced by the potential (\ref{vs}) looks like single field inflation along a quadratic potential, leading to the predictions derived in \ref{sec:ciNG}.\\

\noi Although the potential (\ref{vs}) requires fine-tuning to lead to the desired phenomenology, the idea that light scalar fields might source superhorizon scale perturbations in hybrid models with tachyonic preheating is more general and there might exist less fine-tuned realisations. Therefore the potential (\ref{vs}) should be considered as an attempt to motivate the presence of several light scalar fields at the end of inflation.\\
\noi In the following section we present a calculation of the local contribution to the spectral index $\ns$ and the non-gaussianity parameter $\fnl$, leaving the study of non-equilibrium dynamics for future work.

\section{Non-Gaussianity of Primordial Perturbations}

\subsection{Analytical estimates}
\label{sec:anes}

\noi Now that we derived the potential (\ref{vs}), we calculate the local contribution to the curvature perturbation generated on superhorizon scales when inflation happens along trajectories ending up at the SBV.\\

\noi The inflationary scenario is as follows: initially $\chi=0$ and both $\phi<0$ and $\sigma$ slow-roll towards the minimum of their quadratic potential. When the trajectory hits the tachyonic region, the symmetry breaking transition starts, ending inflation in less than $1$ e-fold. As Fig.~\ref{Vphisigma} emphasizes, the exact amount of inflation happening before the critical surface, defined in Eq.~(\ref{circle}), and the velocity of the phase transition depend on where the trajectory hits the surface. Thus, we have two separate contributions to the curvature perturbation:

\begin{itemize}
\item expansion up to the critical surface $N_{\rm{crit}}(\sigma_*)$.
\item expansion from the critical surface to a final hypersurface of constant energy density in a radiation dominated universe $N_{\rm{tr}}(\sigma_*)$.
\end{itemize} 

\subsubsection{\texorpdfstring{$N_{\rm{crit}}(\sigma_*)$}{TEXT}}

\noi Let us first compute the number of e-folds $N_{\rm{crit}}(\sigma_*)$ from a flat hypersurface at horizon crossing to the critical surface defined by Eq.~(\ref{circle}), i.e. $m^2_\chi=0$.\\

\noi The equations of motion of the slow-rolling fields are
\be \frac{\partial\phi}{\partial N}&=&-\frac{m^2\phi-\alpha mv^2/\sqrt{2}}{3H^2}\nn
    \frac{\partial\sigma}{\partial N}&=&-\frac{m^2\sigma}{3H^2}\label{eom}
\en
\noi where Eqs.~(\ref{eqi:SR}) give 
\be
H^2 = \frac{\alpha^2v^4}{12} + \frac{m^2(\phi^2+\sigma^2)}{6} - \frac{\alpha m v^2\phi}{(3\sqrt{2})}.
\label{hubble}
\en
Taking the ratio between the equations in Eqs.~(\ref{eom}) and solving for $\sigma$ with initial condition $\sigma(\phi_*)=\sigma_*$, we find 

\be
\sigma(\phi) = \sigma_*\frac{\phi - w}{\phi_*-w}\,.
\label{sigmaphi}
\en
We then substitute Eq.~(\ref{sigmaphi}) into Eq.~(\ref{hubble}) to find

\be
N_{\rm crit}(\phi_*,\sigma_*)&=&\frac{1}{4}\left(1+\frac{\sigma_*^2}{(\phi_*-w)^2}\right)\times\nn
&&\left(\left(\phi_*-w\right)^2-\left(\phi_{\rm crit}(\phi_*,\sigma_*)-w\right)^2\right)\,,
\label{ncrit}
\en
where
\be
\phi_{\rm crit}(\phi_*,\sigma_*)&=&-\frac{1}{w\left(\left(\phi_*-w\right)^2+\sigma_*^2\right)}\times\nn
&&\left[\left(\phi^*-w\right)\sqrt{v^2\left(v^2+4w^2\right)\left(\phi_*-w\right)^2-4w^4\sigma_*^2}\right.\nn
&&+\left. v^2\left(\phi_*-w\right)^2-2 w^2\sigma_*^2\right]\,.
\label{fcrit}
\en
is the value of the field $\phi$ at the critical surface (\ref{circle}).
Thus, Eq.~(\ref{ncrit}) gives the amount of inflation happening before the transition starts.

\subsubsection{\texorpdfstring{$N_{\rm{tr}}(\sigma_*)$}{TEXT}}

\noi The second effect of the $\sigma$ field affects the phase transition. We split the evolution in two: $N_{\rm br}$ is the expansion from the critical surface to $\langle \chi^2 \rangle \approx v^2$, the time when the backreaction becomes important; $N_{\rm end}$ is the expansion from $\langle \chi^2 \rangle \approx v^2$ to a hypersurface of constant energy density, assuming that by the time the backreaction sets in, the universe is radiation dominated. Furthermore, we assume that fluctuations of $\phi$ and $\sigma$ are negligible (justified by inflation) so that, near the instability point, we can approximate

\be
\phi(t) &\approx& \phi_{\rm{crit}} - \dot{\phi}t.\nn
\sigma(t) &\approx& \sigma_{\rm{crit}} - \dot{\sigma}t.
\label{linear}
\en
\noi The time dependence of the fluctuations $\ctk$, up to a short time after the transition, is well approximated by the Minkowski linearized equation of motion

\be \partial_t^2\ctk &=& (\msq - 2\alpha^2(\phi^2(t)+\sigma^2(t))-\sqrt{2}\alpha m\phi(t)-k^2)\ctk \nn
                    &\approx & \left[\left((2\alpha^2\phic+\sqrt{2}\alpha m)\dot{\phi}+2\alpha^2\sigmac\dot{\sigma}\right)t-k^2\right]\ctk 
                    \label{chieom}
                   \en
\noi In a quantum field theory, (\ref{chieom}) is valid as an operator equation. The initial state is the vacuum and at tree level it is completely described by the two-point functions Eq.~(\ref{eqi:vac}), which for $\chi$ read

\be \langle \chi^*(\mathbf{k})\chi(\mathbf{k'})\rangle &=& \frac{1}{2|\mathbf{k}|}(2\pi)^3\delta^3(\mathbf{k}-\mathbf{k'}),\nn
    \langle \pi^*(\mathbf{k})\pi(\mathbf{k'})\rangle &=& \frac{|\mathbf{k}|}{2}(2\pi)^3\delta^3(\mathbf{k}-\mathbf{k'}),
    \label{vac}
\en
where $\pi=\delta_t\chi$.\\

\noi The solution is given by Airy functions:

\be \ctk=c_A(\mathbf{k})Ai(\omega t - \frac{k^2}{\omega^2})+c_B(\mathbf{k})Bi(\omega t - \frac{k^2}{\omega^2})
\label{chisol}
\en
\noi where $\omega = \left[(2\alpha^2\phic+\sqrt{2}\alpha m)\dot{\phi}+2\alpha^2\sigmac\dot{\sigma}\right]^{1/3}$.\\

\noi Each mode $k$ starts growing when $m^2_k\equiv\omega^3 t - k^2$ becomes negative, that is at $t_k =\frac{k^2}{\omega^3}$. The coefficients  $c_A(\mathbf{k})$ and $c_B(\mathbf{k})$ can be expressed in terms of $\chi(\mathbf{k})$ and $\pi(\mathbf{k})$ evaluated at time $t_k$ as

\be
c_A(\mathbf{k}) &=& \frac{1}{2}\left[\frac{\chi(t_k,\mathbf{k})}{Ai(0)}+\frac{\pi(t_k,\mathbf{k})}{\omega Ai'(0)}\right],\nn
c_B(\mathbf{k}) &=& \frac{1}{2\sqrt{3}}\left[\frac{\chi(t_k,\mathbf{k})}{Ai(0)}-\frac{\pi(t_k,\mathbf{k})}{\omega Ai'(0)}\right],
\label{coeffs}
\en
\noi where $Ai(0)=3^{-2/3}\Gamma(2/3)^{-1}\approx 0.355$ and $Ai'(0)=-3^{-1/3}\Gamma(1/3)^{-1}\approx -0.259$.\\

\noi It is easy to derive the field and momentum power spectrums  using Eq.~(\ref{chisol}). We find

\be
P_{\chi}(t,k) &\equiv& \frac{\langle \chi^*(t,\mathbf{k})\chi(t,\mathbf{k'})\rangle}{(2\pi)^3\delta^3(\mathbf{k}-\mathbf{k'})}\nn
              &=& \frac{1}{8k}\frac{\left(Ai[\omega(t-t_k)]+\frac{1}{\sqrt{3}}Bi[\omega(t-t_k)]\right)^2}{Ai(0)^2}\nn
              && + \frac{k}{8\omega^2}\frac{\left(Ai[\omega(t-t_k)]-\frac{1}{\sqrt{3}}Bi[\omega(t-t_k)]\right)^2}{Ai'(0)^2}\label{pchi}\\
P_{\pi}(t,k) &\equiv& \frac{ \langle \pi^*(t,\mathbf{k})\pi(t,\mathbf{k'})\rangle}{(2\pi)^3\delta^3(\mathbf{k}-\mathbf{k'})}\nn
             &=& \frac{\omega^2}{8k}\frac{\left(Ai'[\omega(t-t_k)]+\frac{1}{\sqrt{3}}Bi'[\omega(t-t_k)]\right)^2}{Ai(0)^2}\nn
             && + \frac{k}{8}\frac{\left(Ai'[\omega(t-t_k)]-\frac{1}{\sqrt{3}}Bi'[\omega(t-t_k)]\right)^2}{Ai'(0)^2}\label{ppi}.
\en
\noi We are interested in the time at which inflation ends. However, to compute (\ref{chisol}), we linearized the equation of motion. Thus, the solution can only be trusted until the linear approximation breaks down. This happens when $\langle \chi^2 \rangle  = v^2$ and we assume inflation ends at that time.\\ 
\noi We expect $t_{\rm{end}} > \omega^{-1}$, since $\omega$ is the characteristic frequency of the system and reheating should take more than one fluctuation. We are therefore interested in the behaviour of $\ctk$ at late times.\\

\noi Following \cite{Copeland:2002ku} we find that, at late times, the fluctuations of $\chi$ go as 

\be 
\langle \chi^2 \rangle (t) \approx \frac{2.645}{128\pi^3}\frac{\omega}{t}\exp\left[\frac{4}{3}(\omega t)^{3/2}\right].
\en

\noi Assuming inflation ends when $\langle \chi^2 \rangle (t_{\rm{br}}) \approx v^2$, we can solve for $t_{\rm{br}}$. This involves solving a non-linear equation of the form

\be e^{\frac{4}{3}x^{3/2}}=x y, \label{nl}\en
\noi where $x=\omega t$ and $y=\frac{128\pi^3v^2}{2.645\omega^2}$. Note that if $xy<1$, $x$ has to be imaginary, which does not make sense. This is symptomatic of the late time limit we took: if the transition is too fast, the approximation breaks down. Thus, we have the consistency relation:

\be 
t_{\rm{br}} \geq \frac{2.645\omega}{128\pi^3 v^2}.
\label{cons}
\en

\noi To a first approximation, the solution of Eq.~(\ref{nl}) is given by

\be N_{\rm{br}}&=&Ht\approx\frac{H}{\omega}\left(\frac{3}{2}\ln\frac{v}{\omega}\right)^{2/3}.
\en

\noi We assume that at this point, the universe is radiation dominated. The curvature perturbation $\zeta$ is evaluated on a final hypersurface of constant energy density. However, $\langle \chi^2 \rangle \approx v^2$ does not correspond to a constant energy density. Thus, we must include the expansion from then to a constant energy density $\rho_{\rm end}\equiv\rho(\langle \chi^2 \rangle \approx v^2)\exp(-4N_{\rm{end}})$. Find:

\be
\frac{\partial N_{\rm end}}{\partial \sigma_*}=\frac{1}{4}\frac{\partial \log \rho(\langle \chi^2 \rangle \approx v^2)}{\partial \sigma_*}.
\en

\noi The total amount of expansion therefore is

\be 
 N(\phi_*,\sigma_*)&=&N_{\rm{crit}}(\phi_*,\sigma_*)+N_{\rm{br}}(\phi_*,\sigma_*)+N_{\rm{end}}(\phi_*,\sigma_*). \nn
\label{n}
\en

\noi We are now ready to compute the observables of Eqs.~(\ref{eqdn:ampl},\ref{eqdn:spectrum},\ref{fnldn1}). 
The analytical expressions for the observables are long and complicated. However, there exists a limit in which they become simple enough to be presented here. Indeed, for trajectories such that $\sigmac=0$, the derivatives of $N_*^f$ are dominated by the contribution of $N_{\rm crit}(\phi_*,\sigma_*)$ and reduce to

\be 
\frac{\partial N_*^f}{\partial \phi_*} = \frac{\partial N_{\rm crit}}{\partial \phi_*} &=& \frac{1}{2}\left(\phi_* - \frac{v}{\sqrt{2}c}\right)\label{Nphi}\,,\\
\frac{\partial N_*^f}{\partial \sigma_*}\simeq \frac{\partial N_{\rm crit}}{\partial \sigma_*} &=& 0\label{Nsigma}\,.
\en
Eq.~(\ref{Nsigma}) is not exactly zero because of the contribution of $N_{\rm end}$. However, in practice both this term and the loop correction in Eq.~(\ref{fnldn1}) are negligible. Therefore the observables can be expressed using Eq.~(\ref{Nphi}) and its derivative with respect to $\phi_*$ as
 
\be
\ns 
    &=& 1 - \frac{16c^2\mpl^2}{(v-\sqrt{2}c\phi_*)^2}\,,\label{ns0}
\en
and
\be
\fnl
     &=& \frac{10}{3}\frac{c^2\mpl^2}{(v-\sqrt{2}c\phi_*)^2} \,.\label{fnl0}
\en
In the limit $ v/r \gg 2\phi^*$ and for fixed $v$, we find the simple relations $\ns\propto 1-16r^2\mpl^2/v^2$ and $\fnl\propto r^2/v^2$. The opposite regime, when $\phi_*$ dominates the denominator, gives $\ns \propto 1 - 8\mpl^2/\phi_*^2$ and $\fnl \propto \mpl^2/\phi_*^2$.\\
\noi Therefore, the presence of a third scalar field which is light at horizon crossing generates a red spectral index, in contrast to the original hybrid inflation scenario discussed in Section \ref{sec:vdhchi}. In the limit in which the potential~(\ref{vs}) is vacuum dominated at horizon crossing, this is due to the linear term in $\phi$ that implies $\epsilon_*\simeq\eta_*$ and $\ns<1$. As we approach $\phi_*$ domination, $\ns$ decreases.\\

\begin{figure}
\centering
\includegraphics[width= 8cm]{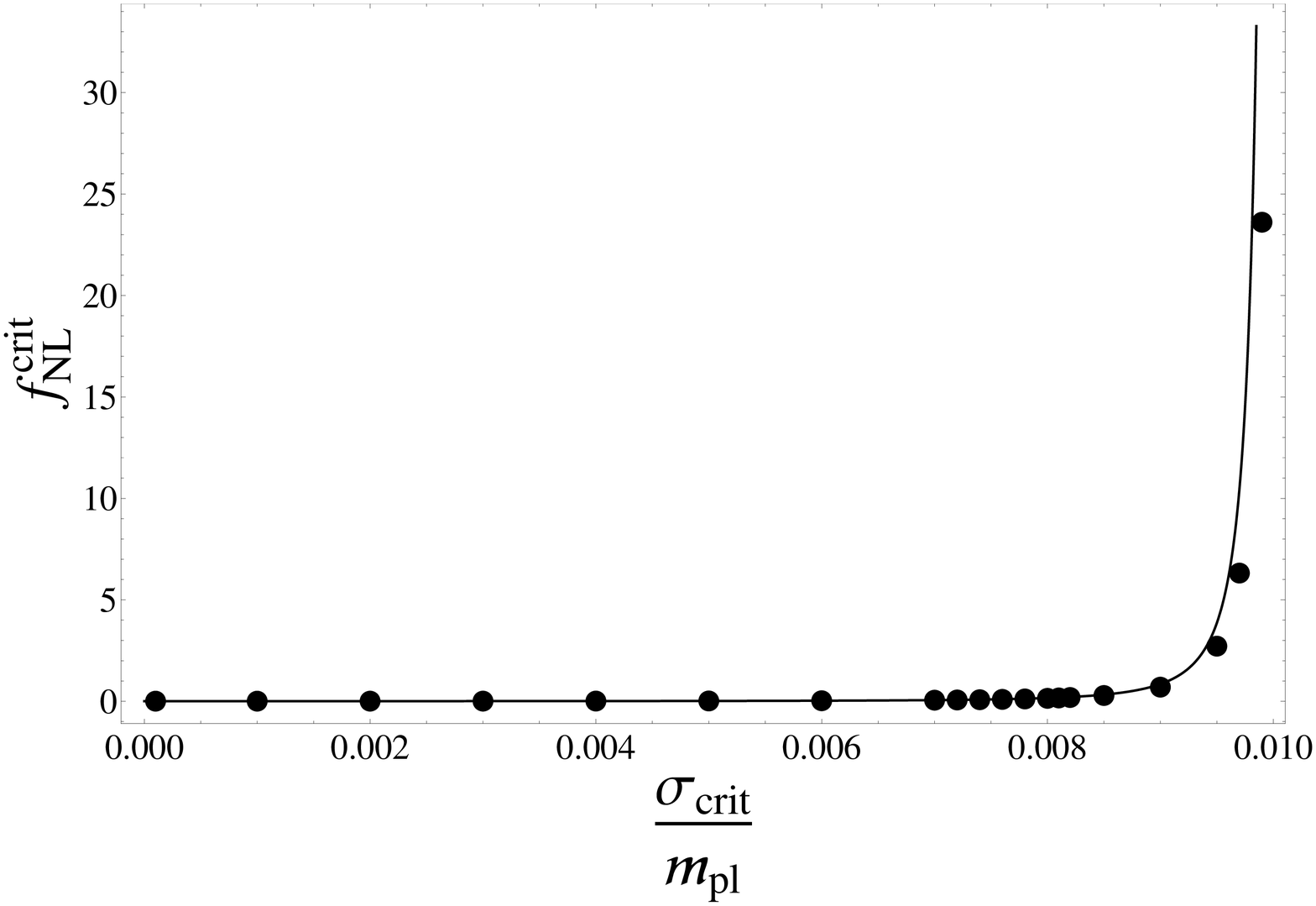}
\caption{$\fnl$ evaluated on the critical surface (\ref{circle}) as a function of $\sigmac$ for the potential~(\ref{vs}). The dots are numerical data points whereas the continuous line is the analytical result. The other parameters are fixed as $v=10^{-2}\mpl$ and $m= 10^{-3}\alpha v$. Besides the expected accuracy of the analytical results, it is interesting to note that the slow-roll contribution is positive and peaks at the edge of the critical surface.}
\label{fnlc3}
\end{figure}

\noi We conclude this section with some comments on the amplitude of the curvature perturbation, $A_\zeta$. PLANCK results give $A_\zeta \simeq 10^{-5}$, constraining the model's parameter space. During the slow-roll phase, the amplitude is the only observable that depends on rescalings of the potential, meaning we can fix it to the observed value a posteriori. However, once inside the critical surface (\ref{circle}), the velocity of the transition depends on the scale of the potential, implying that observables such as $\fnl$ and $\ns$ are not invariant under potential rescalings. Fortunately, both quantities dependence on $\sigmac$ does not change qualitatively as we vary $\alpha$, and they tend to drop as trajectories approach the edge of the critical surface (\ref{circle}). The drop can be understood via the following physical picture. Trajectories that hit the instability close to the edge of the critical surface gather more kinetic energy than the other trajectories, slowing the expansion rate and generating a negative $\delta N$. This explains the drop in $\fnl$ and the spectral index, as a negative $\delta N$ creates a negative contribution to both observables. This effect is in contrast with the slow-roll contribution, which generates a positive contributions that peaks as trajectories approach the edge of the critical surfeace, see Fig.~\ref{fnlc3}.\\
\noi We have to keep in mind that the analytical result (\ref{n}) relies on the linear approximation (\ref{linear}) and is not reliable without confirmation from numerical tests. We present them in the next section. 

\begin{figure}
\centering
\includegraphics[width= 10cm]{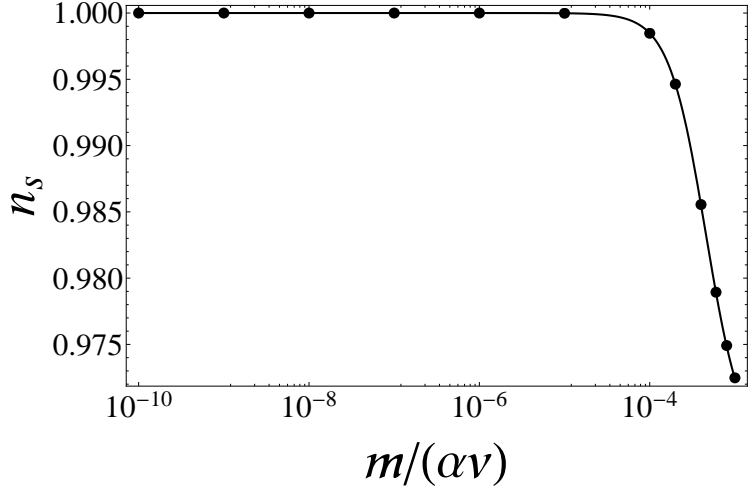}
\caption{$\ns$ as a function of $r\equiv m/(\alpha v)$ for the potential~(\ref{vs}). The continuous line represents the analytical result  Eq.~(\ref{ns0}) and the dots are numerical data points. The other parameters are fixed as $v=10^{-2}\mpl$, $\sigmac = 10^{-4}\mpl$ and $\alpha$ is chosen to satisfy the PLANCK amplitude constraint. Although the lowest data point, $\ns\simeq0.972$ and $c=10^{-3}$, is still outside the PLANCK $68\%$ confidence limit range, the spectral index decreases as the inflaton $\phi$ becomes heavier.}
\label{nsvsc}
\end{figure}

\begin{figure}
\centering
\includegraphics[width= 10cm]{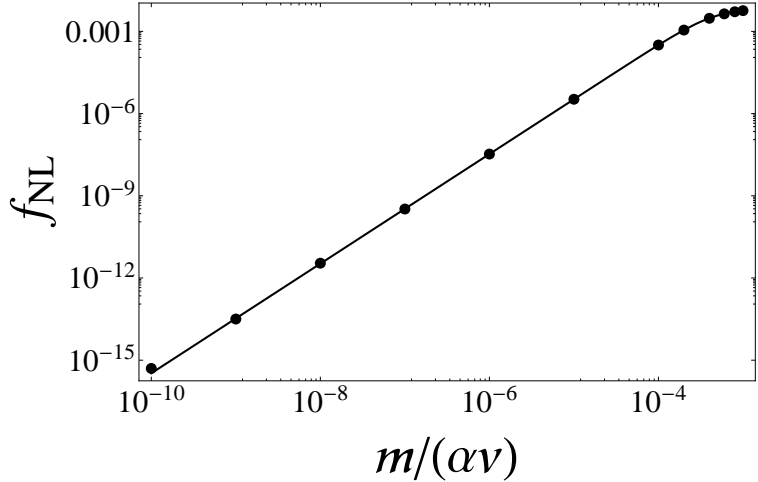}
\caption{$\fnl$ as a function of $c\equiv m/(\alpha v)$ for the potential~(\ref{vs}). The continuous line represents the analytical result Eq.~(\ref{fnl0}) and the dots are numerical data points. The other parameters are fixed as $v=10^{-2}\mpl$, $\sigmac = 10^{-4}\mpl$ and $\alpha$ is chosen to satisfy the PLANCK amplitude constraint.}
\label{fnlvsc}
\end{figure}

\subsection{Numerical analysis}

\noi The analytical results rely on the linear approximation~(\ref{linear}) and therefore are at best indicative. Nevertheless we can use them to identify suitable parameters for detailed numerical study.\\ 
\noi In order to calculate the perturbations numerically, we solved the homogeneous equations for the three fields:

\be
\ddot{\varphi}_\alpha + 3H\dot{\varphi}_\alpha+\frac{\partial V}{\partial \varphi_\alpha} &=& 0\,, \nn
\label{numeom}
\en

\noi where $\varphi_\alpha = (\phi,\sigma,\chi)$ and compute the observables using the method desrcribed in subsection~\ref{ssec:numDeltaN}.\\
\noi Some comments are in order. In order to solve the equations (\ref{numeom}), we have to choose initial conditions for the fields. Inflation happens in the $\phi-\sigma$ plane, therefore both fields follow a classical trajectory with well defined initial conditions. However, $\chi$ is heavy and does not have superhorizon fluctuations (see Section~\ref{ssec:vacFRW} and also subsection~\ref{ssec:obsvdhc}). Its initial conditions are described by the vacuum correlation functions (\ref{vac}), and its initial average value is $\chi_{\rm i}=0$. On the other hand, $\chi=0$ is a classically stable trajectory protected by symmetry, and the transition to the SBV happens because of quantum fluctuations. Therefore, Eq.~(\ref{numeom}) is not an accurate way of describing the quantum field $\chi$. An accurate representation is numerically challenging and unnecessary for our purpose, which is to study the curvature perturbation generated on super-horizon scales by the light scalars $\phi$ and $\sigma$. For this, we just need to know how the time necessary for the phase transition to the SBV depends on the field values at horizon crossing $\phi_*$ and $\sigma_*$. The dependence can be extracted by choosing a fixed initial value  $0<\chi_{\rm i}\ll v$ and varying the initial conditions at horizon crossing in the $\phi-\sigma$ plane, giving the function $N(\phi_*,\sigma_*)$.\\

\noi Therefore, the algorithm is as follows: we choose the initial conditions at horizon crossing $\phi_*$ and $\sigma_*$, and run the simulation until the trajectory oscillates around the SBV. After a few oscillations, we stop the simulation and record the final energy density $\rho_f$. Then, we vary the initial conditions $\phi_*$ and $\sigma_*$ and repeat the simulation, ending it when the energy density equals $\rho_f$. We repeat the process until we have enough points to compute the derivatives of  $N(\phi_*,\sigma_*)$ needed to evaluate the observables $A_\zeta$, $\ns$ and $\fnl$ [Eqs.~(\ref{eqdn:ampl}),~(\ref{eqdn:spectrum}) and (\ref{eqng:fnldn}) respectively], as explained in subsection~\ref{ssec:numDeltaN} . At this point, we compare the numerical $A_\zeta$ to the PLANCK constraint and rescale $\alpha$ as required. Finally, we repeat the process with the correct $\alpha$, obtaining a data point.\\

\begin{figure}
\centering
\includegraphics[width= 10cm]{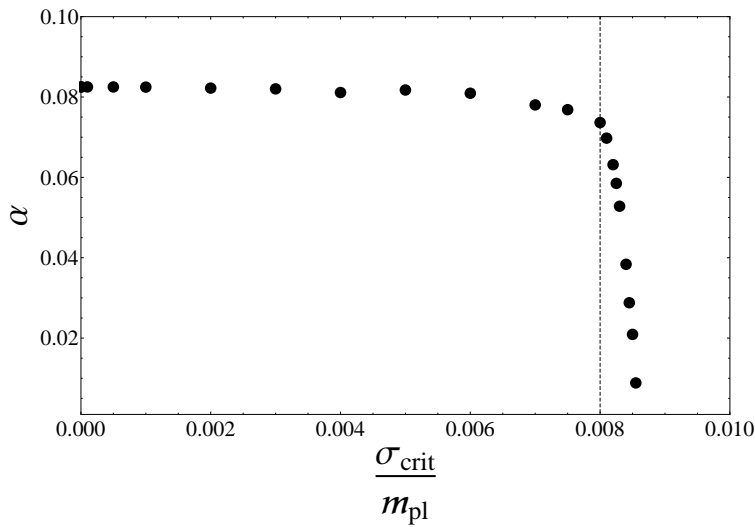}
\caption{$\alpha$ as a function of $\sigmac$ for the potential~(\ref{vs}), after the rescaling necessary to fit the observed amplitude of perturbations. The other parameters are fixed as  $v=10^{-2}\mpl$ and $m= 10^{-3}\alpha v$. The region to the left of the vertical dashed line leads to non-gaussianity parameter $\fnl$ within the PLANCK $68\%$ confidence limit range.}
\label{scalpha3}
\end{figure}

\begin{figure}
\centering
\includegraphics[width= 10cm]{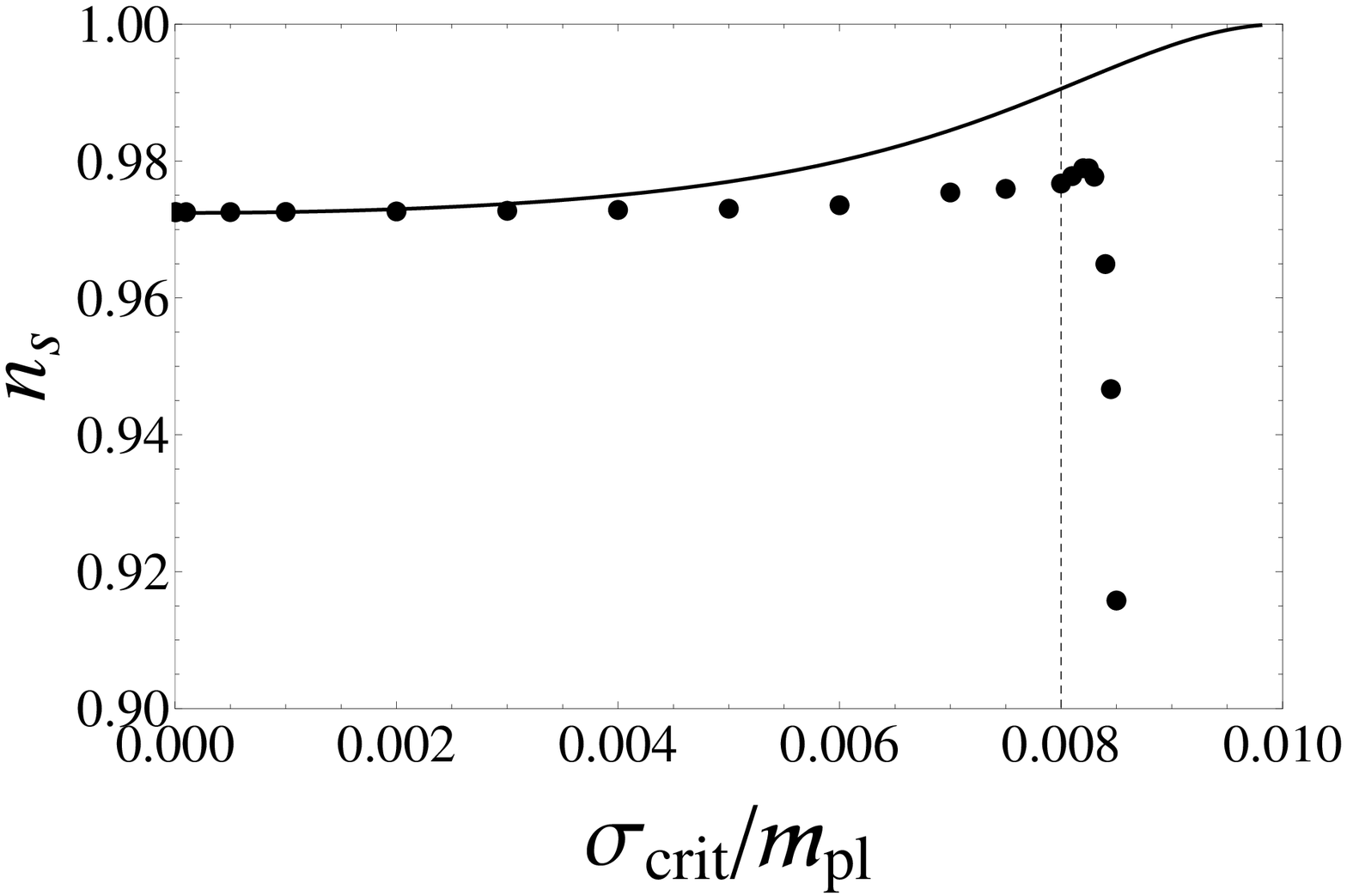}
\caption{$\ns$ as a function of $\sigmac$ for the potential~(\ref{vs}), numerical results (dashed line) and analytical results (continuous line). The other parameters are fixed as $\alpha=10^{-1}$, $v=10^{-2}\mpl$ and $m= 10^{-3}\alpha v$. The region to the left of the vertical dashed line leads to non-gaussianity parameter $\fnl$ within the PLANCK $68\%$ confidence limit range.}
\label{ns3}
\end{figure}

\begin{figure}
\centering
\includegraphics[width= 10cm]{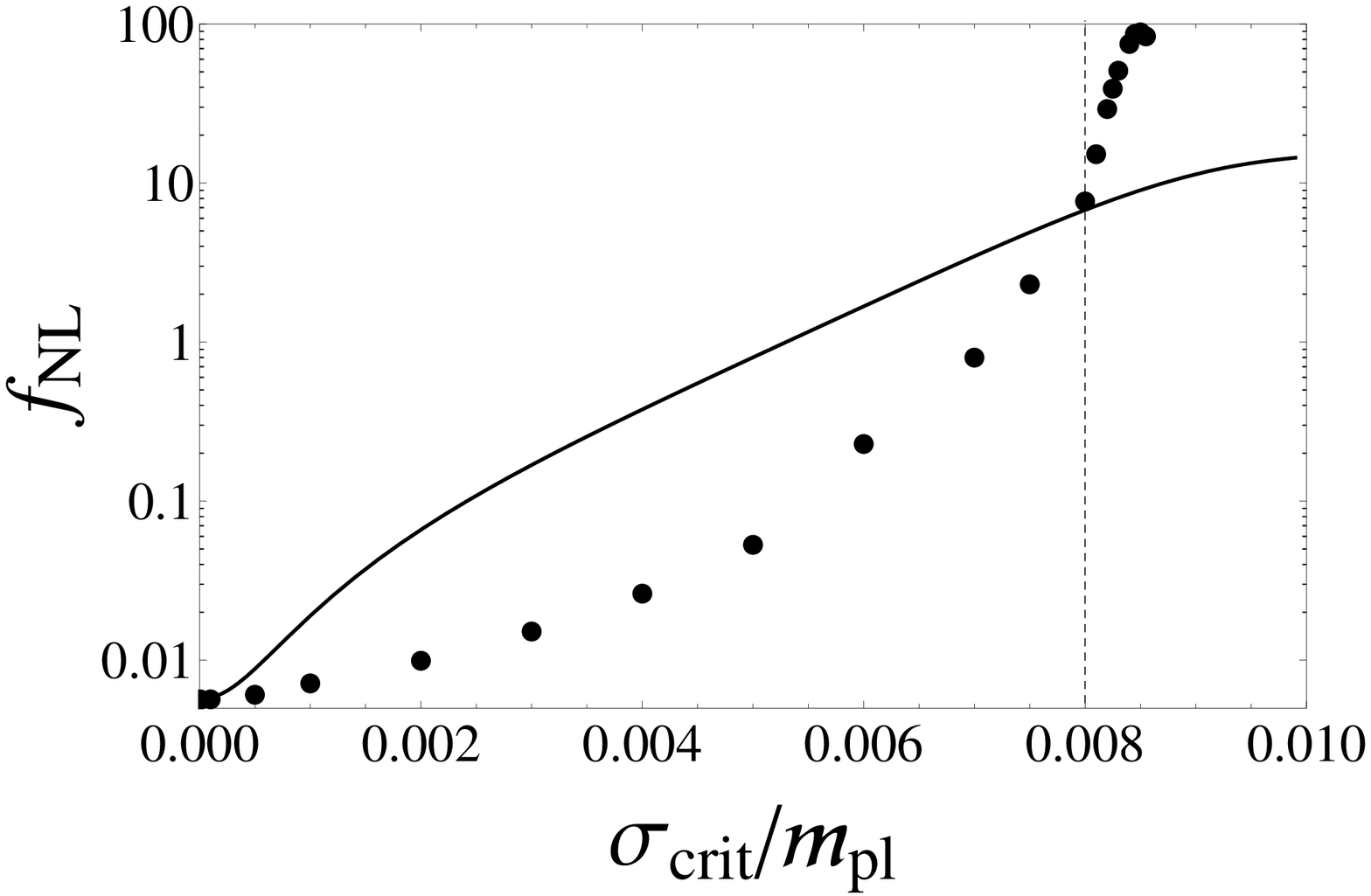}
\caption{$\fnl$ as a function of $\sigmac$ for the potential~(\ref{vs}), numerical results (dashed line) and analytical results (continuous line). The other parameters are fixed as $\alpha=10^{-1}$, $v=10^{-2}\mpl$ and $m= 10^{-3}\alpha v$. The region to the left of the vertical dashed line leads to non-gaussianity parameter $\fnl$ within the PLANCK $68\%$ confidence limit range.}
\label{fnl3}
\end{figure}

\noi The numerical results confirm the validity of the analytic estimates Eqs.~(\ref{ns0}) and (\ref{fnl0}) at small $\sigma_{\rm crit}$, as is shown in Figs.~\ref{nsvsc} and \ref{fnlvsc}. Furtmermore, requiring the spectral index to be within the $95\%$ confidence limit leads to the constraint $c\gtrsim5\times10^{-4}$ for $v=10^{-2}\mpl$. For fixed $c$, considering smaller values of $v$ drives the spectral index closer to unity, as can be seen from Eq.~(\ref{ns0}). On the other hand, larger values lead to smaller $\ns$, increasing the agreement with observations. However, as can be seen from Eq.~(\ref{eom}), increasing $v$ implies that $\phi<0$ rolls faster through the critical surface, decreasing the cross section of field values ending up in the SBV.\\

\noi Eq.~(\ref{gamma}) highlights that as $c$ increases, so does the likelihood of a random trajectory hitting the critical surface. However, increasing $c$ has another undesired effect: the light scalars $\phi$ and $\sigma$ roll faster compared to $\chi$ and the transition to the SBV might not have time to happen if too close to the edge of the critical surface. Indeed, we found that for $c>10^{-3}$, only trajectories significantly far from the edge hit the SBV.\\
   
\noi From Fig.~\ref{nsvsc} we know that for $c<10^{-3}$ the predicted spectral index is in conflict with observations in the limit $\sigma \rightarrow 0$. For $c=10^{-3}$, we find $\ns\simeq0.972$, which is also excluded by the $68\%$ confidence limit bounds, but is well within the $95\%$ bracket. Furthermore, Fig.~\ref{fnlvsc} shows that $\fnl$ grows proportionally to $c$.\\

\noi Considering these constraints we choose $v=10^{-2}\mpl$ and $c=10^{-3}$ for more detailed study. For such parameters, the loop correction in Eq.~(\ref{fnldn1}) is of order $10^{-8}$ in the limit $\sigmac\rightarrow 0$.\\

\noi Fig.~\ref{scalpha3} shows the rescaled dimensionless parameter $\alpha$ as a function of $\sigmac$. It is interesting to note that it is at least of order $10^{-2}$, and of order $10^{-1}$ for trajectories leading to a viable cosmology.\\  

\noi In Figs.~\ref{ns3} and \ref{fnl3} the spectral index $\ns$ and the non-Gaussianity parameter $\fnl$ are plotted as a function of $\sigmac$. The dots represent numerical data points whereas the continuous line shows the analytical results. We can see that the analytical approximation is only valid for small $\sigmac$, and its accuracy decreases with the increasing importance of non-linear dynamics.
For $c=10^{-3}$, we find that the universe reaches the SBV at the relatively low probability $P_{\rm SBV} \approx 4\times10^{-4}$.\\

\noi Requiring the spectral index to be in the observationally compatible range at $95\%$ confidence level constrains the viable range of field values to $|\sigmac| \lesssim 8.3\times10^{-3}\mpl$. The non-Gaussianity parameter $\fnl$ increases the constraint on the initial conditions, giving $|\sigmac| \lesssim 8\times10^{-3}\mpl$ for predictions within the $68\%$ confidence limit bracket. \\

\noi Note that for $\sigmac>8.3\times10^{-3}\mpl$ the spectral index drops sharply, as can be seen in Fig.~\ref{ns3}. 
The drop is due to the increase in magnitude of $N_{,\sigma\sigma}/N_{,\sigma}$ as $\sigmac$ approaches the edge of the circle (\ref{circle}). Since $\dot{\sigma}_*$ is negative, this generates a negative contribution to the spectral index, as can be seen from Eq.~(\ref{eqdn:spectrum}).\\

\section{Discussion}

\noi In this Chapter we have derived a SUSY F-term version of hybrid inflation with $3$ dynamically relevant scalar fields: the inflaton $\phi$, the waterfall field $\chi$ and a light scalar field $\sigma$. It is worth noting that, if we take the view that the inflaton's mass term in the superpotential ({\ref{w}) is added as a soft R-symmetry breaking term, the scalar $\sigma$ naturally has a light mass.\\

\noi The potential (\ref{vs}) has three degenerate global minima, one with unbroken and two with broken symmetry, as illustrated in Fig.~\ref{Vphichi}. Inflationary trajectories that end up in the symmetry breaking vacuum require fine tuning of initial conditions, which could be justified by anthropic arguments. For parameter values leading to observationally compatible predictions, we found that, under reasonable assumptions, the likelihood of an inflationary trajectory ending up in the symmetry breaking vacuum is of order $4\times10^{-4}$.\\

\noi We considered the regime in which the symmetry breaking field $\chi$ is heavy during the inflationary phase, i.e. $m_\chi>H$, and does not affect super-horizon observables. We have shown that the presence of the light scalar $\sigma$ generates a red spectral $\ns<1$, as can be seen in Eq.~(\ref{ns0}), in agreement with observations and in contrast with the original hybrid inflation scenario.\\ 

\noi The analytical analysis carried in Section~\ref{sec:anes} led us to choose parameters $v=10^{-2}\mpl$ and $r\equiv m/(\alpha v)=10^{-3}$ for a detailed study. In Section~\ref{sec:numerical}, we numerically computed the observables $\ns$, $\fnl$ and $A_\zeta$ for such parameters. We found that trajectories hitting the critical surface Eq.~(\ref{circle}) at values such that $|\sigmac|\leq 8\times10^{-3}$ lead to a spectral within the current $2\sigma$ confidence limit range and to non-gaussianity parameter $\fnl$ within the $68\%$ confidence limit range, as shown, respectively, in Figs.~\ref{ns3} and \ref{fnl3}. Furthermore, for such trajectories, the rescaled dimensionless parameter $\alpha$ is of order $10^{-1}$, as can be seen in Fig.~\ref{scalpha3}.\\

\noi It is known that the presence of light scalar fields has non-trivial effects on preheating dynamics~\cite{Bethke:2013aba,preh}. Analogously, light scalars should affect the non-equilibrium dynamics of tachyonic preheating, potentially leading to observational signatures. Their study requires a lattice simulation of the symmetry breaking phase transition, and would complete the work presented in this Chapter.

\chapter{Conclusions}

In this thesis we studied the generation of primordial perturbations in the context of inflationary models. In Chapter \ref{ch:sepuni} we described a a formalism for computing the curvature perturbation seeded on scales corresponding to the observed CMB. These methods were applied to a series of models, using both analytical and numerical techniques.\\

\noi In Chapter \ref{chap:Intro} we emphasized that, given the enormous number of possible realisations of inflation, it is desirable to consider models motivated by particle physics. The class of models that has been the focus of this thesis are hybrid models of inflation. They are the result of the marriage of inflation with spontaneous symmetry breaking, allowing for models of inflation embedded in particle models with Higgs-type mechanisms.\\

\noi In Chapter \ref{ch:hi} we studied in detail three different regimes of the simplest realisation of such class, described by the potential (\ref{eqhi:hi}). The first regime is considered in Section~\ref{sec:vdhchi}. It corresponds to the original scenario of hybrid inflation: inflation is driven by a single scalar field, the inflaton, until the inflating trajectory reaches the critical point, triggering the waterfall phase that abruptly ends inflation. We found that the statistics of the perturbations of super-horizon modes is determined by the single-field inflationary trajectory, the non-linear dynamics of the waterfall phase only affecting sub-horizon modes. The predictions for the spectral index $\ns$ and the reduced bispectrum $\fnl$ for this scenario are given in Eqs.~(\ref{eqvdh:obs}) and are disfavoured by observations.\\
\noi In Sections~\ref{sec:vdlchi} and~\ref{sec:philchi} we considered regimes in which inflation lasts for at least $60$ e-folds after the phase transition, meaning that the observable signatures on CMB scales are solely determined by the dynamics after the symmetry breaking transition has happened. In Section~\ref{sec:vdlchi} the potential is vacuum dominated until the very end of inflation. As shown in Fig.~\ref{vdlfields}, this regime leads to a red spectral index. However, a recently published paper \cite{Clesse:2013jra} analysed the parameter space of waterfall trajectories leading to more than $60$ e-folds of inflation after the critical point with analytical and numerical techniques, reaching the conclusion that it can't lead to a viable cosmology.\\
\noi In Section~\ref{sec:philchi}, inflation is driven by the inflaton's mass term. Considered in this regime, the potential (\ref{eqhi:hi}) is approximately of the separable form and has similar phenomenology to that of hilltop curvaton models. As in such models, a large non-gaussianity can be generated on superhorizon scales via the hilltop mechanism, as shown in Fig.~\ref{confnl}, while satisfying observational constraints for the spectral index (Fig.~\ref{conns}) and the amplitude of perturbations.\\

\noi In Chapter~\ref{ch:susyhi} we took the idea that inflationary models should be motivated by particle physics one step further: we explicitly studied realisations of hybrid inflation derived from SUSY. In Section~\ref{sec:SUSYhi} we described the mininal supersymmetric model of hybrid inflation, arguing that it can lead to observationally compatible predictions in the regime where the phase transition is slow, as discussed in Section \ref{sec:philchi}.\\
\noi In Section~\ref{sec:3HI} we derived from N=$1$ SUSY a F-term hybrid model with $3$ dynamically relevant scalar fields. We studied the properties of the curvature perturbation on CMB scales and found that the presence of a third light scalar field generates a red spectral index in the regime where the symmetry breaking transition is fast, with values compatible with the observationally favoured range, see Fig.~\ref{ns3}. This result is in contrast to the prediction of the analogous scenario with no light fields discussed in Section~\ref{sec:vdhchi}. Furthermore, large non-gaussianity can be generated in this model, see Fig.~\ref{fnl3}, constraining viable initial conditions. The scenario requires fine-tuning of the initial conditions to a high degree, with most of the trajectories ending up in a symmetry preserving vacuum. Assuming a homogeneous probability distribution of the initial conditions, approximately one trajectory in $10^{4}$ leads to the desired phenomenology. If backed by future observations, one might invoke the anthropic principle to justify such phenomenology.\\
\noi Independently of the particular details of the model, the phenomenology of hybrid inflation with additional light scalar fields is interesting. As we saw, the presence of light scalar fields can drastically change the predictions of the model, possibly increasing the compatibility with observations.


\begin{thebibliography}{tbds}

\bibitem{mpHP}
  D.~Mulryne, S.~Orani, A.~Rajantie,
  '\emph{Non-Gaussianity from the hybrid potential}',
  Phys.Rev. {\bf D84} (2011) 123527,
  [hep-th/11074739]
  
 \bibitem{mpHPS}
  S.~Orani, A.~Rajantie,
  '\emph{Supersymmetric hybrid inflation with a light scalar}',  
	Phys.Rev. {\bf D88} (2013) 043508,
 [astro-ph.CO/13048041]. 

  \bibitem{planck}
  P.~A.~R.~Ade {\it et al.}  [ Planck Collaboration],
  [astro-ph.CO/13035084];
  P.~A.~R.~Ade {\it et al.}  [ Planck Collaboration],
  [astro-ph.CO/13035082].
 
\bibitem{wmap} 
  E.~Komatsu {\it et al.},
  [astro-ph.CO/12125226];
  E.~Komatsu {\it et al.},
  [astro-ph.CO/10014538].
	
	\bibitem{observations}
  D.~N.~Spergel {\it et al.}  [WMAP Collaboration],
  Astrophys.\ J.\ Suppl.\  {\bf 170}, 377 (2007).
  [astro-ph/0603449];
  D.~N.~Spergel {\it et al.}  [WMAP Collaboration],
  Astrophys.\ J.\ Suppl.\  {\bf 148}, 175 (2003),
  [astro-ph/0302209];
H. V. Peiris {\em et al.}, Astrophys. J. Suppl. {\bf 148}, 213 (2003), 
[astro-ph/0302225]. 


\bibitem{inflation}
A.~A.~Starobinsky, Phys.\ Lett. {\bf 91B}, 99 (1980);
A.~H.~Guth, Phys.\ Rev.\ D {\bf 23}, 347 (1981);
A.~Albrecht, P.~J.~Steinhardt, Phys. Rev. Lett. {\bf 48}, 
1220 (1982);
S.~W.~Hawking, I.~G.~Moss, Phys.\ Lett.\ {\bf 110B}, 
35 (1982);
A.~D.~Linde, Phys.\ Lett.\ {\bf 108B}, 389 (1982);
A.~D.~Linde, Phys.\ Lett.\ {\bf 129B}, 177 (1983). 		

\bibitem{Maldacena:2002vr}
  J.~M.~Maldacena,
  JHEP {\bf 0305}, 013 (2003),
  [astro-ph/0210603].
		
		
\bibitem{Linde:1993cn}
  A.~D.~Linde,
  Phys.\ Rev.\  {\bf D49}, 748-754 (1994),
  [astro-ph/9307002].
			
\bibitem{higgsINF}
  J.~L.~Cervantes-Cota, H.~Dehnen,
  Nucl.\ Phys.\ {\bf B442}, 391-412 (1995);
  M.~Atkins, X.~Calmet,
  Phys.\ Lett.\ B {\bf 697} (2011) 37,
  [hep-ph/10114179];
  F.~L.~Bezrukov, A.~Magnin, M.~Shaposhnikov,
  Phys.\ Lett.\ B {\bf 675} (2009) 88,
  [hep-ph/08124950].
	
	\bibitem{Wald} 
  R.~M.~Wald,
  `\emph{General Relativity}',
  Chicago, Usa: Univ. Pr. ( 1984).
	
  \bibitem{FRLW}
  A.~Friedman,
  Zeitschrift für Physik A 10: 377–386 (1922);
  G.~Lema$\hat{\rm i}$tre, 
  Annales de la Société Scientifique de Bruxelles {\bf A47}: 49–56 (1927);
  H.~P.~Robertson,
  Astrophysical Journal {\bf 82}: 284-301 (1935);
  A.~G.~Walker,
  Proceeding of the London Mathematical Society 2 {\bf 42}: 91-127 (1937).
  
  \bibitem{Alpher:1948ve}
  R.~A.~Alpher, H.~Bethe, G.~Gamowk,
  Phys.\ Rev.\  {\bf 73} (1948) 803.
	
	 \bibitem{nucleosyn}
  D.~D.~Clayton, W.~A.~Fowler, T.~Hull, B.~Zimmerman,
  Ann.\ Phys.\ {\bf 12}, 331-408 (1961).
	
	\bibitem{reviews}
J.~E.~Lidsey, A.~R.~Liddle, E.~W.~Kolb, E.~J.~Copeland, T.~Barreiro, M.~Abney,
Rev.\ Mod.\ Phys.\  {\bf 69}, 373-410 (1997),
[astro-ph/9508078];
D.~H.~Lyth, A.~Riotto,
Phys.\ Rep. {\bf 314}, 1 (1999),
[hep-ph/9807278];
  B.~A.~Bassett, S.~Tsujikawa, D.~Wands,
  Rev.\ Mod.\ Phys.\  {\bf 78}, 537-589 (2006),
  [astro-ph/0507632].
	
	\bibitem{Linde:1983gd}
  A.~D.~Linde,
  Phys.\ Lett.\ B {\bf 129} (1983) 177.
	
	\bibitem{eternal}
  A.~D.~Linde,
  Phys.\ Scripta {\bf T15}, 169 (1987);
    A.~D.~Linde,
  Phys.\ Lett.\  {\bf B175}, 395-400 (1986).

	
	\bibitem{Lyth:2009zz}
  D.~H.~Lyth, A.~R.~Liddle,
  '\emph{The primordial density perturbation: Cosmology, inflation and the origin of structure}',
  Cambridge, UK: Cambridge Univ. Pr. (2009).
	
	
\bibitem{Linde:2005ht}
  A.~D.~Linde,
 `\emph{Particle physics and inflationary cosmology}',
  Contemp.\ Concepts Phys.\  {\bf 5} (1990),
  [hep-th/0503203].
	
	\bibitem{Gibbons:1976ue}
  G.~W.~Gibbons, S.~W.~Hawking,
  Phys.\ Rev.\ D {\bf 15} (1977) 2752.
	
	\bibitem{Gibbons:1977mu}
  G.~W.~Gibbons, S.~W.~Hawking,
  Phys.\ Rev.\ D {\bf 15} (1977) 2738.
	
	\bibitem{Bunch:1978yq}
  T.~S.~Bunch, P.~C.~W.~Davies,
  Proc.\ Roy.\ Soc.\ Lond.\ A {\bf 360} (1978).
	
	\bibitem{Peskin:1995ev}
  M.~E.~Peskin, D.~V.~Schroeder,
 `\emph{An Introduction to quantum field theory}',
  Reading, USA: Addison-Wesley (1995).
	
 \bibitem{liddlelyth}
  A.~R.~Liddle, D.~H.~Lyth,
  '\emph{Cosmological Inflation and Large-Scale Structure}',
 Cambridge, UK: Cambridge Univ. Pr. (2000). 
	
	  \bibitem{Mukhanov:1990me}
  V.~F.~Mukhanov, H.~A.~Feldman, R.~H.~Brandenberger,
  Phys.\ Rept.\  {\bf 215} (1992) 203.

 
\bibitem{Weinberg:2008zzc}
  S.~Weinberg,
  `\emph{Cosmology}',
  Oxford, UK: Oxford Univ. Pr. (2008)
	
	\bibitem{zetaConv}
 D.~H.~Lyth, K.~A.~Malik, M.~Sasaki,
  JCAP {\bf 0505}, 004 (2005),
  [astro-ph/0411220];
    G.~I.~Rigopoulos, E.~P.~S.~Shellard,
  Phys.\ Rev.\  {\bf D68}, 123518 (2003),
  [astro-ph/0306620];
  D.~Langlois, F.~Vernizzi,
  Phys.\ Rev.\  D {\bf 72}, 103501 (2005),
  [astro-ph/0509078].
	
	\bibitem{Starobinsky:1986fxa}
  A.~A.~Starobinsky,
  JETP Lett.\  {\bf 42}, 152-155 (1985).

 \bibitem{Sasaki:1995aw}
  M.~Sasaki, E.~D.~Stewart,
  Prog.\ Theor.\ Phys.\  {\bf 95}, 71-78 (1996),
  [astro-ph/9507001].
	
	\bibitem{largeNG1b}
 C.~T.~Byrnes, K.~-Y.~Choi, L.~M.~H.~Hall,  
JCAP {\bf 0902}, 017 (2009),
 [astro-ph/08120807]. 
	
	
	\bibitem{Lyth:2005fi}
  D.~H.~Lyth, Y.~Rodriguez,
  Phys.\ Rev.\ Lett.\  {\bf 95}, 121302 (2005),
  [astro-ph/0504045].
	
	\bibitem{Seery:2008qj}
  D.~Seery, K.~A.~Malik, D.~H.~Lyth,
  JCAP {\bf 0803} (2008) 014,
  [astro-ph/08020588].

\bibitem{otherMethods}
   G.~I.~Rigopoulos, E.~P.~S.~Shellard, B.~J.~W.~van Tent,
  Phys.\ Rev.\  {\bf D73}, 083521 (2006),
  [astro-ph/0504508];
 G.~I.~Rigopoulos, E.~P.~S.~Shellard, B.~J.~W.~van Tent,
  Phys.\ Rev.\  {\bf D76}, 083512 (2007),
  [astro-ph/0511041];
  S.~Yokoyama, T.~Suyama, T.~Tanaka,
  JCAP {\bf 0707}, 013 (2007),
  [astro-ph/07053178];
  S.~Yokoyama, T.~Suyama, T.~Tanaka,
  JCAP {\bf 0707}, 013 (2007), 
  [astro-ph/07053178];
 S.~Yokoyama, T.~Suyama, T.~Tanaka,
  Phys.\ Rev.\  D {\bf 77}, 083511 (2008), 
  [astro-ph/07112920];
D.~J.~Mulryne, D.~Seery, D.~Wesley,
  JCAP {\bf 1104}, 030 (2011),
  [astro-ph.CO/10083159];
  D.~J.~Mulryne, D.~Seery, D.~Wesley,
  JCAP {\bf 1001}, 024 (2010),
  [astro-ph.CO/09092256].
	
	\bibitem{Komatsu:2000vy}
  E.~Komatsu, D.~N.~Spergel,
  [astro-ph/0012197].
	
	\bibitem{Creminelli:2006rz} 
  P.~Creminelli, L.~Senatore, M.~Zaldarriaga, M.~Tegmark,
  JCAP {\bf 0703}, 005 (2007),
  [astro-ph/0610600].
	
	\bibitem{Senatore:2009gt} 
  L.~Senatore, K.~M.~Smith, M.~Zaldarriaga,
  JCAP {\bf 1001}, 028 (2010),
  [arXiv:09053746].
	
	\bibitem{Elliston:2011dr}
  J.~Elliston, D.~J.~Mulryne, D.~Seery, R.~Tavakol,
  [astro-ph.CO/11062153];
  J.~Elliston, D.~Mulryne, D.~Seery, R.~Tavakol,
  [astro-ph.CO/11072270].
	
	
	\bibitem{Leung:2013rza}
  G.~Leung, E.~R.~M.~Tarrant, C.~T.~Byrnes, E.~J.~Copeland,
  [astro-ph.CO/13034678].
  G.~Leung, E.~R.~M.~Tarrant, C.~T.~Byrnes, E.~J.~Copeland,
  JCAP {\bf 1209}, 008 (2012),
  [astro-ph.CO/12065196].
	
	\bibitem{consist} 
  P.~Creminelli, M.~Zaldarriaga,
  JCAP {\bf 0410}, 006 (2004),
  [astro-ph/0407059];
  C.~Cheung, A.~L.~Fitzpatrick, J.~Kaplan, L.~Senatore,
  JCAP {\bf 0802}, 021 (2008),
  [hep-th/07090295].
	
	\bibitem{SUSY}
	P.~Binetruy, J.~M.~Cline, C.~Grojean,
  Phys.\ Lett.\ B {\bf 489}, 403 (2000),
  [hep-th/0007029];
	E.~J.~Copeland, A.~R.~Liddle, D.~H.~Lyth, E.~D.~Stewart, D.~Wands,
  Phys.\ Rev.\ D {\bf 49}, 6410 (1994),
  [astro-ph/9401011].
	
	  \bibitem{dmarg}
 D.~Mulryne, D.~Seery, D.~Wesley,
  [astro-ph.CO/09113550].
	
	\bibitem{Copeland:2002ku}
  E.~J.~Copeland, S.~Pascoli, A.~Rajantie,
  Phys.\ Rev.\  {\bf D65}, 103517 (2002),
  [hep-ph/0202031].

  \bibitem{Kibble:1976sj}
  T.~W.~B.~Kibble,
  J.\ Phys.\ A {\bf 9} (1976) 1387.

\bibitem{topDef}
 E.~.P.~S.~Shellard, A.~Vilenkin,
 '\emph{Cosmc Strings and Other Topological Defects}',
 Cambridge Monographs on Mathematical Physics, Cambridge, (2000).

\bibitem{minSUSY}
  G.~R.~Dvali, Q.~Shafi, R.~K.~Schaefer,
  Phys.\ Rev.\ Lett.\  {\bf 73} (1994) 1886,
  [hep-ph/9406319];
  G.~R.~Dvali, Q.~Shafi, R.~K.~Schaefer,
  Phys.\ Rev.\ Lett.\  {\bf 73} (1994) 1886,
  [hep-ph/9406319].
 
 \bibitem{Contaldi:1998qs}
  C.~Contaldi, M.~Hindmarsh, J.~Magueijo,
  Phys.\ Rev.\ Lett.\  {\bf 82} (1999) 2034,
  [astro-ph/9809053].
	\bibitem{GarciaBellido:1996qt}
  J.~Garcia-Bellido, A.~D.~Linde, D.~Wands,
  Phys.\ Rev.\  {\bf D54}, 6040-6058 (1996),
  [astro-ph/9605094].
	
	 \bibitem{bholes}
   H.~Firouzjahi, A.~Green, K.~Malik, M.~Zarei,
  [astro-ph.CO/12092652];
  I.~Zaballa, A.~M.~Green, K.~A.~Malik, M.~Sasaki,
  JCAP {\bf 0703} (2007) 010,
  [astro-ph/0612379];
  D.~H.~Lyth, K.~A.~Malik, M.~Sasaki, I.~Zaballa,
  JCAP {\bf 0601} (2006) 011,
  [astro-ph/0510647];
   A.~M.~Green, K.~A.~Malik,
  Phys.\ Rev.\ D {\bf 64} (2001) 021301,
  [hep-ph/0008113].
  
  \bibitem{lythBH}
  D.~H.~Lyth,
[astro-ph.CO/11071681];
  D.~H.~Lyth,
  JCAP {\bf 1205} (2012) 022,
  [astro-ph.CO/12014312].
	
	\bibitem{Amin:2011hj}
  M.~A.~Amin, R.~Easther, H.~Finkel, R.~Flauger, M.~P.~Hertzberg,
  Phys.\ Rev.\ Lett.\  {\bf 108} (2012) 241302,
  [astro-ph.CO/11063335].
  
  \bibitem{Broadhead:2003ni}
  M.~Broadhead, J.~McDonald,
  Phys.\ Rev.\ D {\bf 69} (2004) 063510,
  [hep-ph/0309298].
  

	
	
	\bibitem{Clesse:2010iz}
  S.~Clesse,
  Phys.\ Rev.\  {\bf D83}, 063518 (2011),
  [gr-qc/10064522].
  
	
	\bibitem{Clesse:2013jra} 
  S.~Clesse, B.~Garbrecht, Y.~Zhu,
  [astro-ph.CO/13047042].
	
	\bibitem{Vernizzi:2006ve}
  F.~Vernizzi, D.~Wands,
  JCAP {\bf 0605}, 019 (2006),
  [astro-ph/0603799].

	\bibitem{GarciaBellido:1995qq}
  J.~Garcia-Bellido, D.~Wands,
  Phys.\ Rev.\  {\bf D53}, 5437-5445 (1996),
  [astro-ph/9511029].
	

	\bibitem{Kim:2010ud}
  S.~A.~Kim, A.~R.~Liddle, D.~Seery,
  Phys.\ Rev.\ Lett.\  {\bf 105}, 181302 (2010),
  [astro-ph.CO/10054410].

	\bibitem{curv}
  S.~Mollerach,
  Phys.\ Rev.\  D {\bf 42} (1990) 313;
  A.~D.~Linde, V.~F.~Mukhanov,
  Phys.\ Rev.\  D {\bf 56}, 535 (1997), 
  [astro-ph/9610219];
  D.~H.~Lyth, D.~Wands,
  Phys.\ Lett.\ B {\bf 524}, 5 (2002). 
  [hep-ph/0110002];
  T.~Moroi, T.~Takahashi,
  Phys.\ Lett.\ B {\bf 522}, 215 (2001),
  [Erratum-ibid.\ B {\bf 539}, 303 (2002)],
  [hep-ph/0110096];
  K.~Enqvist, S.~Nurmi,
  JCAP {\bf 0510}, 013 (2005),
  [astro-ph/0508573];
  A.~Linde, V.~Mukhanov,
  JCAP {\bf 0604}, 009 (2006),
  [astro-ph/0511736];
  K.~A.~Malik, D.~H.~Lyth,
  JCAP {\bf 0609}, 008 (2006),
  [astro-ph/0604387];
  M.~Sasaki, J.~Valiviita, D.~Wands,
  Phys.\ Rev.\  D {\bf 74}, 103003 (2006),
  [astro-ph/0607627];
  A.~Chambers, S.~Nurmi, A.~Rajantie,
  JCAP {\bf 1001}, 012 (2010),
  [astro-ph.CO/09094535].

	
	\bibitem{barnaby}
  N.~Barnaby, J.~M.~Cline,
  Phys.\ Rev.\  {\bf D73}, 106012 (2006),
  [astro-ph/0601481];
 N.~Barnaby, J.~M.~Cline,
  Phys.\ Rev.\  {\bf D75}, 086004 (2007),
  [astro-ph/0611750].

   \bibitem{Abolhasani:2011yp} 
A.~A.~Abolhasani, H.~Firouzjahi, M.~Sasaki,
  [astro-ph.CO/11066315];
A.~A.~Abolhasani, H.~Firouzjahi, M.~H.~Namjoo,
[astro-ph.CO/10106292].

\bibitem{largeNG1}
  L.~Alabidi,
  JCAP {\bf 0610}, 015 (2006),
  [astro-ph/0604611].
	
	\bibitem{inhom2}
M.~Sasaki,
  Prog.\ Theor.\ Phys.\  {\bf 120}, 159 (2008),
  [astro-ph/08050974];
  A.~Naruko, M.~Sasaki,
  Prog.\ Theor.\ Phys.\  {\bf 121}, 193 (2009),
  [astro-ph/08070180].
	

 \bibitem{largeNG2}
C.~T.~Byrnes, K.~-Y.~Choi, L.~M.~H.~Hall,
 JCAP {\bf 0810}, 008 (2008),
  [astro-ph/08071101];
C.~M.~Peterson, M.~Tegmark,
    [astro-ph.CO/10116675].
		
		\bibitem{Bethke:2013aba} 
  L.~Bethke, D.~G.~Figueroa and A.~Rajantie,
   [astro-ph.CO/13042657].
	
	\bibitem{preh}
  A.~Chambers, A.~Rajantie,
  Phys.\ Rev.\ Lett.\  {\bf 100}, 041302 (2008),
  [astro-ph/07104133];
  A.~Chambers, A.~Rajantie,
  JCAP {\bf 0808 } (2008)  002,
  [astro-ph/08054795];
  J.~R.~Bond, A.~V.~Frolov, Z.~Huang, L.~Kofman,
  Phys.\ Rev.\ Lett.\  {\bf 103}, 071301 (2009).
  [astro-ph.CO/09033407].
		
		  \bibitem{Suyama:2013dqa}
  T.~Suyama, S.~Yokoyama,
  [astro-ph.CO/13031254].
 

\bibitem{SUSYinf}
  A.~Mazumdar, J.~Rocher,
  Phys.\ Rept.\  {\bf 497} (2011),
  [hep-ph/10010993].
	
	
  


\end{thebibliography}
\end{document}